\documentclass[twocolumn,twocolappendix]{aastex63}

\usepackage{amsmath} 
\usepackage[style=iso]{datetime2} 
\usepackage{booktabs} 
\usepackage{ragged2e} 
\usepackage{needspace} 
\usepackage[figuresleft]{rotating} 

\usepackage{color} 
\definecolor{green}{rgb}{0.0,0.725,0.024}
\definecolor{red}{rgb}{1.0,0.0,0.0}
\definecolor{RED}{rgb}{1.0,0.0,0.0}

\newcommand{\hersc}{{\it Herschel}}
\newcommand{\planck}{{\it Planck}}

\newcommand{\sne}{supernov\ae}

\newcommand{\HI}{H{\sc i}}

\newcommand{\dustbff}{{\texttt{DustBFF}}}
\newcommand{\SigmaDeproj}{$\Sigma_{H}^{\it(deproj)}$}

\renewcommand{\th}{\textsuperscript{th}}

\defcitealias{CJRClark2021A}{Paper\,I} 

\renewcommand{\baselinestretch}{0.925} 

\received{\DTMDate{2022-05-27}}
\revised{\DTMDate{2022-12-03}}
\accepted{\DTMDate{2023-01-20}}
\submitjournal{(and accepted for publication in) The Astrophysical Journal}

\makeatletter 
\newcommand\footnoteref[1]{\protected@xdef\@thefnmark{\ref{#1}}\@footnotemark}
\makeatother

\shorttitle{The Quest for the Missing Dust: II}
\shortauthors{Clark et al.}

\begin{document}

\title{The Quest for the Missing Dust: II -- Two Orders of Magnitude of Evolution in the Dust-to-Gas Ratio Resolved Within Local Group Galaxies}

\correspondingauthor{Christopher J. R. Clark}
\email{cclark@stsci.edu}

\author[0000-0001-7959-4902]{Christopher J. R. Clark}
\affiliation{Space Telescope Science Institute, 3700 San Martin Drive, Baltimore, Maryland 21218-2463, United States of America}

\author[0000-0001-6326-7069]{Julia C. Roman-Duval}
\affiliation{Space Telescope Science Institute, 3700 San Martin Drive, Baltimore, Maryland 21218-2463, United States of America}

\author[0000-0001-5340-6774]{Karl D. Gordon}
\affiliation{Space Telescope Science Institute, 3700 San Martin Drive, Baltimore, Maryland 21218-2463, United States of America}

\author[0000-0001-6118-2985]{Caroline Bot}
\affiliation{Observatoire Astronomique de Strasbourg, Universit\'e de Strasbourg, UMR 7550, 11 rue de l’Universit\'e, F-67000 Strasbourg, France}

\author[0000-0002-3532-6970]{Matthew W. L. Smith}
\affiliation{School of Physics and Astronomy, Cardiff University, Queen's Buildings, The Parade, Cardiff, Wales, CF24 3AA, United Kingdom}

\author[0000-0001-8918-1597]{Lea M. Z. Hagen}



\begin{abstract}

We explore evolution in the dust-to-gas ratio with density within four well-resolved Local Group galaxies -- the LMC, SMC, M\,31, and M\,33. We do this using new \hersc\ maps, which restore extended emission that was missed by previous \hersc\ reductions. This improved data allows us to probe the dust-to-gas ratio across 2.5 orders of magnitude in ISM surface density. We find significant evolution in the dust-to-gas ratio, with dust-to-gas varying with density within each galaxy by up to a factor 22.4. We explore several possible reasons for this, and our favored explanation is dust grain growth in denser regions of ISM. We find that the evolution of the dust-to-gas ratio with ISM surface density is very similar between M\,31 and M\,33, despite their large differences in mass, metallicity, and star formation rate; conversely, we find M\,33 and the LMC to have very different dust-to-gas evolution profiles, despite their close similarity in those properties. Our dust-to-gas ratios address previous disagreement between UV- and FIR-based dust-to-gas estimates for the Magellanic Clouds, removing the disagreement for the LMC, and considerably reducing it for the SMC -- with our new dust-to-gas measurements being factors of 2.4 and 2.0 greater than the previous far-infrared estimates, respectively. We also observe that the dust-to-gas ratio appears to fall at the highest densities for the LMC, M\,31, and M\,33; this is unlikely to be an actual physical phenomenon, and we posit that it may be due to a combined effect of dark gas, and changing dust mass opacity.

\end{abstract}



\section{Introduction} \label{Section:Introduction}

\begin{figure*}
\centering
\includegraphics[width=0.895\textwidth]{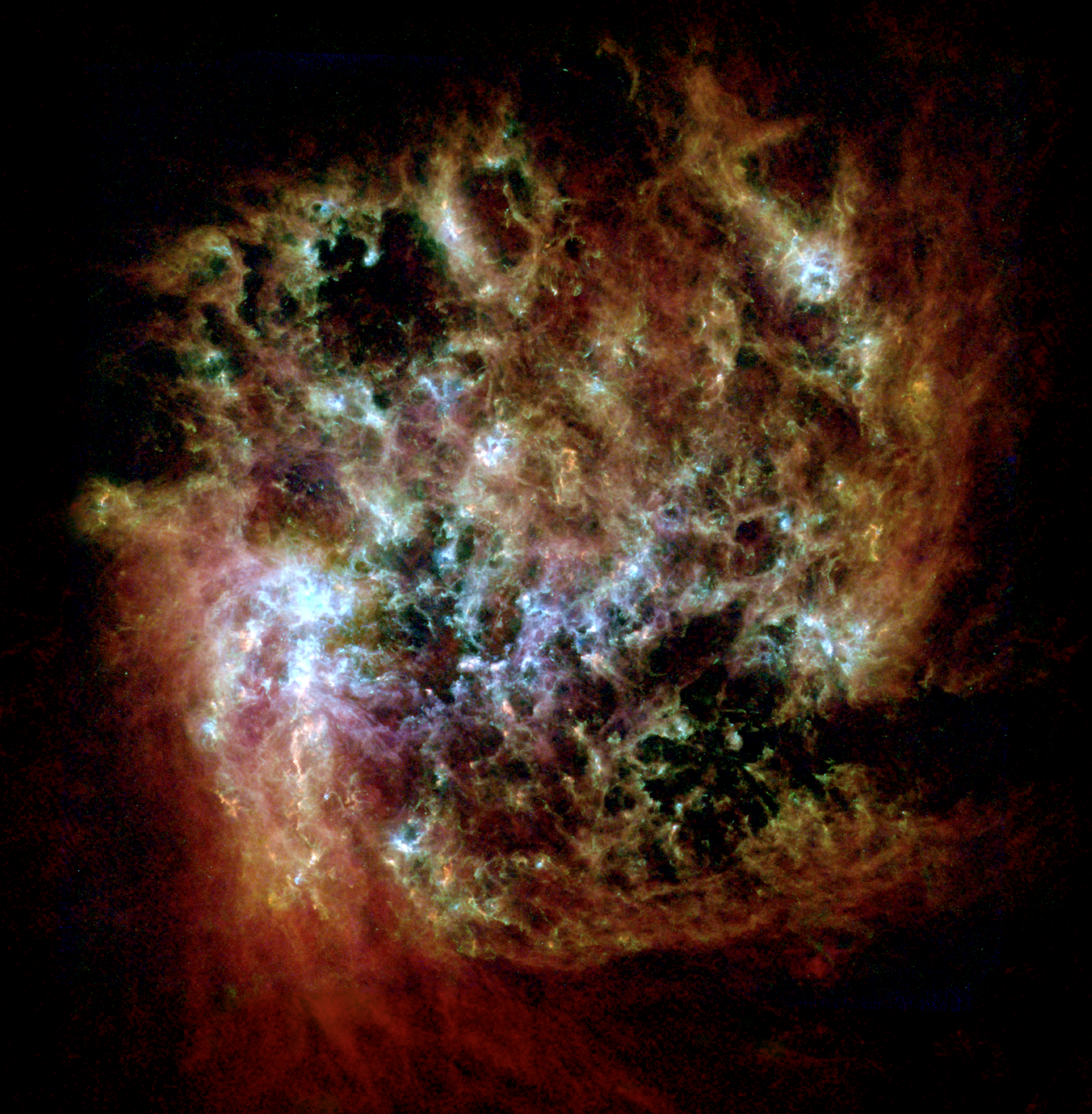}\\
\vspace{0.5cm}
\includegraphics[width=0.275\textwidth]{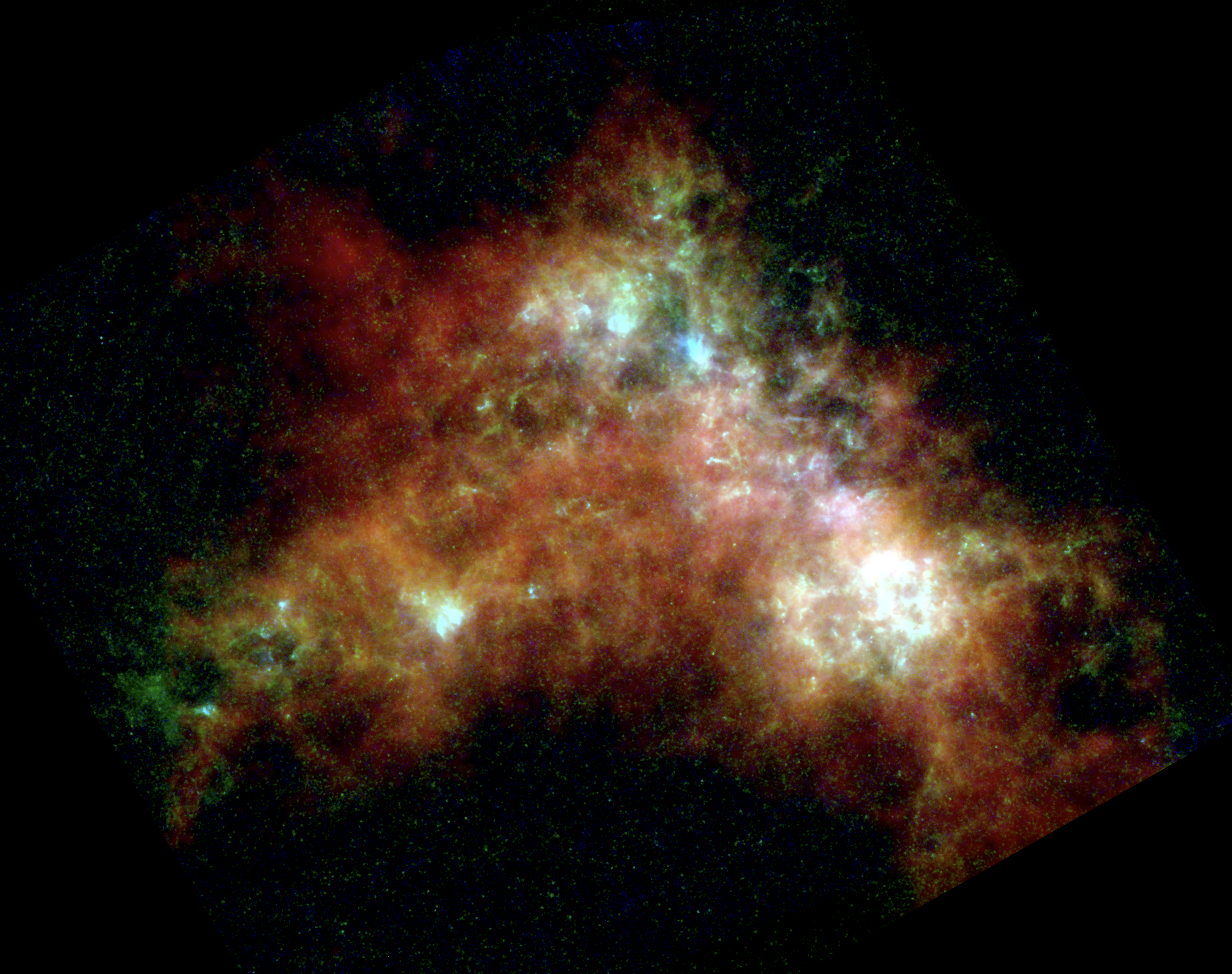}
\includegraphics[width=0.025\textwidth]{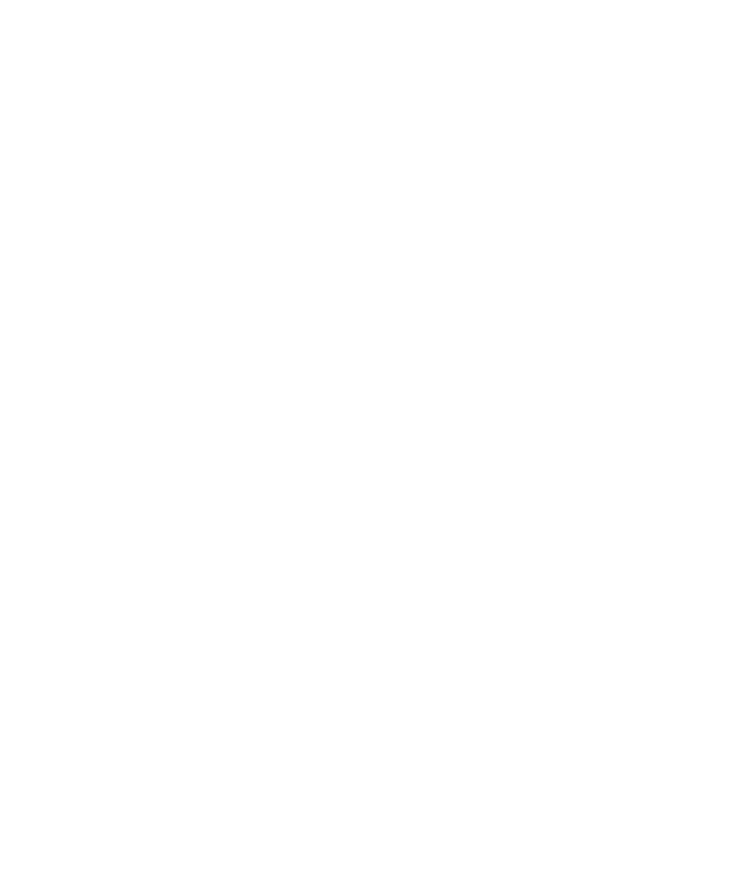}
\includegraphics[width=0.275\textwidth]{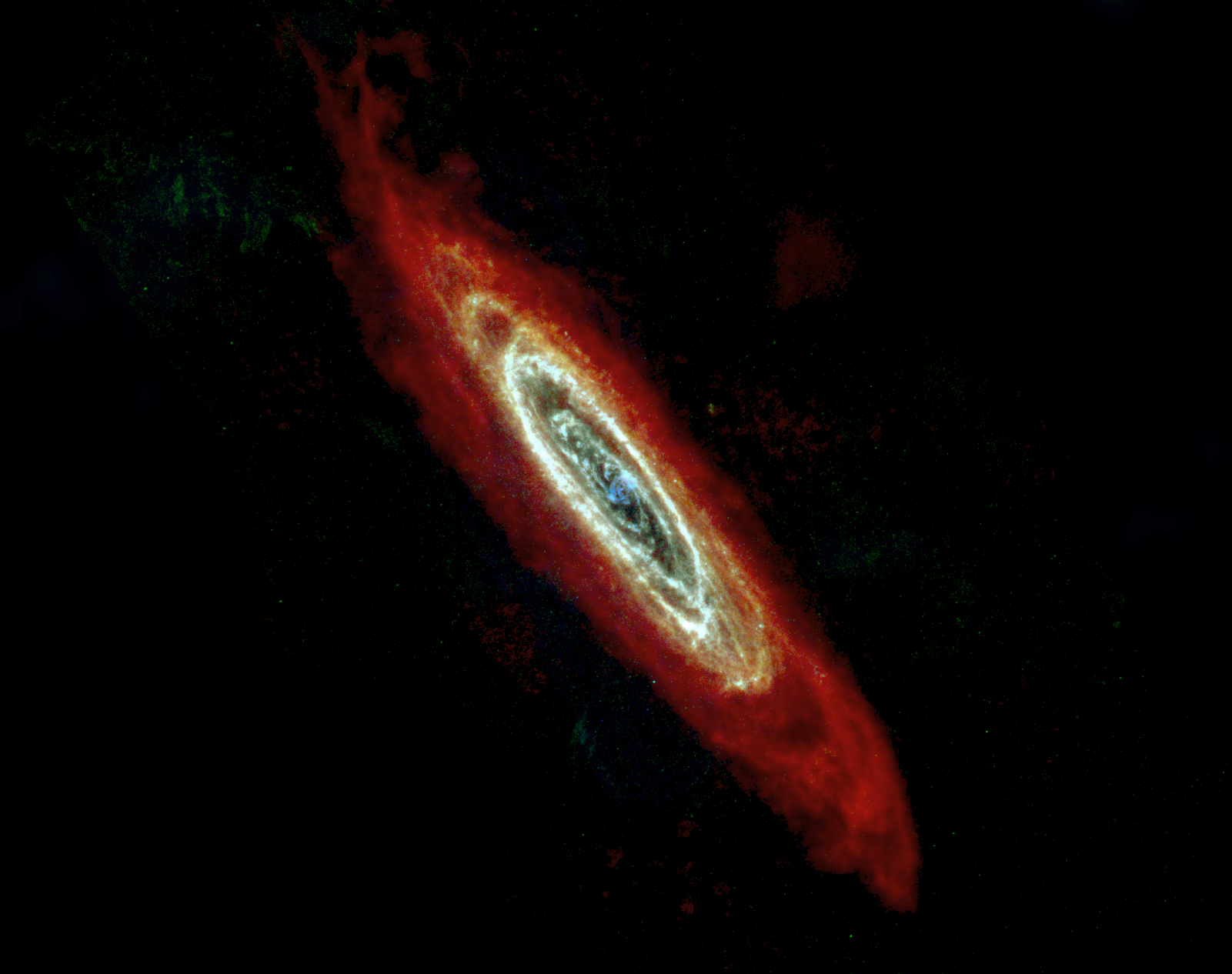}
\includegraphics[width=0.025\textwidth]{Blank.png}
\includegraphics[width=0.275\textwidth]{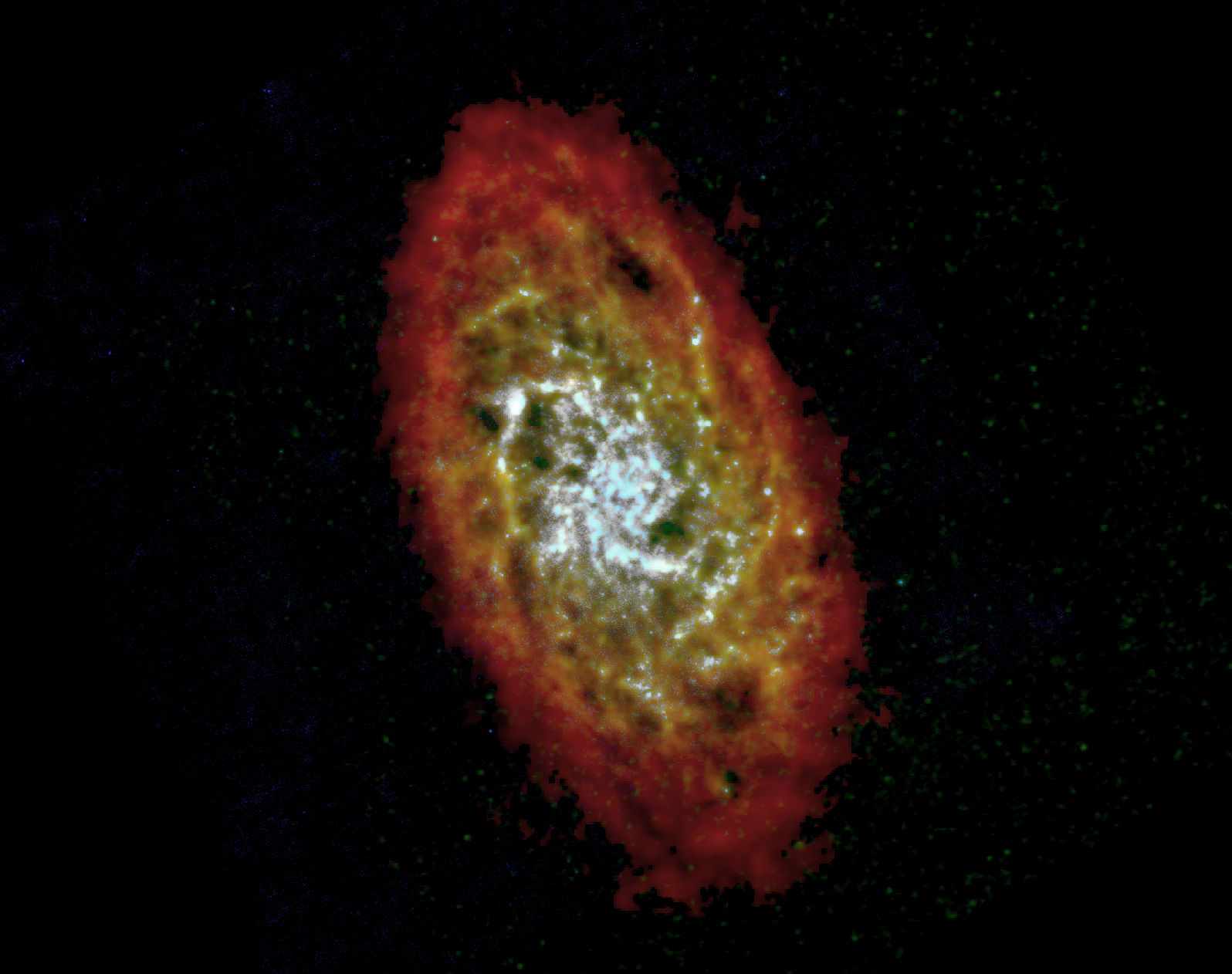}
\caption{Three-color images of the galaxies in our target sample, showing hydrogen gas in red (as traced by 21\,cm and CO emission), \hersc-SPIRE 350\,\micron\ cooler dust emission in green, and \hersc-PACS 100\,\micron\ warmer dust emission in blue. The \hersc\ images are foreground-subtracted maps; see Section~\ref{Section:Sample_and_Data}. The LMC, the best-resolved galaxy of the sample, is shown at the top. The lower row shows the SMC ({\it left}), M\,31 ({\it center}), and M\,33 ({\it right}). Each channel of each image uses a logarithmic color scale that displays the map's structure over the its full value range. Simple visual inspection of the changing colors within these images make clear that there is significant evolution in the dust and gas properties in each of our sample galaxies.}
\label{Fig:Glamour_RGB}
\end{figure*}

The relationship between dust and gas in the InterStellar Medium (ISM) is complex. There are various mechanisms by which dust is added to the ISM, and others which remove dust from it.

For instance, newly-created dust grains can be added to the ISM via the deaths of stars -- either by core-collapse \sne\ \citep{Barlow2010,Matsuura2011E,Gomez2012B}, or by asymptotic giant branch stars \citep{Hofner2018A}. The dust mass of the ISM can also increase in situ, through dust grains accreting gas-phase metals in higher-density environments \citep{Kohler2015A,Zhukovska2016A,Jones2017A}. Similarly, dust can be removed from the ISM in a number of different ways. Dust can undergo photo-destruction by high-energy photons, especially in poorly-shielded low-density environments; or dust can be sputtered by forward \& reverse \sne\ shocks (\citealp{Jones2004B,Bocchio2014B,Slavin2015C}; making the status of \sne\ as net sources or sinks of dust uncertain). 

The key metric for assessing the dust-richness of the ISM is the dust-to-gas ratio (D/G). The various processes that add or remove dust from the ISM will change D/G, as will other processes, such as the preferential removal of the dust-richest ISM via the formation of new stars and planets \citep{Hjorth2014A,Forgan2017G}, and the dilution of the ISM by the infall of low-metallicity extragalactic gas \citep{Edmunds1998} -- both of which will drive down D/G.

Clearly, the balance between dust creation and destruction in galaxies, and therefore D/G, is sensitive to many factors. And unavoidably, the relative importance of these factors will vary greatly between different environments; both between, and within, galaxies. The balance between dust creation and destruction is especially important in light of the `dust budget crisis', arising from the fact that observed sources of dust appear insufficient to account for observed dust masses in some galaxies \citep{Matsuura2009B}, particularly at high redshift \citep{Rowlands2014B}, unless significant grain-growth occurs in the ISM \citep{DeVis2017B,Galliano2021B}. 

Strong evidence for grain growth comes from the fact that the D/G ratio is observed to increase in higher density environments, even when controlling for metallicity. This has been seen in spectroscopic absorption measurements of the depletion of various metals from the gas phase \citep{Jenkins2009B,Tchernyshyov2015B,Roman-Duval2021B}, in measurements of extinction curve evolution \citep{Gordon2003B,Fitzpatrick2005C}, and in resolved Far-InfraRed (FIR) observations of the dust emission within nearby galaxies \citep{Mattsson2014,Roman-Duval2014D,Roman-Duval2017B}. 

It should be noted that work by \citet{Nanni2020C} finds that grain growth is not necessarily required to explain the chemical evolution of low-metallicity and high-$z$ galaxies, if the initial mass function is top-heavy, and if outflow rates are high. Similarly, work by \citet{DeLooze2020A} indicates that low-redshift dust masses could be explained primarily by dust from stellar sources, without the need for efficient interstellar grain-growth, if dust lifetimes are long (1--2\,Gyr) and if the survival fraction of fresh supernova dust passing through supernov\ae\ reverse shocks is high (37--89\%). However, even this framework requires that 20--50\% of the present-day dust mass of galaxies has grown in the ISM. These works highlight the necessity of constraining the role and relative importance of grain growth in galaxies' chemical evolution.

In recent years, one of the key avenues for understanding the chemical evolution of galaxies has been the relationship between D/G and metallicity\footnote{This relationship is plotted later, in Figure~\ref{Fig:DtH_vs_Z}, to display results presented in subsequent sections of this work.} (see Figure~9 in \citealp{Galliano2018C}; see also \citealp{Remy-Ruyer2014A} and \citealp{DeVis2019B}). Understanding how D/G evolves with metallicity is especially important with respect to the `critical metallicity' – the knee in the D/G\,--$Z$ relationship, at approximately 0.2\,${\rm Z_{\odot}}$ \citep{Remy-Ruyer2014A,DeVis2019B}. Above the critical metallicity, D/G increases linearly with $Z$, while below it, D/G drops sharply. A linear trend in D/G\,--$Z$ is expected in a regime where the dust-to-metals ratio remains constant; or in other words, where a constant fraction of metals is locked up in dust grains. 

The critical metallicity is generally understood to reflect a threshold above which ISM grain-growth becomes efficient, and starts to dominate over less-efficient stellar sources of dust, and is expected from theoretical models \citep{Asano2013,Feldmann2015C,Zhukovska2016A}. Understanding the critical metallicity has important ramifications for understanding the ISM in general, particularly given the role of dust as the formation site of molecular hydrogen \citep{Gould1963}, as a cooling pathway during star formation \citep{Dopcke2011A}, and in shielding CO from photodissociation -- thereby dictating the CO-to-H$_{2}$ conversion needed for using CO as a tracer of molecular gas \citep{Wolfire2010A,Clark2015A}.

There are, however, discrepancies in our current understanding of the D/G\,--$Z$ relationship. The critical metallicity break is only observed for (mostly nearby) galaxies for which dust mass measurements are derived from FIR observations, and for which gas masses are measured via \HI\ and CO observations \citep{Remy-Ruyer2014A,DeVis2019B}. In contrast, the critical metallicity break is {\it not} observed for galaxies for which D/G is derived from depletions, specifically in Damped Lyman Alpha systems (DLAs)\footnote{DLAs are neutral gas absorption systems with $N_{\it H{\sc I}} > 2 \times 10^{20}\,{\rm cm^{-2}}$, found over a wide range of redshifts, and observed via quasar absorption spectroscopy \citep{Wolfe2005C,DeCia2016A}.}. Rather, for DLAs, the D/G\,--$Z$ trend remains linear, over the entire $0.01 < Z < 1\,{\rm Z_{\odot}}$ range for which measurements exist \citep{Galliano2018C,Roman-Duval2022B}, with no critical metallicity break. Previously, this linear trend in D/G\,--$Z$ for DLAs was found to overlap the trend for galaxies with D/G determined via FIR observations; however, recent recalibration indicates that the DLA trend is actually offset to lower D/G \citep{Roman-Duval2022A}.

Even within the Local Group (the largest members of which are shown in Figure~\ref{Fig:Glamour_RGB}, excepting the Milky Way), problems remain in the D/G\,--$Z$ relationship. Specifically, the D/G estimates for the Large Magellanic Cloud (LMC) and Small Magellanic Cloud (SMC) derived from FIR observations are considerably lower than D/G values derived from UV absorption line spectroscopy measurements of depletions \citep{Roman-Duval2017B}. The FIR D/G values are smaller than the UV D/G values by a factor of $\sim$2 for the LMC, and a factor of $\sim$5 for the SMC. These offsets between D/G are persistent over the full range of gas densities studied within both Magellanic Clouds to date \citep{Roman-Duval2017B,Roman-Duval2022A}.

Are these D/G discrepancies in the Magellanic Clouds due to previous FIR measurements being affected by temperature mixing, or other systematic problems? Or are the D/G measurements from the UV depletions in error, meaning that the lower D/G values from the FIR might be tracing the beginning of the critical metallicity transition? The LMC and SMC have metallicities of 0.5 and 0.2\,${\rm Z_{\odot}}$ respectively, so the SMC in particular might be in the critical metallicity regime.

There are challenges that need to be addressed when making D/G measurements with FIR data. If the physical resolution of FIR observations is not good enough, it will systematically bias dust mass estimates due to beam smearing. High-density regions get blurred into the lower-density regions that surround them, and temperature mixing can cause higher-luminosity emission from warm dust to dominate the FIR Spectral Energy Distribution (SED) over fainter emission from greater masses of cold dust. Additionally, limited physical resolution also diminishes the range of ISM densities that can be sampled. Moreover, galaxies sufficiently nearby to overcome such resolution limits (like those in the Local Group) often suffer from large-angular-scale emission being filtered out of FIR data during the reduction process, systematically biasing the data against properly sampling dust in diffuse environments, where ISM densities will be lowest. Fortunately, by using high-resolution data and the latest reduction techniques, it is possible to tackle these issues.


The specifics of these obstacles, and how we are able to overcome them for the sample of Local Group galaxies we study in this work, are discussed fully in Section~\ref{Section:Sample_and_Data}. In Section~\ref{Section:SED_Fitting}, we describe our pixel-by-pixel SED fitting, and the dust results obtained from it. In Section~\ref{Section:DtG_Ratio}, we examine D/G in our target galaxies, and how it evolves with ISM density. In Section~\ref{Section:Causes_of_Turnover}, we investigate the surprising turnover in D/G exhibited by some of the galaxies in our sample. In Section~\ref{Section:DtH_Reconcile}, we discuss how the FIR-derived D/G measurements from this (and previous) work compares to UV-derived D/H measurements.  In Section~\ref{Section:Data_Products}, we provide specific details of the various maps being publicly released alongside this paper.

Note that in this paper, we specifically consider D/G in terms of the dust-to-hydrogen ratio, D/H. We do this for several reasons. Firstly, hydrogen mass (or density) is the actual observable quantity. Frequently, authors will use a measured hydrogen mass to infer the total gas mass, by applying a correction factor to account for the faction of the gas mass made up of elements heavier than hydrogen. However, this fails to consider the fact that the mass fraction of elements heaver than hydrogen evolves as a function of ISM metallicity \citep{Balser2006D} -- varying from 1.33 (for low-metallicity galaxies where $Z$$\to$0) to 1.45 (for high-metallicity giant ellipticals where $Z = 1.5Z_{\odot}$). Moreover, this evolution with metallicity is not purely monotonic, and will have intrinsic scatter due to differences in nucleosynthesis from different stellar populations. Directly working with D/H allows us to sidestep the question of this correction, and the ambiguity it can cause. Additionally, later in the paper we compare our dust-to-gas measurements to values calculated from depletions, where the observed hydrogen column is always used as the denominator; so here, once again, use of D/H allows for more clarity.


\needspace{2\baselineskip} \section{Sample \& Data} \label{Section:Sample_and_Data}

\begin{table}
\centering
\caption{Basic properties of the Local Group galaxies studied in this work. Values taken from the NASA/IPAC Extragalactic Database, except where otherwise noted. Portions of this table are reproduced from Table~1 of \citetalias{CJRClark2021A}, provided again here for convenience.}
\label{Table:Galaxy_Properties}
\begin{tabular}{lrrrr}
\toprule \toprule
\multicolumn{1}{c}{} &
\multicolumn{1}{c}{M\,31} &
\multicolumn{1}{c}{M\,33} &
\multicolumn{1}{c}{LMC} &
\multicolumn{1}{c}{SMC} \\
\cmidrule(lr){2-5}
$\alpha$ (J2000) & 10.69\degr & 23.46\degr & 80.89\degr & 13.16\degr \\
$\delta$ (J2000) & +41.27\degr &  +30.66\degr & -69.76\degr & -72.80\degr \\
Distance (kpc) & 790 & 840 & 50 & 62 \\
Hubble Type & SAb & SAcd & SBm & Irr \\
$R_{25}$ & 89\arcmin & 32\arcmin & 323\arcmin & 151\arcmin \\
$R_{25}$ (kpc) & 20.5 & 7.5 & 5.0 & 2.5 \\
$^{\rm a}$Pos. Angle & 35\degr & 23\degr & 170\degr & 45\degr \\
$^{\rm a}$Axial Ratio & 2.57 & 1.70 & 1.17 & 1.66 \\
Inclination & 77\degr & 56\degr & 26\degr & -- \\
$M_{\star}$ (${\rm M_{\odot}}$) & $^{\rm b}10^{11.1}$  & $^{\rm c}10^{9.7}$ & $^{\rm d}10^{9.4}$ & $^{\rm e}10^{8.3}$  \\
$^{\rm f}Z$ (${\rm Z_{\odot}}$) & $^{\rm g}$1.3 & $^{\rm h}$0.5 & $^{\rm i}$0.5 & $^{\rm i}$0.2 \\
\bottomrule
\end{tabular}
\footnotesize
\justify
$^{\rm a}$ From the HyperLEDA database \citep{Makarov2014A}. \quad \\
\mbox{$^{\rm b}$ \citet{Tamm2012A}} \quad \\
$^{\rm c}$ \citet{Corbelli2014B} \quad \\
$^{\rm d}$ \citet{VanDerMarel2006A} \quad \\
$^{\rm e}$ \citet{Rubele2018A} \quad \\
$^{\rm f}$ All metallicities are from `direct' method determinations using auroral lines in the ISM. For M\,31 and M\,33, we quote the `characteristic metallicity', as measured at 0.4$R_{25}$ \citep{Pilyugin2004C,Moustakas2010B}. \quad \\
\mbox{$^{\rm g}$ \citet{Zurita2012A}}; we have corrected for the -0.3\,dex offset they report due to the low-metallicity bias in which H{\sc ii} regions have detectable auroral lines in M\,31.  \quad  \\
$^{\rm h}$ \citet{Magrini2016C} \quad \\
$^{\rm i}$ \citet{ToribioSanCipriano2017B}
\end{table}

Physical resolution is key for investigating D/H, and other ISM properties. Superior physical resolution translates into superior {\it density} resolution. With better resolution, compact higher-density regions of ISM will appear less `smeared' together with surrounding lower-density material, allowing a wider range of environments to be probed. Achieving the best physical and density resolution therefore requires observing galaxies that are suitably nearby, using telescopes that have sufficiently high angular resolution.

Physical resolution is especially important for observations of dust emission. Attempts to measure dust masses (and dust mass surface densities) of nearby galaxies generally suffer from the problem that multiple dust components, each with different temperatures (and other properties), will be confused together in the observed FIR SED. This will bias the dust properties inferred from that SED; specifically, this temperature mixing tends to cause dust masses to be underestimated \citep{Galliano2011B,Priestley2020B}. The best way to overcome temperature mixing is to observe galaxies using sufficient resolution that the dust in any given resolution element is mostly homogeneous. In practice, \citet{Galliano2011B} found that this bias becomes minimal only when physical resolution is better than $\sim$150\,pc. Specifically, \citet{Galliano2011B} found that increasing physical resolution leads to increased dust masses from SED fitting, with an asymptote at $\approx$\,20\,pc (above this, improving the resolution leads to no further change in modeled mass). However, physical resolution better than $\approx$\,150\,pc results in dust masses for which the error is \textless10\% divergent from masses obtained at the 20\,pc asymptote; see Figure~6 in \citet{Galliano2011B}. 

These requirements for high spatial resolution mean that we are effectively limited to observations of galaxies in the Local Group, if we want to obtain D/H estimates free of temperature mixing bias.

\needspace{2\baselineskip} \subsection{Sample Galaxies} \label{Subsection:Sample_Galaxies}

\begin{figure}
\centering
\includegraphics[width=0.475\textwidth]{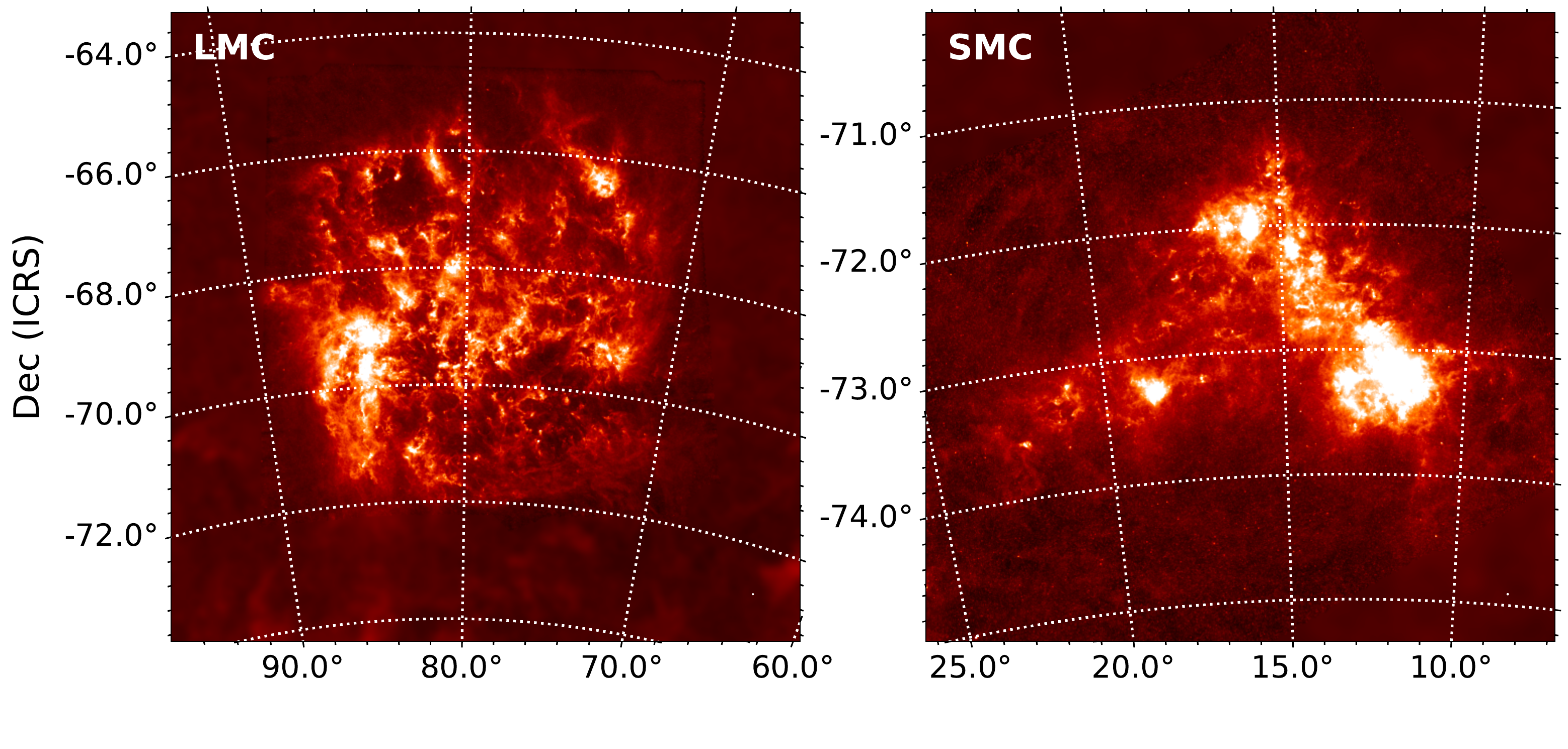}
\includegraphics[width=0.475\textwidth]{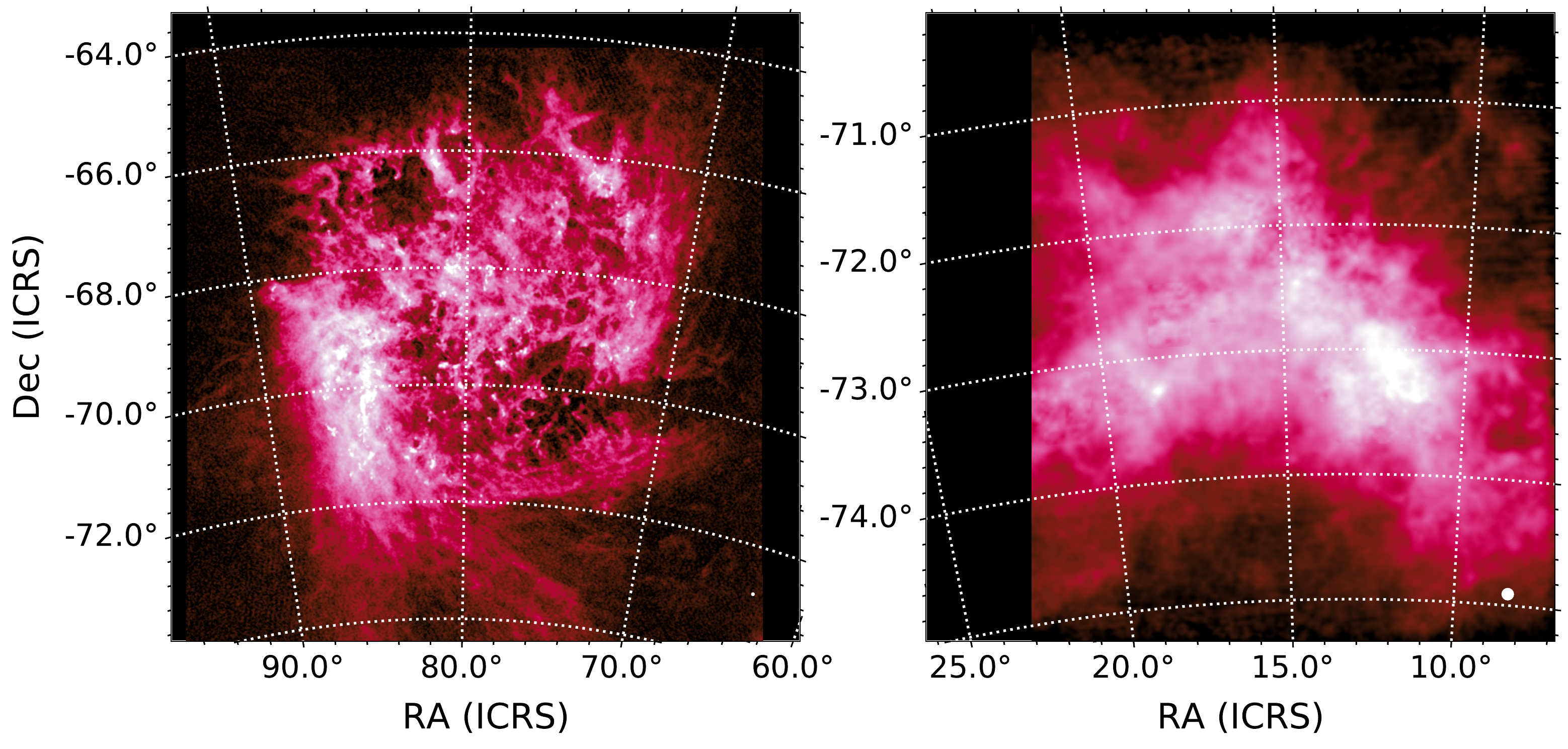}
\caption{{\it Upper:} Feathered \hersc-SPIRE 250\,\micron\ maps of the LMC ({\it left}) and SMC ({\it right}), as produced by \citetalias{CJRClark2021A}; these images show the foreground-subtracted versions of the maps. {\it Lower:} $\Sigma_{H}$ maps for the LMC ({\it left}) and SMC ({\it right}), made by combining atomic and molecular gas surface density maps produced from \HI\ and CO data, respectively; these maps are shown at the limiting resolutions at which we conduct our analyses (see Section~\ref{Subsection:Limiting Resolution}; FWHM shown by white circles in lower right corners).}
\label{Fig:LMC+SMC_Overview}
\end{figure}

For our investigation into the resolved properties of D/H in nearby galaxies, our sample consists of the four very well-resolved galaxies of the Local Group -- the Large Magellanic Cloud (LMC), Small Magellanic Cloud (SMC), M\,31 (the Andromeda Galaxy), and M\,33 (the Triangulum Galaxy). 

These galaxies exhibit a broad range of characteristics. The SMC is a 0.2\,Z$_{\odot}$ \citep{ToribioSanCipriano2017B} dwarf galaxy exhibiting the distinct ISM traits that are characteristic of low-metallicity systems, such as: the absence of the 2175\,\textup{~\AA} extinction bump along most (but not all) sightlines \citep{Gordon2003B,Murray2019C}; being extremely irregular, having likely been disturbed by a recent collision with the LMC \citep{VanDerMarel2001D,Murray2019C,Choi2022A}; and being highly elongated along our line-of-sight \citep{Scowcroft2016B}. The LMC has a metallicity 0.5\,Z$_{\odot}$ \citep{ToribioSanCipriano2017B}, and is on the spiral / dwarf-irregular transition; it is experiencing intense star formation, especially on its eastern edge, where gas from the SMC appears to be infalling \citep{Bekki2007L,Fukui2017B,Tsuge2019A}. M\,33 has a similar 0.5\,Z$_{\odot}$ metallicity to the LMC \citep{Koning2015A,Magrini2016C}, but has almost twice the stellar mass \citep{Corbelli2014B,VanDerMarel2006A}, and a much more orderly spiral structure; it is an interacting companion of M\,31 \citep{Bekki2008H,Putman2009D}, and has the highest star-forming efficiency in the Local Group \citep{Gardan2007A}. M\,31 is a high-mass, $L^{\ast}$, spiral/ring galaxy \citep{Gordon2006A} that reaches super-Solar metallicity \citep{Zurita2012A}, and is currently passing through the green valley \citep{Mutch2011B}. 

This wide range of properties gives excellent scope to examine what traits may affect D/H. The sample is illustrated in Figure~\ref{Fig:Glamour_RGB}, where the variation in the relative abundance of dust and gas can be clearly seen from visual inspection alone. All four galaxies have sufficiently low radial velocities ($-300<v<300\,{\rm km\,s^{-1}}$) that redshift/violetshift is entirely negligible, allowing us to safely treat all emission as being rest-frame.

\needspace{2\baselineskip} \subsection{New Herschel Data} \label{Subsection:New_Herschel_Data}

The natural choice of telescope for observing dust in Local Group galaxies is the \hersc\ Space Observatory \citep{Pilbratt2010D}. \hersc\ provides exquisite sensitivity (able to reach the extragalactic confusion limit in most bands), over a 100--500\,\micron\ wavelength range which samples the vast majority of dust emission, with a 36\arcsec\ limiting resolution at 500\,\micron\ that corresponds to a physical scale of \textless\,150\,pc in M\,33 (the most distant of the Local Group's highly extended galaxies), and to \textless\,9\,pc in the LMC. 

However, FIR observations of Local Group targets come with their unique complications. Because the galaxies of the Local Group are so extended, they are vulnerable to the well-known problem of emission on large angular scales being filtered out of data produced by scanning detector arrays across the sky, as \hersc\ does \citep{Meixner2013A,Roussel2013A,MWLSmith2017A,MWLSmith2021A}. The diffuse emission lost because of this effect will naturally tend to correspond to the diffuse, lower-density dust in a galaxy. This will systematically bias any analysis concerning ISM density.

All-sky FIR surveys, such as those by the InfraRed Astronomical Satellite (IRAS; \citealp{Neugebauer1984}) and \planck\ \citep{Planck2011I}, are free from the filtering of large scale emission that affects \hersc. However, the resolution of IRAS and \planck\ is a factor of $\sim$10 worse than \hersc, severely limiting their ability to resolve the densest portions of the ISM and overcome the temperature mixing problem. Plus, these facilities provide sparser coverage of the dust SED than \hersc. These factors severely limit their use in studying the ISM in nearby galaxies.


Clearly, answering the open questions about D/H will require reliable \hersc\ observations of Local Group galaxies. Therefore, in the first paper of this series, \citet[][hereafter \citetalias{CJRClark2021A}]{CJRClark2021A}, we produced new versions of the \hersc\ maps for the Local Group galaxies M\,31 (Andromeda), M\,33 (Triangulum), the LMC, and the SMC. These new maps were produced by combining \hersc\ observations, in Fourier space, with data from lower-resolution FIR telescopes that did not suffer from filtering, to restore the missing large-scale emission. These new maps also incorporate significant calibration corrections (up to 30\% in some cases).

The new maps from \citetalias{CJRClark2021A} cover two \hersc\ Photodetector Array Camera and Spectrometer (PACS; \citealp{Poglitsch2010B}) bands, at 100 and 160\,\micron, and all three \hersc\ Spectral and Photometric Imaging REceiver (SPIRE; \citealp{Griffin2010D}) bands, at 250, 350, and 500\,\micron. As described in Section~6 of \citetalias{CJRClark2021A}, we also produced Galactic-foreground-subtracted versions of the new maps; it is these that we use throughout this paper. 

To illustrate the feathered FIR data, \hersc\ 250\,\micron\ maps for each galaxy are shown in the upper panels of Figure~\ref{Fig:LMC+SMC_Overview} for the LMC and SMC, and Figure~\ref{Fig:M31+M33_Overview} for M\,31, and M\,33. 

\needspace{2\baselineskip} \subsection{Molecular \& Atomic Gas Data} \label{Subsection:Gas_Data}

\begin{figure}
\centering
\includegraphics[width=0.475\textwidth]{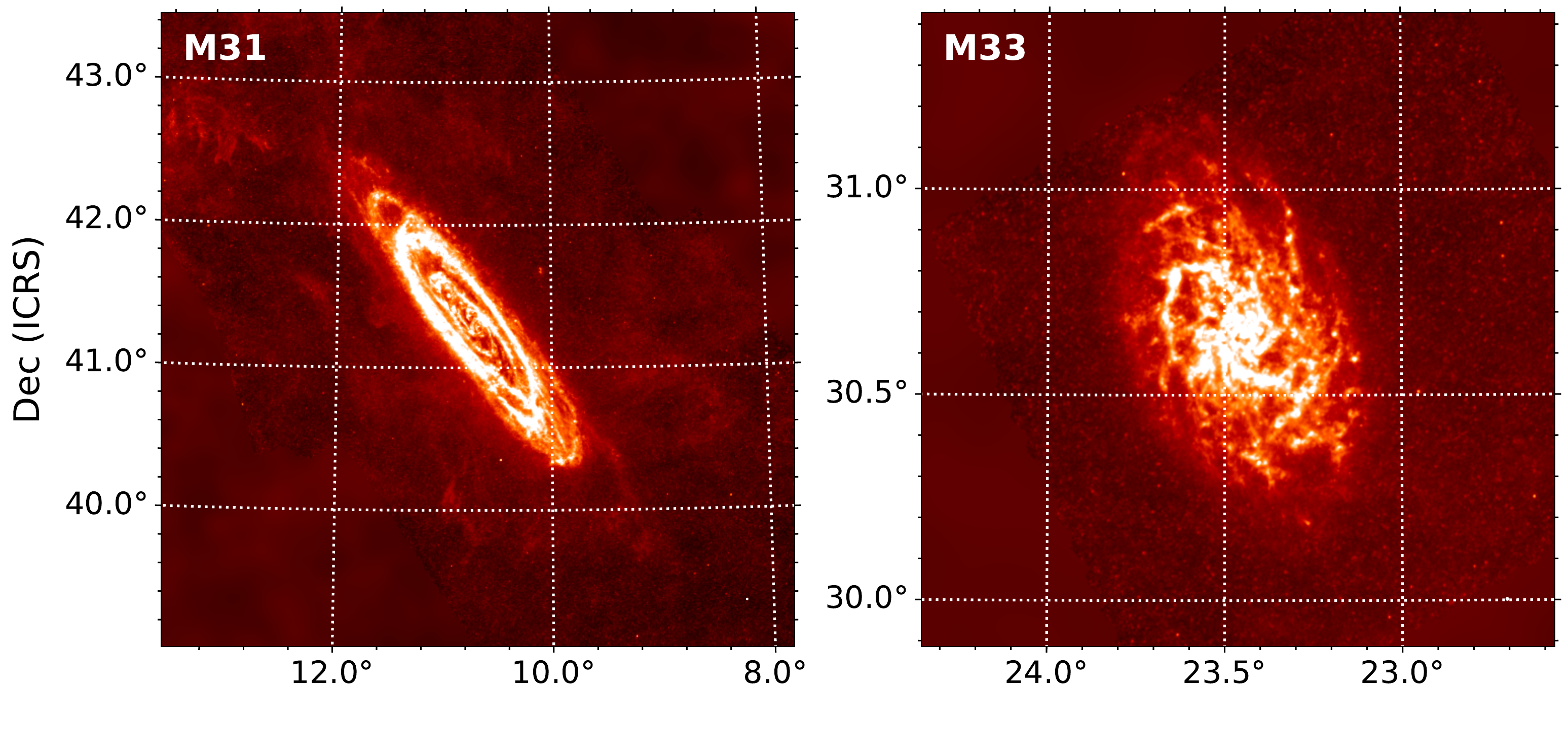}
\includegraphics[width=0.475\textwidth]{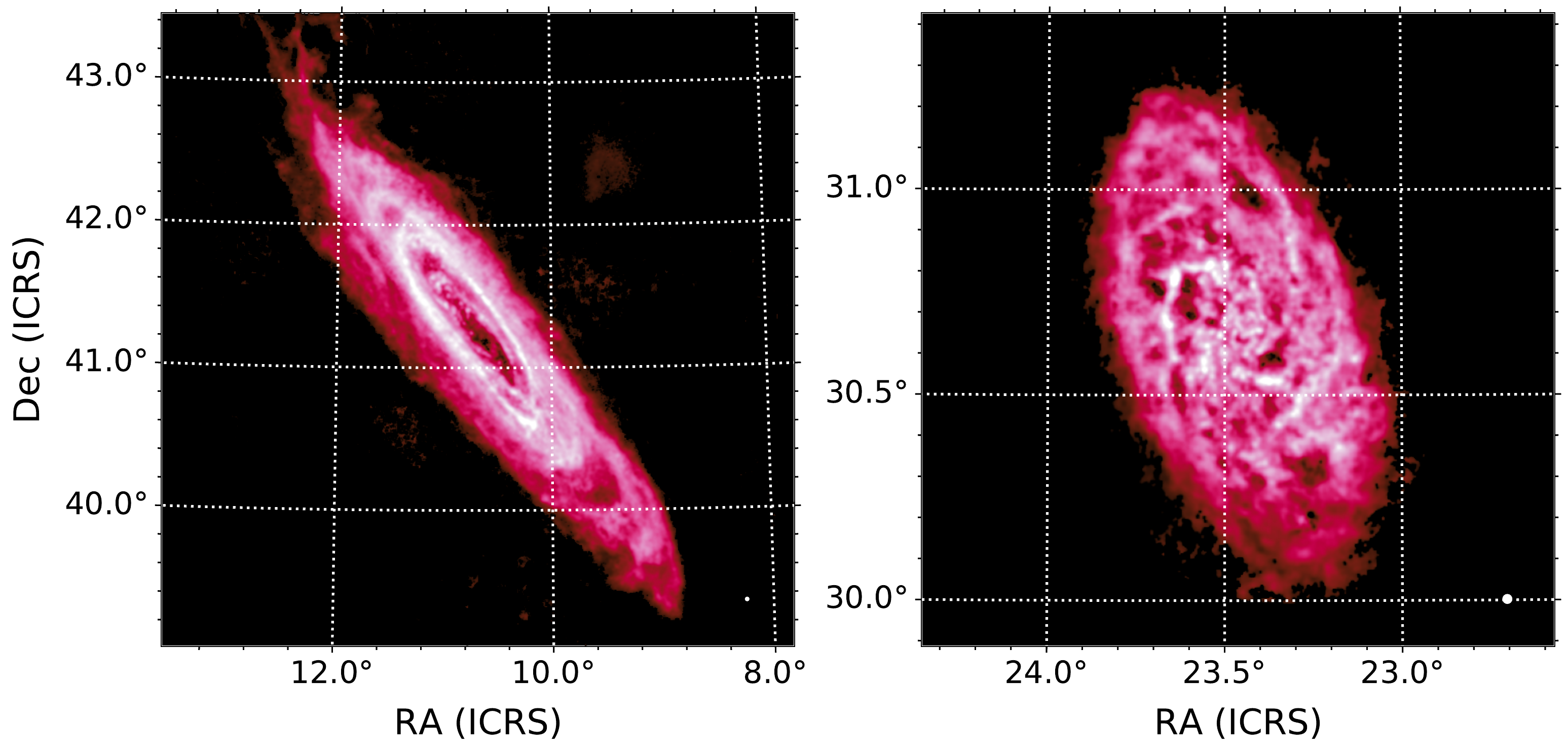}
\caption{Same as for Figure~\ref{Fig:LMC+SMC_Overview}, but depicting M\,31 ({\it left}) and M\,33 ({\it right}).}
\label{Fig:M31+M33_Overview}
\end{figure}

For the gas component of our target galaxies' ISM, we created maps of the hydrogen surface density, $\Sigma_{H}$, by adding together maps of the atomic and molecular hydrogen, produced from \HI\ 21\,cm hyperfine line and $^{12}$C$^{16}$O(1-0) rotational line (hereafter CO) observations, respectively. In this work, we use the same $\Sigma_{H}$ maps as described in Section~7.1 of \citetalias{CJRClark2021A}. Here we briefly recap the input data, and how the $\Sigma_{H}$ maps were produced.

The data we used to create the maps of $\Sigma_H$ was: for the LMC, the \HI\ data of \citet{Kim2003A} and CO data of \citet{Wong2011C}; for the SMC, the \HI\ data of \citet{Stanimirovic1999A} and the CO data of \citet{Mizuno2001C}; for M\,31 the \HI\ data of \citet{Braun2009A} and the CO data of \citet{Nieten2006A}; and for M\,33,the \HI\ data of \citet{Koch2018C} and the CO map of \citet{Gratier2010C,Druard2014A}. The atomic and molecular gas emission on all scales should be captured by these observations; the CO data is all from single-dish observations, and the \HI\ data is a feathered combination of high-resolution interferometric observations and low-resolution single-dish observations. Our handling of limiting resolutions is discussed in Section~\ref{Subsection:Limiting Resolution}

We calculated the molecular gas surface density using the standard relation $\Sigma_{\rm H2}= \alpha_{\rm CO} I_{\rm CO(1-0)}$, where $I_{\rm CO(1-0)}$ is the CO(1-0) line velocity-integrated main-beam brightness temperature (in ${\rm K\,km\,s^{-1}}$), and $\alpha_{\rm CO}$ is the CO-to-${\rm H_{2}}$ conversion factor (in ${\rm K^{-1}\,km^{-1}\,s\,M_{\odot}\,pc^{-2}}$). For the spirals M\,31 and M\,33, we used the standard Milky Way $\alpha_{\rm CO}$ value of $3.2\,{\rm K^{-1}\,km^{-1}\,s\,pc^{-2}}$; for the LMC and SMC dwarf galaxies, we used higher $\alpha_{\rm CO}$ values of 6.4 and 21 ${\rm K^{-1}\,km^{-1}\,s\,pc^{-2}}$, respectively \citep{Bolatto2013B}. 

{As described in Section\,7.1 of \citetalias{CJRClark2021A}, some authors prefer an $\alpha_{\rm CO}$ for M\,33 that is higher than the Milky Way value \citep{Bigiel2010D,Druard2014A}. However, \citet{Gratier2010C} find a value within 10\% of that of the Milky Way. Plus \citeauthor{Rosolowsky2003B} (\citeyear{Rosolowsky2003B} -- who do not measure an an absolute $\alpha_{\rm CO}$, only its apparent relative variation) find that $\alpha_{\rm CO}$ in the highest-metallicity regions of M\,33, at \textgreater\,1.4\,Z$_{\odot}$, is not systematically different from that in regions at \textless\,0.5\,Z$_{\odot}$; given that $\alpha_{\rm CO}$ at such high metallicities should be comparable to that of the Milky Way, it suggests this is not significantly different elsewhere in M\,33. Regardless, Appendix~\ref{AppendixSubsection:Differences_in_alpha-CO} shows that our primary results are mostly insensitive to changes in $\alpha_{\rm CO}$.}

To add together a galaxy's maps of atomic and molecular hydrogen, to create the combined $\Sigma_{H}$ map, we first convolved the higher-resolution map to the resolution of the lower-resolution map, assuming a Gaussian PSF, as per the beam size information in the header of each map.

Our $\Sigma_{H}$ maps for each galaxy are shown in the lower panels of Figure~\ref{Fig:LMC+SMC_Overview} for the LMC and SMC, and Figure~\ref{Fig:M31+M33_Overview} for M\,31, and M\,33. 

\needspace{2\baselineskip} \subsection{Convolution to Limiting Resolution} \label{Subsection:Limiting Resolution}

For all analyses, the \hersc\ and $\Sigma_{H}$ maps were convolved to a common resolution, to match whatever the worst resolution was amongst all the data for each galaxy. We created conversion kernels for this with the \texttt{Python} package \texttt{photutils}. For the input \hersc\ PSFs, we used the azimuthally-averaged beams from \citet{Aniano2011A}\footnote{\label{Footnote:Aniano}\url{https://www.astro.princeton.edu/~draine/Kernels.html}}. For the input $\Sigma_{H}$ PSFs, we assumed Gaussian beams, as per Section~\ref{Subsection:Gas_Data}.

For M\,31 and M\,33, the limiting resolution was the 36\arcsec\ of the 500\,\micron\ maps; for the LMC and SMC, the limiting resolutions were dictated by the gas observations (see Section~\ref{Subsection:Gas_Data} above), being 1\arcmin\ for the LMC, and 2.6\arcmin\ for the SMC. We also re-bin all maps to use a pixel width equal to that galaxy's limiting resolution, so that every pixel is statistically independent.

These limiting angular resolutions correspond to {\it physical} resolutions of 14\,pc for the LMC, 47\,pc for the SMC, 137\,pc for M\,31, and 147\,pc for M\,33.

\needspace{2\baselineskip} \section{SED Fitting} \label{Section:SED_Fitting}

To constrain the dust properties of our target galaxies, while also taking full advantage of the resolution provided by our data, we carried out pixel-by-pixel SED-fitting of the FIR maps. For this, we applied a Modified BlackBody (MBB) model; specifically, a Broken-Emissivity Modified BlackBody (BEMBB) model \citep{Gordon2014B}.

A MBB is essentially the simplest model that can be used to reliably fit FIR dust emission. It takes the following form, where the surface brightness, $S(\lambda)$, of dust emission at a given wavelength, $\lambda$, can be expressed by:

\begin{equation}
S(\lambda) = \kappa(\lambda)\ B(\lambda, T_{d})\ \Sigma_{d}
\label{Equation:SED_Surface_Brightness}
\end{equation}

\noindent where $B$ is the Planck function evaluated at wavelength $\lambda$ and dust temperature $T_{d}$, $\Sigma_{d}$ is the dust mass surface density, and $\kappa(\lambda)$ is the dust mass absorption coefficient at wavelength $\lambda$; $\kappa(\lambda)$ varies with wavelength according to an emissivity law:

\begin{equation}
\kappa(\lambda) = \frac{\kappa(\lambda_{\it ref})}{\lambda_{\it ref}^{-\beta}}\lambda^{-\beta}
\label{Equation:SED_MBB}
\end{equation}

\noindent where $\kappa(\lambda_{\it ref})$ is the value of $\kappa(\lambda)$ at a reference wavelength $\lambda_{\it ref}$, and $\beta$ is the dust emissivity spectral index.

The MBB model is less directly physically motivated than dust grain models based on the radiative transfer and optical properties of different potential dust species (eg, \citealp{Draine2014A,Jones2017A,DeLooze2019A}). However, these various dust grain models are not currently able to incorporate submillimeter (submm) excess, wherein more emission is observed in the Rayleigh-Jeans regime than would be otherwise expected from models, with the excess increasing towards longer wavelengths. The phenomena of submm excess is observed primarily in dwarf galaxies, such as the Magellanic Clouds \citep{Galliano2003A,Bot2010A,Remy-Ruyer2013A,Gordon2014B}, but is also seen in lower-density regions of more massive galaxies, including M\,33 \citep{Paradis2012B,Relano2018A}.

It is therefore clearly preferable that we use an SED model that is able to incorporate submm excess emission\footnote{As a test, we repeated the main analyses we present later in this work, but using a single MBB model instead of BEMBB. We found that all of the trends in D/H we discuss later in this work also appear when using single MBB fitting, albeit with much worse quality fits in the lower-density and lower-metallicity environments.}. The BEMBB model achieves this by having the value of $\beta$ change at some break wavelength $\lambda_{\it break}$; a shallower $\beta$ at longer wavelengths captures the excess emission. As such, the BEMBB emissivity law takes the form:

\begin{equation}
\kappa(\lambda) = \frac{\kappa(\lambda_{\it ref})}{\lambda_{\it ref}^{-\beta}} E(\lambda)
\label{Equation:SED_BEMBB}
\end{equation}

\noindent where

\begin{equation}
E(\lambda) =
\begin{cases}
\lambda^{-\beta_{1}} & \text{for}\ \lambda < \lambda_{break}\\
\lambda^{\beta_{2}-\beta{1}}\,\lambda^{-\beta_{1}} & \text{for}\ \lambda \geq \lambda_{break}
\end{cases}       
\label{Equation:SED_BEMBB_Break}
\end{equation}

\noindent for which $\beta$ takes a value of $\beta_{1}$ at wavelengths $< \lambda_{\it break}$, while it takes a value of $\beta_{2}$ at wavelengths $\geq \lambda_{\it break}$. 

\begin{table}
\centering
\caption{SED model grid parameter ranges and step sizes, used for pixel-by-pixel SED fitting with \dustbff, for our BEMBB model. For the logarithmically-spaced parameters $\Sigma_{d}$ and $T_{d}$, we also give the percentage difference between grid steps.}
\label{Table:DustBFF_Grid}
\begin{tabular}{lrrr}
\toprule \toprule
\multicolumn{1}{c}{Parameter} &
\multicolumn{1}{c}{Minimum} &
\multicolumn{1}{c}{Maximum} &
\multicolumn{1}{c}{Step} \\
\cmidrule(lr){1-4}
$\Sigma_{d}$ (${\rm M_{\odot}\,pc^{-2}}$) & $10^{-6}$ & $10^{1}$ & 0.05\,dex (12\%) \\
$T_{d}$ (K) & 10 & 70 & 0.04\,dex (5.9\%) \\
$\beta_{1}$ & 0 & 3.5 & 0.175 \\
$\lambda_{\it break}$ (\micron) & 125 & 525 & 33.3 \\
$e_{500}$ & -0.5 & 2.0 & 0.2 \\
\bottomrule
\end{tabular}
\footnotesize
\justify
\end{table}

We perform our SED fitting using the Dust Brute Force Fitter (\dustbff), a grid-based Bayesian SED-fitting code; \dustbff, and the mathematical formalism from which it {operates}, are presented in \citet{Gordon2014B}. We employ \dustbff\ in the same manner as in \citetalias{CJRClark2021A}, so we refer the reader there for a full description of our implementation; the only differences in this work are that we used a slightly different setup of the parameter grid (to account for the wider range of densities our \hersc\ data can probe), and that the data we fitted were of course the \hersc\ 100--500\,\micron\ bands (and we therefore used covariance matrices specific to these bands, given below). We adopt a value of the dust mass absorption coefficient, at a reference wavelength of 160\,\micron, of $\kappa_{160} = 1.24\,{\rm m^{2}\,kg^{-1}}$, following \citet{Roman-Duval2017B}; this allows for ease of comparison with their previous work investigating D/H in the Local Group.

The \dustbff\ parameter grid we use for all of our sample galaxies is given in Table~\ref{Table:DustBFF_Grid}. Note that BEMBB implementation in \dustbff\ does not parameterize $\beta_{2}$ directly. Rather, the break in $\beta$ is parameterized via the 500\,\micron\ excess, $e_{500}$, which gives the relative excess in the 500\,\micron\ flux, above what would arise from a standard MBB model (with negative values indicating a 500\,\micron\ deficit); $e_{500}$ is thus defined:

\begin{equation}
e_{500} = \left( \frac{\lambda_{\it break}}{\rm 500\,\mu m} \right)^{\beta_{2}-\beta_{1}} - 1
\label{Equation:DustBFF_e500}
\end{equation}

\dustbff\ uses a full covariance matrix in its model evaluations. This covariance matrix, $\mathcal{C}$, is given by:

\begin{equation}
\mathcal{C} = \mathcal{C_{\it calib}} + \mathcal{C_{\it instr}}
\label{Equation:DustBFF_Covariance}
\end{equation}

\noindent where $\mathcal{C_{\it calib}}$ is a matrix that contains the calibration uncertainty for each band, and $\mathcal{C_{\it instr}}$ is a matrix incorporating the effect of the map's instrumental noise on the probability for each model. 

The calibration covariance matrix $\mathcal{C_{\it calib}}$ is calculated by multiplying the proposed fluxes for a given model by the relative calibration uncertainty matrix $\mathcal{U}_{\it calib}$, which is itself given by summing $\mathcal{U}_{\it uncorr}$ and $\mathcal{U}_{\it corr}$, which are the matrices containing the fractional uncertainties that are uncorrelated and correlated between each band. 

The diagonal values of {the} instrumental noise covariance matrix, $\mathcal{C_{\it instr}}$, were calculated for each galaxy in each band, by taking the median pixel value of the uncertainty map of each, added in quadrature to the uncertainty on the foreground subtraction (see Section~6 of \citetalias{CJRClark2021A}). The uncertainty maps, as presented in \citetalias{CJRClark2021A}, propagate the uncertainty arising from the feathering process used to restore large-scale emission to the \hersc\ data, {along with other the instrumental noise contributions found in standard \hersc\ uncertainty maps}. The off-diagonals elements of $\mathcal{C_{\it instr}}$ are all zero. The diagonal elements of $\mathcal{C_{\it instr}}$ for each galaxy and band are given in Table~\ref{Table:Herschel_C_Instr}.

\begin{table}
\centering
\caption{Diagonal elements of $\mathcal{C_{\it instr}}$ instrumental noise covariance matrix for our SED fitting (off-diagonals are all zero). All values are in map units of MJy\,sr$^{-1}$.}
\label{Table:Herschel_C_Instr}
\begin{tabular}{lrrrrr}
\toprule \toprule
\multicolumn{1}{c}{Galaxy} &
\multicolumn{1}{c}{100\,\micron} &
\multicolumn{1}{c}{150\,\micron} &
\multicolumn{1}{c}{250\,\micron} &
\multicolumn{1}{c}{350\,\micron} &
\multicolumn{1}{c}{500\,\micron} \\
\cmidrule(lr){1-6}
LMC & 11.98 & 13.85 & 2.64 & 0.91 & 0.43 \\
SMC & 10.51 & 9.30 & 1.23 & 0.45 & 0.29 \\
M\,31 & 6.25 & 4.24 & 2.03 & 0.71 & 0.39 \\
M\,33 & 6.06 & 4.01 & 2.11 & 0.75 & 0.44 \\
\bottomrule
\end{tabular}
\end{table}

The uncorrelated uncertainty of a band essentially reflects the repeatability of its photometric measurements. The correlated uncertainties, on the other hand, arise from uncertainties in the photometric calibration. For instance, if all the bands of an instrument were calibrated using a certain stellar model, but that model is only constrained to within a certain percentage, then all of that instruments' bands will share an uncertainty of that percentage; whatever the actual underlying error is, it will be the same between bands. Not accounting for correlated uncertainties can cause severe biases in model results \citep{Galliano2011B,Kelly2012B,Veneziani2013B}. 

For the \hersc-PACS 100 and 160\,\micron\ bands, the overall calibration uncertainty is taken as 7\%; of this, we treat 2\% as being correlated between bands, as this is the upper end of the quoted uncertainty on the continuum model of the 5 late-type giant stars used as the \hersc-PACS photometric calibrator sources \citep{Balog2014B,Decin2007B}. We therefore assume the uncorrelated component is the quadrature subtraction of the correlated uncertainty from the total uncertainty, being $\sqrt{7\%^{2} - 2\%^{2}} = 6.7\%$. 

For the \hersc-SPIRE 250, 350, and 500\,\micron\ bands, we use an overall calibration uncertainty of 5.5\%, of which 4\% is taken to be correlated between bands \citep{Griffin2010D,Griffin2013A,Bendo2013A} -- ie, the \hersc-SPIRE photometric uncertainty is in fact {\it dominated} by the correlated component, highlighting the importance of correctly accounting for this effect. This gives an uncorrelated component of $\sqrt{5.5\%^{2} - 4\%^{2}} = 3.8\%$, again applying quadrature subtraction. 

{Whilst using the full covariance matrices for fitting allows us to account for the often dramatic impacts of correlated uncertainties, a shortcoming is that it necessarily requires all uncertainties to be treated according to the same likelihood function -- eg, a Gaussian (see Section~4 of \citealp{Gordon2014B}, specifically their Equation~16). However, the uncertainties on absolute calibrations tend {\it not} to be Gaussian. Rather, the true value of the calibration error will be confidently contained within the stated bounds; this is the case for \hersc-PACS \citep{Decin2007B} and \hersc-SPIRE \citep{Bendo2013A}. In other words, unlike with a Gaussian tail, there is not a 32\% chance of the true value being outside the stated $\pm$ range. Ordinarily, this would result in the total combination of the uncorrelated statistical (Gaussian) uncertainties and correlated absolute (non-Gaussian) uncertainties being smaller than if they were both Gaussian and added in quadrature. However, we {\it also} wish the total uncertainties, as stated in the diagonal elements of $\mathcal{U}_{\it calib}$, to reflect the canonical 7\% and 5.5\% uncertainties for PACS and SPIRE. To reconcile these two requirements, we use slightly larger uncorrelated uncertainties than the standard 2\% for PACS \citep{Balog2014B} and 1.5\% for SPIRE \citep{Bendo2013A}, such that the quadrature sum of the uncorrelated and correlated components gives the canonical 7\% and 5.5\% values. These larger uncorrelated uncertainties also provide some leeway to account for additional sources of uncertainty (such as that caused by uncertainty on the beam area, which is at least 1\%; \citealp{Bendo2013A}).}

We therefore use the correlated and uncorrelated relative uncertainty matrices:

\begin{equation}
\mathcal{U}_{\it corr} = 
\begin{bmatrix}
0.02 & 0.02 & 0 & 0 & 0 \\
0.02 & 0.02 & 0 & 0 & 0 \\
0 & 0 & 0.04 & 0.04 & 0.04 \\
0 & 0 & 0.04 & 0.04 & 0.04 \\
0 & 0 & 0.04 & 0.04 & 0.04 \\
\end{bmatrix}
\label{Equation:DustBFF_U_Matrix_Corr}
\end{equation}
 
\noindent and:

\begin{equation}
\mathcal{U}_{\it uncorr} = 
\begin{bmatrix}
0.067 & 0 & 0 & 0 & 0 \\
0 & 0.067 & 0 & 0 & 0 \\
0 & 0 & 0.038 & 0 & 0 \\
0 & 0 & 0 & 0.038 & 0 \\
0 & 0 & 0 & 0 & 0.038 \\
\end{bmatrix}
\label{Equation:DustBFF_U_Matrix_Uncorr}
\end{equation}

\noindent to give a final $\mathcal{U}_{\it calib}$ of:

\begin{equation}
\mathcal{U}_{\it calib} = 
\begin{bmatrix}
0.07 & 0.02 & 0 & 0 & 0 \\
0.02 & 0.07 & 0 & 0 & 0 \\
0 & 0 & 0.055 & 0.04 & 0.04 \\
0 & 0 & 0.04 & 0.055 & 0.04 \\
0 & 0 & 0.04 & 0.04 & 0.055 \\
\end{bmatrix}
\label{Equation:DustBFF_U_Matrix_Calib}
\end{equation}

\noindent for which the columns and rows contain the values for the \hersc\ bands in ascending order of wavelength.

\begin{figure}
\centering
\includegraphics[width=0.475\textwidth]{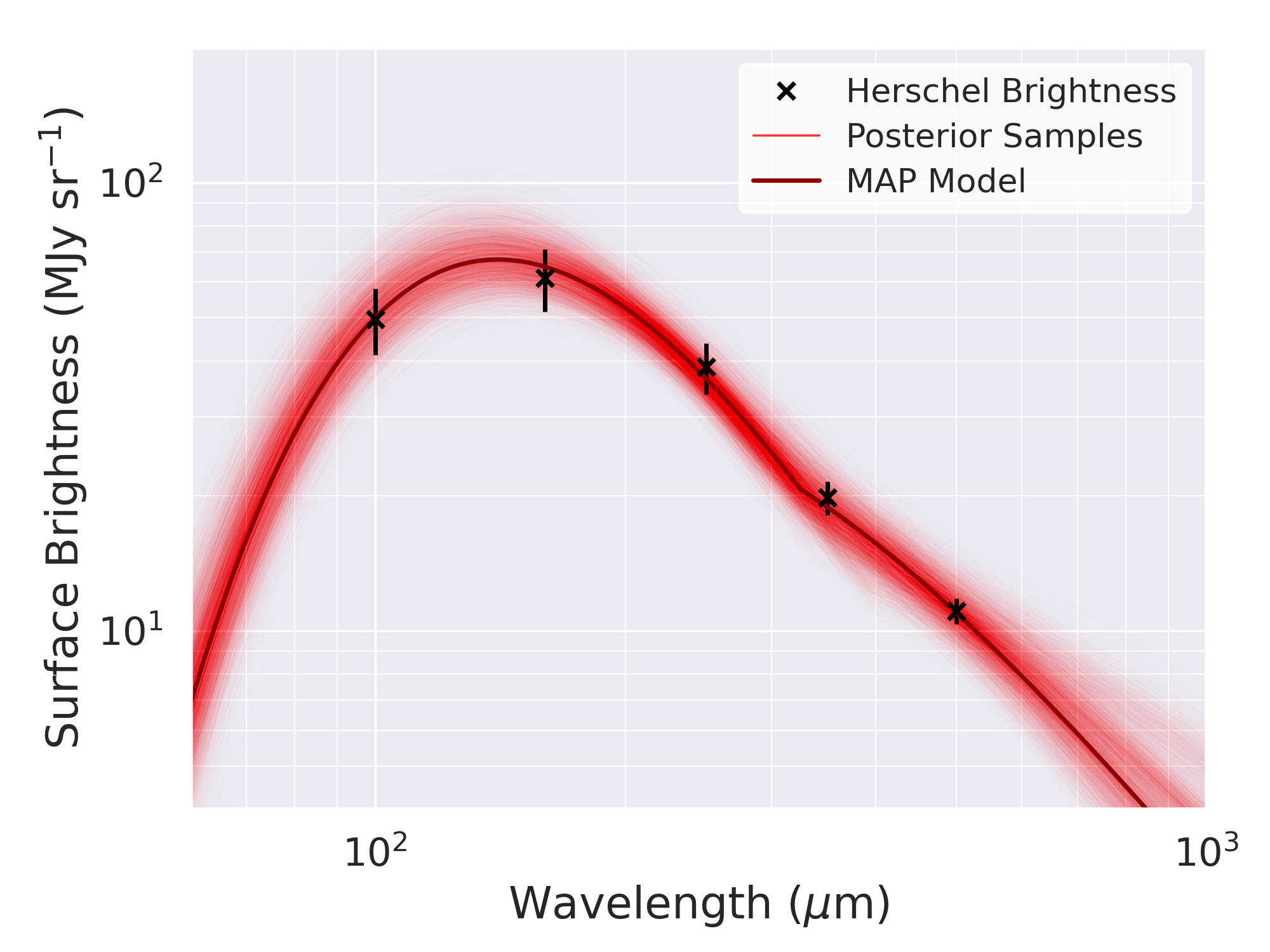}
\caption{SED for example pixel in M\,33 located at $\alpha = 23.6486\degr$, $\delta = 30.6592\degr$. The observed surface {brightness values} from our feathered \hersc\ data are plotted, along with the 1000 model SEDs from each set of \dustbff\ posterior sample parameters, and the \dustbff\ Maximum A-Posteriori (MAP) fit SED.}
\label{Fig:SED_Example}
\end{figure}

In total, the parameter grid contains $\approx$9 million models, and across our galaxies we fit it to $\sim$1 million pixels. Computing the parameter posterior probability distributions for all of these pixels for all of our galaxies took approximately one month using a 32\,$\times$\,3.2\,GHz thread computer. Propagating the full posterior distribution, with likelihoods for every model, for every pixel, throughout our analyses, would have been impractical. For each pixel, we therefore drew 1000 random samples (with replacement) from the full posterior distribution, with the probability of a given model being drawn being proportional to its likelihood. This set of random samples was then propagated as our posterior for all analyses. We show an example SED for a pixel in M\,33, with the 1000 posterior samples illustrated, in Figure~\ref{Fig:SED_Example}, with a corner plot of that pixel's posterior in Figure~\ref{Fig:SED_Corner}.

\begin{figure*}
\includegraphics[width=0.32\textwidth,trim={0 0 20ex 0},clip]{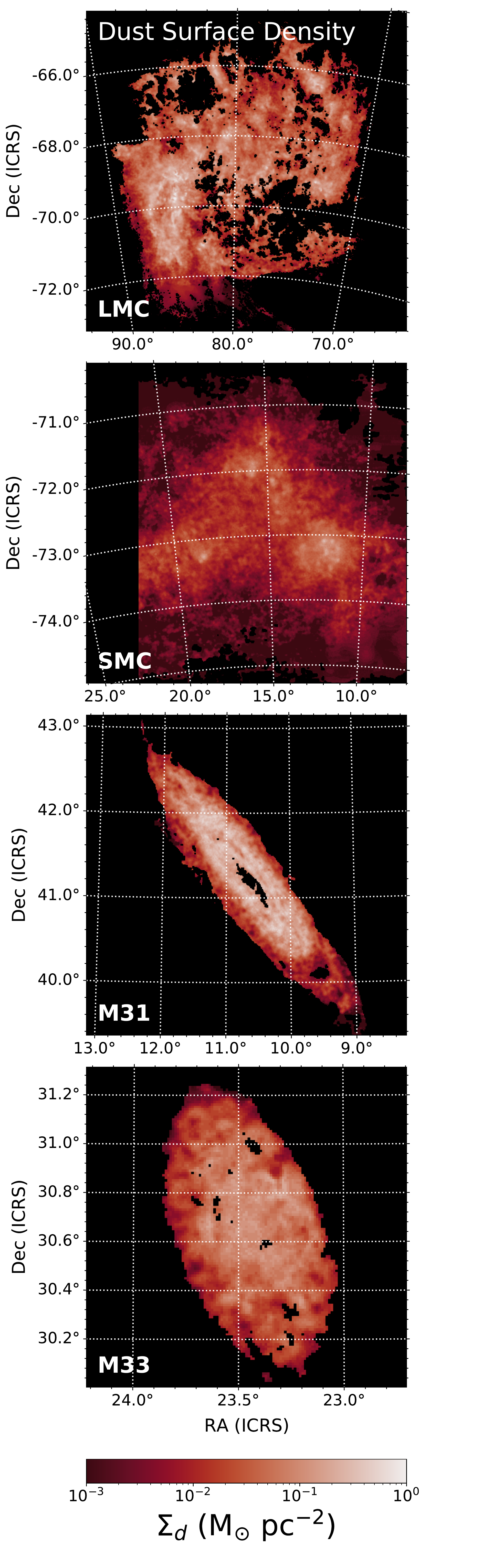}
\includegraphics[width=0.32\textwidth,trim={0 0 20ex 0},clip]{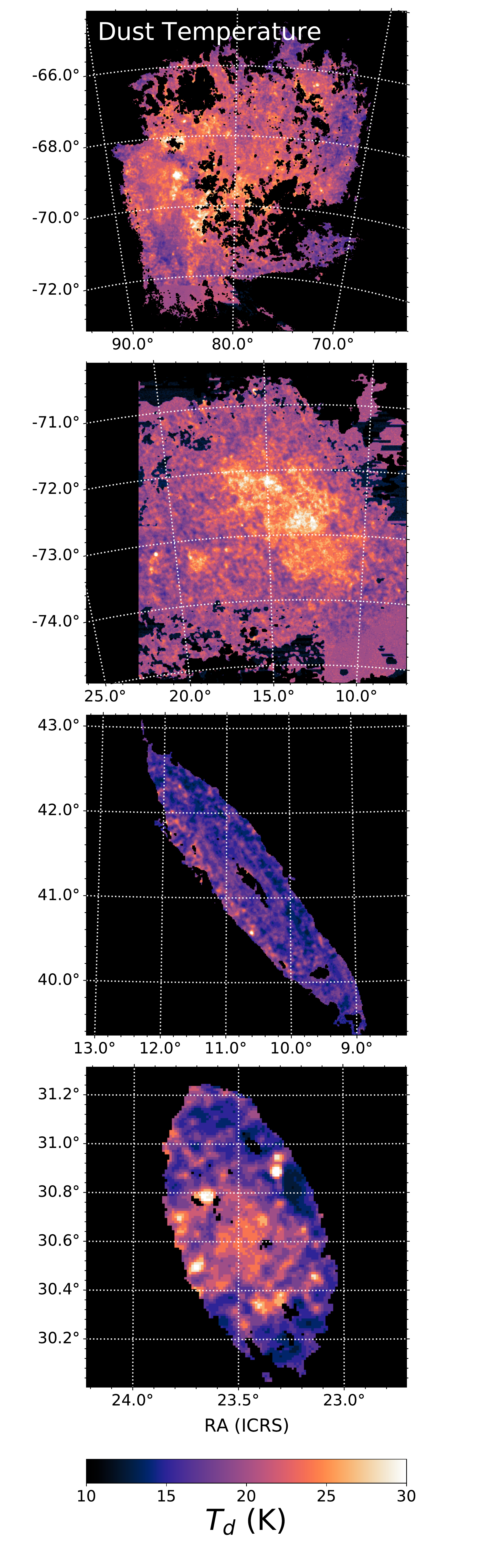}
\includegraphics[width=0.32\textwidth,trim={0 0 20ex 0},clip]{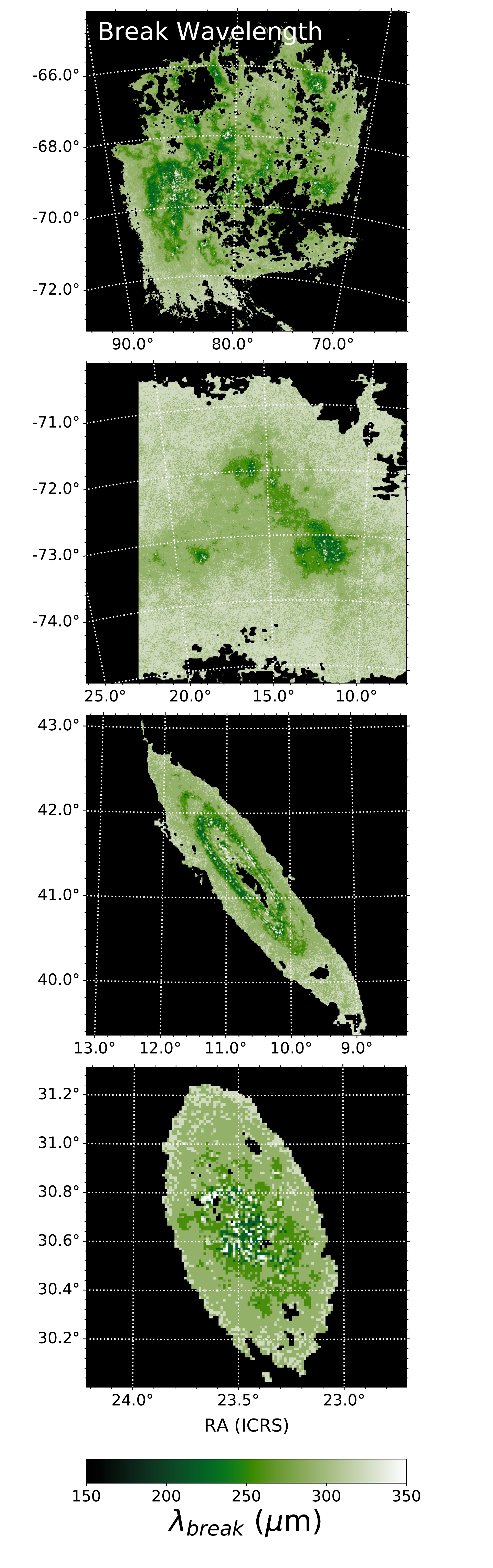}
\caption{Results of our pixel-by-pixel SED fitting for our target galaxies, with the median value for each parameter shown in each pixel; This figure shows $\Sigma_{d}$, $T_{d}$, and $\lambda_{\it break}$; Figure~\ref{Fig:SED_Params_Grid_2} shows $\beta_{1}$, $\beta{2}$, and $e_{500}$. SED fitting was only performed for pixels where our $\Sigma_{H}$ maps had S/N\,\textgreater\,4. For each parameter, the same color scale is used for all galaxies, to allow direct comparison.}
\label{Fig:SED_Params_Grid_1}
\end{figure*}

\begin{figure*}
\includegraphics[width=0.32\textwidth,trim={0 0 20ex 0},clip]{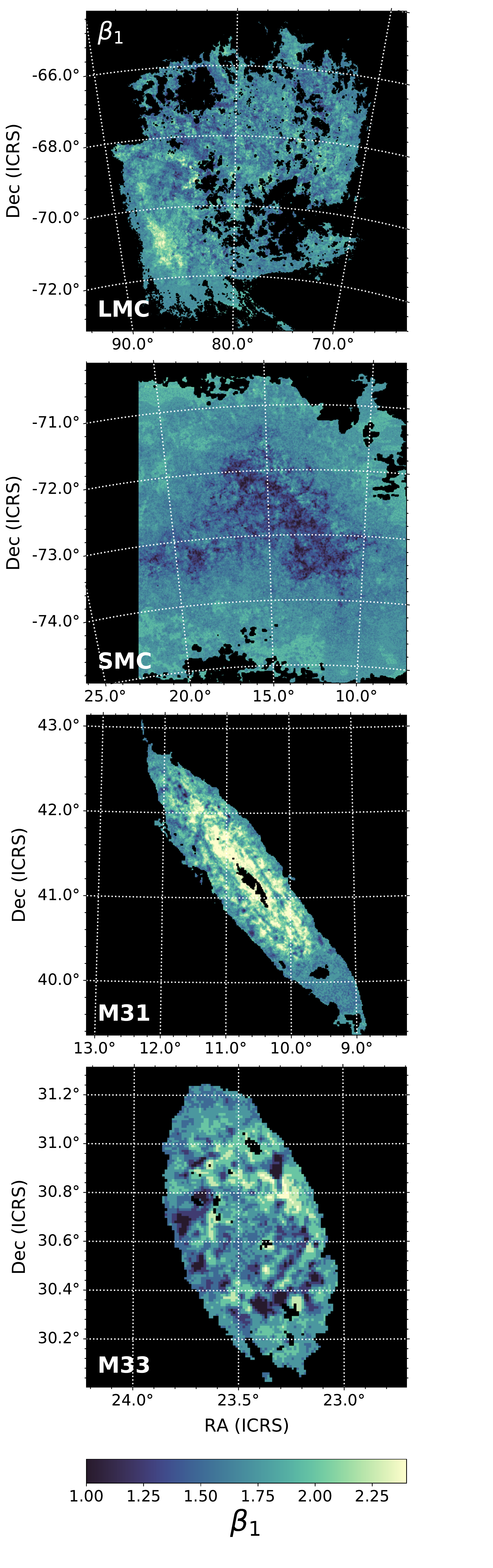}
\includegraphics[width=0.32\textwidth,trim={0 0 20ex 0},clip]{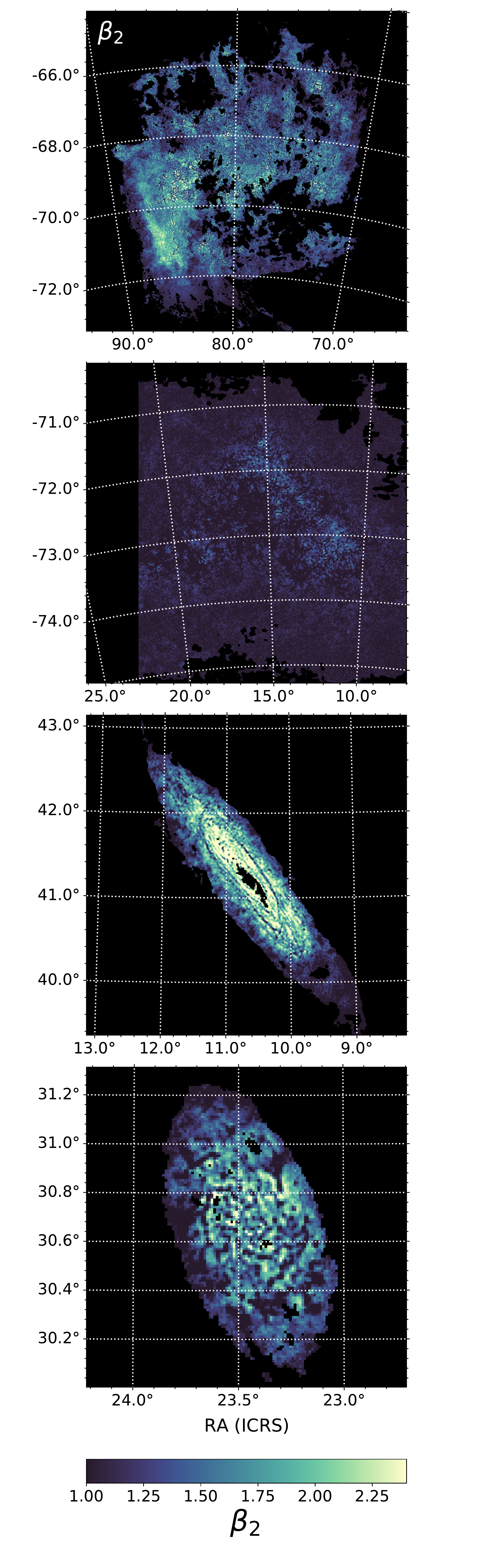}
\includegraphics[width=0.32\textwidth,trim={0 0 20ex 0},clip]{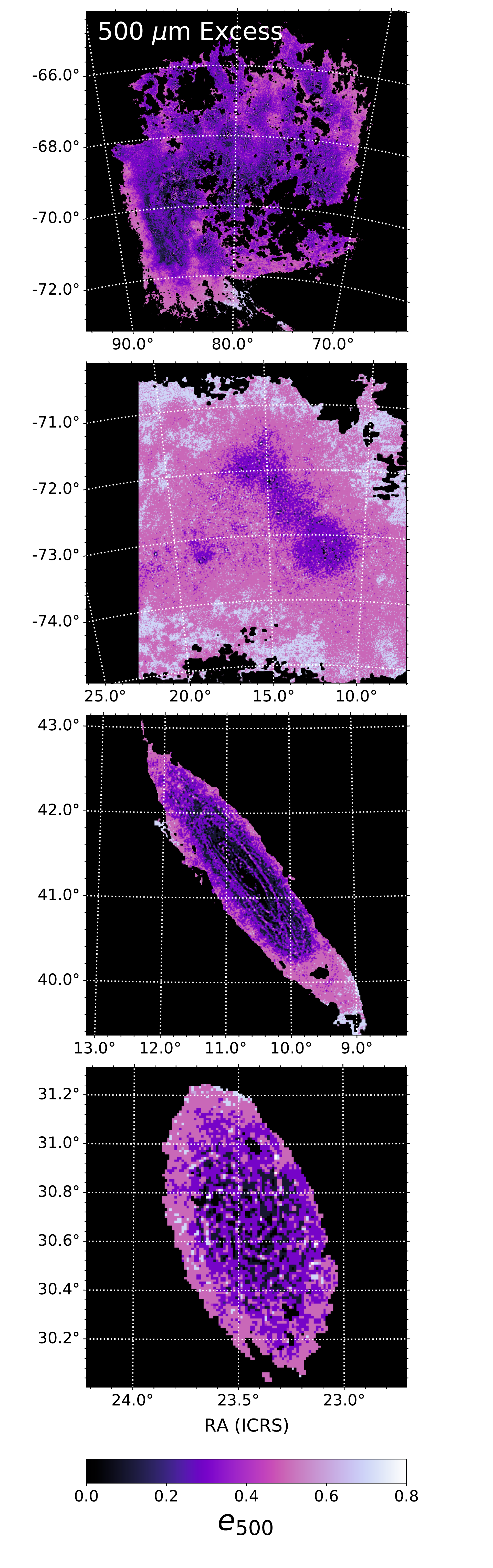}
\caption{As per Figure~\ref{Fig:SED_Params_Grid_1}, except for SED parameters $\beta_{1}$, $\beta_{2}$, and $e_{500}$. Identical color scales are used for both $\beta_{1}$ and $\beta_{2}$.}
\label{Fig:SED_Params_Grid_2}
\end{figure*}

When deciding what pixels of our \hersc\ data to fit, we imposed a Signal-to-Noise ratio (S/N) threshold dictated by the $\Sigma_{H}$ data. Because the motivation for our SED-fitting here is to explore variation in the D/H ratio, there was no point fitting pixels for which there was not reliable gas data. In particular, we found that the very peripheries of the $\Sigma_{H}$ maps, particularly for M\,31 and M\,33, seem to exhibit unphysically low gas surface density measurements, dropping abruptly to $\sim$0 at their edges. This seems to be caused by the sensitivity tailing off at the edge of the input maps' coverage areas. And in the Magellanic cloud $\Sigma_{H}$ data, there is conspicuous instrumental striping in very low-density areas; systematic bias from these artefacts seems likely to dominate over astrophysical emission (even when binning). We therefore imposed a sensitivity cut of S/N\,\textgreater\,4 using our $\Sigma_{H}$ maps. Because our $\Sigma_{H}$ maps are a combination of the re-projected input maps of atomic and molecular gas, we calculated the noise ourselves, by measuring the RMS noise within noise-dominated regions of the maps, where there was no apparent astrophysical signal. Only pixels above the S/N\,\textgreater\,4 threshold underwent SED fitting and subsequent analysis. Conveniently, our $\Sigma_{H}$ maps are very closely matched in sensitivity, with the S/N\,\textgreater\,4 threshold corresponding to $\Sigma_{H} > 2.8\,{\rm M_{\odot}\,pc^{-2}}$ for the LMC, $\Sigma_{H} > 2.6\,{\rm M_{\odot}\,pc^{-2}}$ for M\,31, and $\Sigma_{H} > 2.1\,{\rm M_{\odot}\,pc^{-2}}$ for the SMC and M\,33 (note these thresholds are before accounting for inclination projection effects, as discussed in Section~\ref{Subsection:DtH_versus_Sigma_H}).

\begin{figure}
\centering
\includegraphics[width=0.475\textwidth]{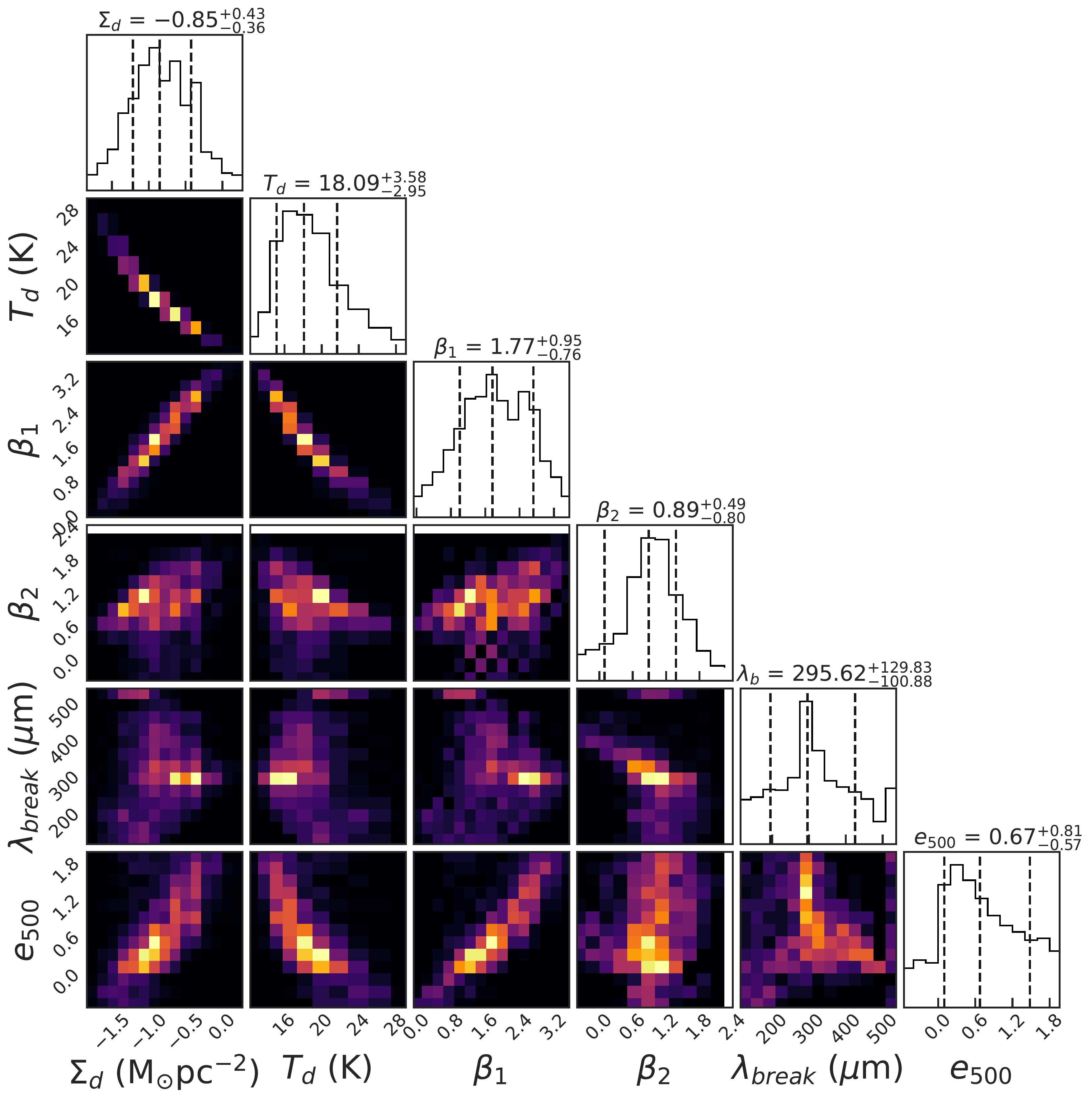}
\caption{Corner plot, showing the marginalized posterior distribution of each parameter, and their covariances, for an example pixel in M\,33 located at $\alpha = 23.6486\degr$, $\delta = 30.6592\degr$.}
\label{Fig:SED_Corner} 
\end{figure}

\subsection{SED Fitting Results} \label{Subsection:SED_Fitting_Results}

The results of our \dustbff\ SED fitting are shown in Figures~\ref{Fig:SED_Params_Grid_1} and \ref{Fig:SED_Params_Grid_2}, which shows maps of the median (ie, 50\th\ percentile) value of each parameter for each pixel, in each galaxy. Note that maps of the parameter medians for each pixel tend to be noisier than maps of each pixels's Maximum A-Posteriori (MAP) SED parameter values; however they also tend to be more representative of the marginalized distribution for each individual parameter, hence we chose to display these. While the SED fitting is not the focus of this work in-and-of itself, these outputs provide a good check of \dustbff, and show some interesting features, so we do make a few comments here.

{Our SED fitting appears to do a good job of modeling the data. For every pixel, we computed the residuals between the MAP SED model and the data, and measured the median residual across all pixels for each band, for each galaxy, and found it to be \textless2.4\% in every case. No particular sub-region or environment appears to exhibit noticeable residuals, either.}

Overall, our maps of SED parameters are in decent agreement with those of previous authors who have done similar resolved SED fitting of these galaxies. For the rest of this subsection, we perform comparisons to several specific works\footnote{Several authors, such as \citet{Draine2014A} and \citet{Chastenet2017A}, have done resolved SED fitting of galaxies in our sample using physical dust models, with different parameter sets to the MBB-based model we use. Therefore, other than $\Sigma_{d}$, we do not have parameters in common that can be compared directly.}. 

\citet{Gordon2014B} previously performed resolved SED-fitting of the Magellanic Clouds as part of the original {\it Herschel} Inventory of The Agents of Galaxy Evolution (HERITAGE; \citealp{Meixner2013A}) key program. Although the general ranges of values and broad morphology of our parameter maps agree well, it is difficult to make more detailed comparisons; the striping/cross-hatching artefacts found in the original HERITAGE data are also strongly apparent in their parameter maps; plus, the S/N limits of the HERITAGE data, especially for the SMC, renders the modeled areas much smaller than for our data, with its restored diffuse emission and reduced noise.

{We do note that like \citet{Gordon2014B}, we find very low values of $\beta_{1}$ and $\beta_{2}$ in the SMC, with many pixels in the galaxy's outskirts having values only slightly above 1. We do not believe that this is due to significant temperature-mixing causing a flat SED shape to be favored. In order for temperature-mixing to cause $\beta_{2}$ to be significantly flatter than $\beta_{1}$, the colder dust components must have a Wein's Law peak at a wavelength \textgreater\,$\lambda_{\it break}$. However, \textgreater\,94\%\ of SMC pixels have $\lambda_{\it break}$\,\textgreater\,290\,\micron. For dust to have a Wein's Law peak beyond this wavelength requires a temperature of \textless\,10\,K. Temperatures this cold are hard to explain physically, and \citet{Bot2010A} report that in order to explain SED flattening in the SMC, the temperatures would have to be so cold as to approach the Cosmic Microwave Background. Additionally, \citet{Gordon2014B} show that the required mass of dust this cold would exceed the available mass of ISM metals in the SMC. Similarly, if temperature mixing of dust at temperatures {\it above} 10\,K were artificially flattening the SED, we would expect significant residuals between the models and observations. This is because even an SED with a shallow $\beta$ will fail to closely fit the flat-peaked SED arising from significant temperature mixing, resulting in residuals (especially when averaged over large numbers of pixels, such as in our data). However, as discussed earlier in this section, there are no meaningful residuals evident in our SED fits.}

The work of \citet{Utomo2019B} provides for a particularly useful comparison to our own, as they carried out pixel-by-pixel SED fitting of \hersc\ data for all of the galaxies in our sample, using an algorithm substantially similar to \dustbff\ -- albeit without the restoration of extended emission available in our data, and using an SED model that assumed a fixed $\beta = 1.8$ with no break at longer wavelengths. Their $\Sigma_{d}$ and $T_{d}$ parameter maps (see their Figure~1) are a good match to our own, for the areas where we both have results. The \citet{Utomo2019B} maps of $\Sigma_{d}$, in particular, are almost a perfect match in morphology to ours. Our maps also agree with their general structure for $T_{d}$, too, although we differ on some of the specifics. For instance, we find the same areas of very high $T_{d}$ in M\,33 around star-forming regions, and we also find the `ridge' of highest temperature in the SMC to be slightly offset northwest of the `ridge' of maximal $\Sigma_{d}$. On the other hand, the morphology of this SMC temperature structure differs somewhat between our respective maps, and we also find M\,31 to have consistently lower $T_{d}$ than the other galaxies in the sample (dust-mass-weighted median of 15.9\,K in M\,31, compared to 21.3, 22.0, and 18.4\,K for the LMC, SMC, and M\,33 respectively; see Table~\ref{Table:DustBFF_Params}), while \citet{Utomo2019B} do not report such a difference\footnote{We do not fit the SED of many of the pixels in the gas-poor centre of M\,31, where dust will be warmest, because they fell below the S/N threshold in $\Sigma_{H}$. However, outside this central region there is {extensive} area where both \citet{Utomo2019B} and ourselves modeled the SED, where we still find $\approx$5\,K difference.}. The fact that we allow $\beta$ to vary, whereas \citet{Utomo2019B} keep it fixed, could account for this difference in temperature, given the strong degeneracy between those two parameters.

The colder dust temperatures we find for M\,31 do, however, agree well with \citet{MWLSmith2012B}, \citet{Viaene2014B}, and \citet{Whitworth2019A}\footnote{\citet{MWLSmith2012B} and \citet{Whitworth2019A} use an unbroken variable $\beta$ for their MBB models, while \citet{Viaene2014B} adopt a fixed-$\beta$ MBB model.}, which find $\approx$16\,K dust temperatures over most of the disc, with a much smaller warmer region in the center, small enough that it would mostly fall within the low-$\Sigma_{H}$ region we do not fit. Our results also agree with the finding {from} \citet{MWLSmith2012B} and \citet{Whitworth2019A} that $\beta$ decreases with radius in M\,31. Indeed, whereas both of those studies used an unbroken-$\beta$ model, we find that both $\beta_{1}$ and $\beta_{2}$ fall with radius. Additionally, thanks to our restoration of the diffuse emission, we are able to perform our SED modeling out to a radius of over 32\,kpc, compared to the 18\,kpc of \citet{MWLSmith2012B}, and 20\,kpc of \citet{Whitworth2019A}, finding that the trends of decreasing $T_{d}$ and $\beta$ continue out to larger extreme radii. We are also able to successfully perform SED fitting in low S/N pixels within diffuse regions between the star-forming rings of M\,31, that those previous studies did not model.

For M\,33, although we find that $T_{d}$ and $\beta_{1}$ generally fall with radius, in line with the findings of \citet{Tabatabaei2014A} (who also used a single-MBB, unbroken-$\beta$ model), we find that $\beta_{2}$ does not fall conspicuously with radius, instead being depressed around regions of heightened star formation (the same regions with elevated dust temperature), suggestive of possible grain processing in these environments.

\needspace{2\baselineskip} \subsubsection{Global Dust and Gas Values} \label{Subsubsection:Global_Values}

\begin{table}
\centering
\caption{Dust and gas parameters for each of our galaxies. The total dust mass, $M_{d}$, is the sum of the median posterior mass value of every pixel; the D/H is this value divided by the sum of the same pixels in the $\Sigma_{H}$ map. For the other parameters, we give the dust-mass-weighted average of the pixel medians.}
\label{Table:DustBFF_Params}
\begin{tabular}{lrrrr}
\toprule \toprule
\multicolumn{1}{c}{} &
\multicolumn{1}{c}{M\,31} &
\multicolumn{1}{c}{M\,33} &
\multicolumn{1}{c}{LMC} &
\multicolumn{1}{c}{SMC} \\
\cmidrule(lr){2-5}
$M_{d}$ ($10^{6}\,{\rm M_{\odot}}$) & 45.0 & 5.41 & 1.35 & $0.165$ \\
$T_{d}$ (K) & 15.9 & 18.4 & 21.3 & 22.0 \\
$\beta_{1}$ & 2.12 & 1.81 & 1.73 & 1.49 \\
$\beta_{2}$ & 1.96 & 1.57 & 1.54 & 1.09 \\
$\lambda_{\it break}$ (\micron) & 277 & 283 & 272 & 283 \\
$e_{500}$ & 0.21 & 0.28 & 0.28 & 0.41 \\
D/H  & $10^{-2.11}$ & $10^{-2.13}$ & $10^{-2.39}$ & $10^{-3.38}$ \\
\bottomrule
\end{tabular}
\footnotesize
\justify
\end{table}

In Table~\ref{Table:DustBFF_Params}, we provide global SED parameters for our galaxies. We report the total dust mass, $M_{d}$, where for each galaxy we take the sum of the posterior median dust mass for each pixel. For the D/H values (discussed more fully in Section~\ref{Section:DtG_Ratio}), we divide this by the total hydrogen mass computed from the same pixels in our $\Sigma_{H}$ maps. For the other SED parameters, we give the dust-mass-weighted average of the posterior medians for each pixel. In all cases, we only consider pixels that exceed our previously-discussed $\Sigma_{H}$ S/N criterion\footnote{For M\,31, the application of the $\Sigma_{H}$ S/N cut  does mean that there are pixels in the center of the galaxy for which the dust content is not incorporated into Table~\ref{Table:DustBFF_Params}, due to their low hydrogen column density. However as these pixels comprise \textless\,2\% of the total dust mass of M\,31, their omission has minimal impact on the global values.}. 

To make sure that our imposition of the S/N criterion does not significantly affect the global D/H values, we also recalculated them using all pixels in the dust and gas maps for which the $\Sigma_{H}$ map had S/N\,\textgreater\,1; this much weaker threshold is intended to make sure we are not excluding significant amounts of diffuse emission, while also still avoiding artefacts and other image errors in low S/N regions introducing significant bias into the measurements. To prevent contamination from any Galactic foreground emission that had not been fully subtracted, we also only counted pixels for M\,31 and M\,33 that are within the elliptical apertures employed for photometry in Section~6.3 of \citetalias{CJRClark2021A}. The resulting D/H were all with in 3\% of those we found when using the S/N\,\textgreater\,4 criterion, indicating a very minimal contribution from the masked pixels.

\citet{Gordon2014B} also performed their SED fitting with \dustbff, using a BEMBB model. This allows for a good direct comparison of our global results to results obtained using the older, unfeathered \hersc\ maps of the Magellanic Clouds. The dust masses we find for the LMC and SMC are factors of 1.59 and 1.48 {\it smaller}, respectively, than the total dust masses found by \citet{Gordon2014B}, after correcting for our differing values of $\kappa(\lambda_{\it ref})$. Similarly, in comparison to the dust masses from the resolved SED fitting of \citeauthor{Utomo2019B} (\citeyear{Utomo2019B}; in which they model the illuminating interstellar radiation field, as per \citealp{Draine2007A}), ours are smaller by factors of 3.3 for the LMC, 3.9 for the SMC, 1.46 for M\,31, and 3.8 for M\,33 (again, after scaling to our $\kappa(\lambda_{\it ref})$)

The fact we tend to find find lower masses may seem surprising at first, given that the new \hersc\ maps from \citetalias{CJRClark2021A} {\it restored} dust emission that was missing in older maps. However, there was more emission that needed restoring in the shorter wavelength bands (especially in the \hersc-PACS 100 and 160\,\micron\ data) than at longer wavelengths. As a result, essentially all the dust emission from our galaxies is rendered significantly bluer in the new maps (see Figure~17 of \citetalias{CJRClark2021A}). Bluer emission corresponds to warmer dust temperatures; and as the luminosity of a given mass of dust goes approximately $\propto T_{d}^{4+\beta}$, this reduces the modeled dust mass for a given FIR brightness. 

To check this, we directly compared the dust-mass-weighed dust temperatures we find, to those of \citet{Gordon2014B}, by reprojecting their parameter maps to our pixel grid, and only comparing pixels for which both had coverage. We found that our mass-weighted-average global dust temperatures were 2.0\,K warmer for the LMC, and 5.5\,K warmer for the SMC. These correspond to factor 1.11 and 1.29 differences in temperature, respectively. Assuming there were no other difference, then the simple $\propto T_{d}^{4+\beta}$ relation would suggest that we should {\it expect} our dust masses to be factors of 2.1 and 5.5 reduced due to this effect, assuming typical $\beta = 1.7$. We don't perform the same direct comparison with the \citet{Utomo2019B} maps, due to {differences between the foreground subtractions \&\ effective apertures we use, and differences in method. For instance, by using a broken-emissivity approach (as compared to their fixed $\beta$=1.8 method), it is possible for flux at longer wavelengths to be accounted for by the $\beta$ becoming flatter, instead of by driving up the total SED normalization -- and hence mass}. The new \hersc\ maps from \citetalias{CJRClark2021A} generally increase the total FIR flux measured in each pixel, with per-band averages increases of 21\%, which will counteract some of the mass reduction due to higher temperature. Plus, \citetalias{CJRClark2021A} applied calibration corrections to the \hersc\ data, which will also have knock-on effects to the SED parameters (generally increasing 100--160\,\micron\ fluxes, but increasing 250--500\,\micron\ fluxes. So overall, the reduction in dust mass we find is of order the difference that should be expected. 

On the other hand, comparing to the total dust masses reported by \citet{Chastenet2017A}, who used the THEMIS physical dust grain model \citep{Jones2017A}, our LMC dust mass is 20\% {\it greater} than theirs, although our SMC dust mass is a factor of 1.67 less, after $\kappa(\lambda_{\it ref})$ corrections\footnote{The ratio of silicate to carbonaceous dust is different in every pixel modeled by \citet{Chastenet2017A}; as per their Section 5.2, we assume an average 2:1 ratio of carbonaceous to silicate dust, which agrees with the range of average ratios they quote for both galaxies. Using the average THEMIS emissivity slope of $\beta = 1.78$ \citep{Nersesian2019A} to convert via Equation~\ref{Equation:SED_MBB}, this gives an average $\kappa_{160} = 2.84\,{\rm m^{2}\,kg^{-1}}$. We compare to their favored multi-ISRF masses.}. \citet{Chastenet2017A} note that their masses are significantly lower than other authors, which they ascribe to the high fraction of more-emissive carbonaceous grains in their modeling results (which may be reasonable, given recent results from \citealp{Roman-Duval2022A} finding LMC and SMC dust to be more carbon-rich than that of the Milky Way, at a given column density).

\needspace{2\baselineskip} \subsubsection{Submillimeter Excess} \label{Subsubsection:Submillimeter Excess}

\begin{table}
\centering
\caption{Spearman rank correlation coefficients between the submm excess at 500\,\micron, $e_{500}$, and various other parameters, for the pixels in our maps. All relations have $\mathcal{P}_{\it null} < 10^{-16}$.}
\label{Table:Excess_Correlations}
\begin{tabular}{lrrrr}
\toprule \toprule
\multicolumn{1}{c}{$e_{500}$ vs} &
\multicolumn{1}{c}{M\,31} &
\multicolumn{1}{c}{M\,33} &
\multicolumn{1}{c}{LMC} &
\multicolumn{1}{c}{SMC} \\
\cmidrule(lr){1-1}
\cmidrule(lr){2-5}
$\Sigma_{H}$ & $-$0.54 & $-$0.16 & $-$0.47 & $-$0.39 \\
Deprojected Radius & 0.41 & 0.74 & 0.57 & -- \\
D/H & $-$0.80 & $-$0.71 & $-$0.57 & $-$0.60 \\
$\Sigma_{d}$ & $-$0.77 & $-$0.65 & $-$0.67 & $-$0.68 \\
\bottomrule
\end{tabular}
\footnotesize
\justify
\end{table}

Significant submm excess is apparent in the outskirts of all of our galaxies, especially the lowest-metallicity SMC. {This is in line with previous work showing that greater submm excess tends to be found in lower-metallicity environments \citep{Galliano2003A,Bot2010A,Remy-Ruyer2013A,Gordon2014B}. However, several works have also found good evidence that submm excess can be driven by ISM density, with more excess found in lower-density environments \citep{Planck2011XXV,Relano2018A}. Our improved data put us in a good position to compare these possibilities.} 

Thanks to our restoration of diffuse emission, we are able to conduct SED fitting out to larger radii than previously performed. For instance, we are able to model dust emission in the \HI\ filament to the southeast of the LMC, where it is interacting with the lower-metallicity Magellanic Stream \citep{Nidever2008B}, in which we find elevated levels of submm excess. Similarly, in M\,31, we are able to perform our SED modeling out to a radius of over 32\,kpc, compared to 18\,kpc in \citet{MWLSmith2012B}, and 20\,kpc in \citet{Whitworth2019A} and \citet{Draine2014A} At these previously-unprobed extreme radii, we again find that submm excess increases to higher levels. 

{Note that with our BEMBB model, submm excess corresponds to $\beta_{2} < \beta_{1}$. Whilst our model grid does allow for $\beta_{2} \geq \beta_{1}$ (ie, a submm deficit), this only happens for \textless\,3.7\% of pixels in the median $e_{500}$ maps across all four of our galaxies (being as few as 0.03\% of pixels for the SMC). In other words, the dust SED effectively always becomes shallower towards longer wavelengths (agreeing with \citealp{Planck2013XVII}). We find that the break wavelength is  in the range $260\,\micron < \lambda_{\it break} < 325\,\micron$ for 80\% of pixels, with $\lambda_{\it break}$ generally being lower in areas of greater ISM density, albeit with elevated $\lambda_{\it break}$ often seen immediately around star-forming regions (see Figure~\ref{Fig:SED_Params_Grid_1}).}

{Spearman rank correlation tests find that the strength of the submm excess in M\,31 and M\,33 is more strongly correlated with deprojected radius than with $\Sigma_{H}$; for the LMC, we find that the correlation is stronger with density than with radius, although the difference is smaller. However we note that the LMC is much more disturbed than either of the spirals. Correlation coefficients are given in Table~\ref{Table:Excess_Correlations}. We do not test correlation with radius for the SMC, due to its high degree of disturbance preventing us from meaningly quantifying radii.}

{The fact the correlation with radius is stronger than with ISM density for M\,31 and M\,33, but not the LMC (which lacks a significant metallicity gradient; see Section~\ref{Subsubsection:Effect_of_Metallicity}), potentially suggest that metallicity is playing a role, either through different dust composition, or through improved shielding. These possibilities are supported by the fact that we find even stronger correlation of $e_{500}$ with D/H and $\Sigma_{d}$ (Table~\ref{Table:Excess_Correlations}). We caution against over-interpretation of these final two correlations, however, as both depend upon $\Sigma_{d}$, which is a parameter in our SED fitting model along with $e_{500}$, and the two can be degenerate (eg, Figure~\ref{Fig:SED_Corner}). We also see in M\,31 and M\,33 that there are narrow regions of elevated $e_{500}$ tracing the spiral arms (see Figure~\ref{Fig:SED_Params_Grid_2}), despite these being higher-density regions. If shielding in dust-rich regions can drive down submm excess, then it is conceivable that proximity to radiation and shocks from ongoing star formation elevates $e_{500}$ in certain parts of the spiral arms, possibly in lower-density regions carved by young stellar winds, on smaller scales than what our data can resolve in these galaxies.}

We note that the S/N of the FIR emission at larger radii tends to be very low, leading to large uncertainties on the modeled parameters for any individual pixel. However, the fact that the trend of increasing submm excess is evident at all azimuth, across thousands of independent pixels, gives us reason to be confident in the trend, and that it is not an artifact of our foreground subtraction, or arising from localized cosmic microwave background fluctuations, etc.

\begin{figure*}
\centering
\includegraphics[width=0.975\textwidth]{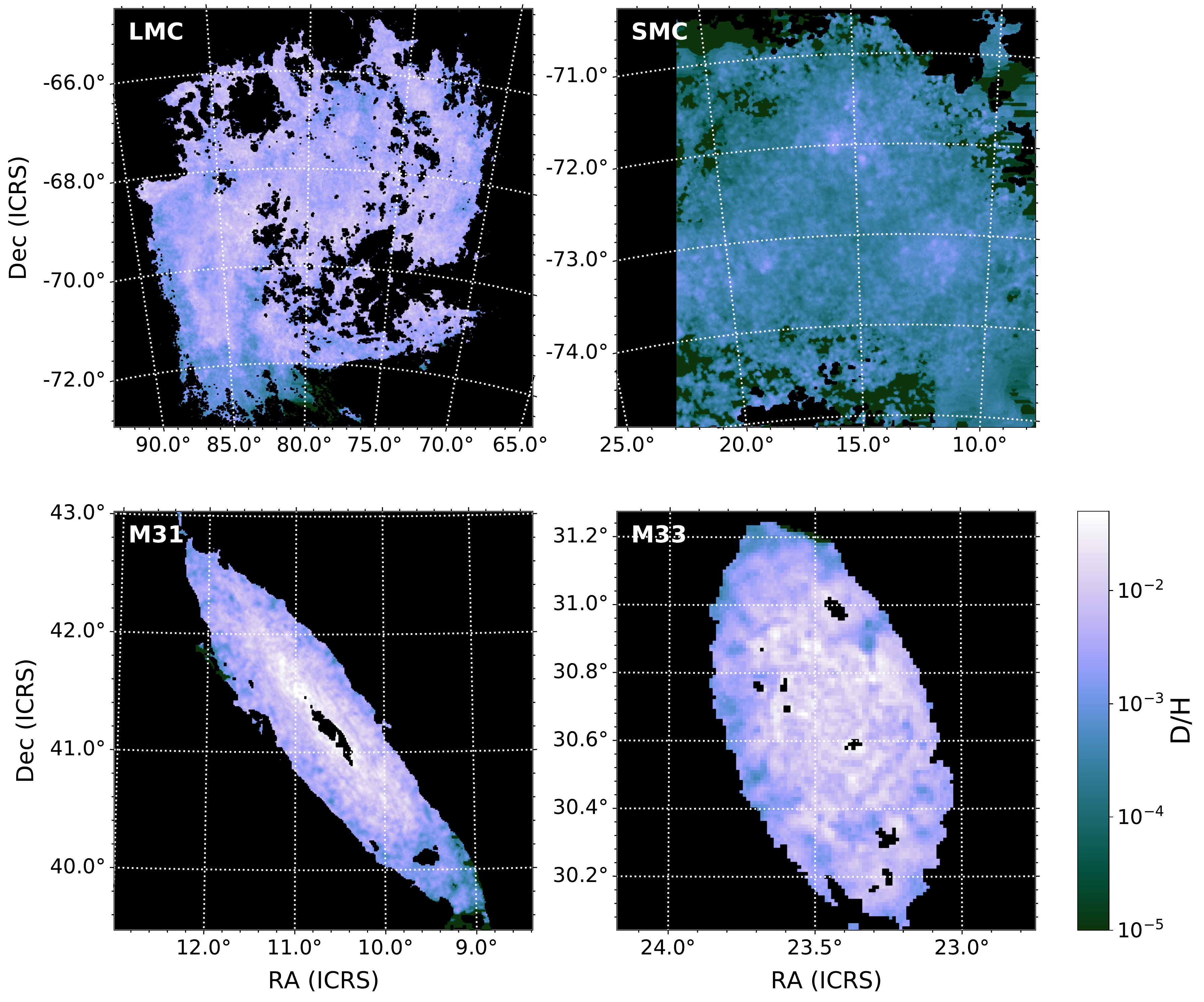}
\caption{Maps of the dust-to-hydrogen ratio for our target galaxies. All four are displayed using the same color scale, to allow for direct comparison.}
\label{Fig:DtH_Overview_Grid}
\end{figure*}

\needspace{2\baselineskip} \section{Evolution in the Dust-to-Gas Ratio} \label{Section:DtG_Ratio}

With maps of both the dust and hydrogen content of our galaxies, we can now explore the behavior of the dust-to-gas ratio within them. In Figure~\ref{Fig:DtH_Overview_Grid}, we show maps of D/H for our targets. 

For the two spirals, we find that D/H is elevated in the central regions, and tends to fall with increasing radius, in line with the broad trends observed by previous authors \citep{Draine2014A,Gratier2017B}. This trend is particularly pronounced in M\,31, while in M\,33 elevated D/H more closely follows the spiral structure, rather than being purely radial. 

For the LMC, the structure of the D/H map broadly traces the morphology of the dust mass (see Figure~\ref{Fig:DtH_Overview_Grid}), but with some other noteworthy features. The highest D/H is often found at the very edges of supershells excavated by recent star formation\footnote{As we discuss later in Section~\ref{Subsection:DtH_Residuals}, with relation to Figure~\ref{Fig:LMC_Depletions_vs_Residuals_Map}, D/H at the edge of these supershells is even elevated above what would be expected for their $\Sigma_{H}$, based on the LMC's global relationship between D/H and $\Sigma_{H}$.}. These supershells are apparent especially in poor \HI\ \citep{Dawson2013A}; the low gas density within the shells means that they mostly fall below our S/N threshold in $\Sigma_{H}$. The high D/H at the edges of these shells might indicate that the gas is being blown away more efficiently than the dust \citep{Draine2011D}. Alternatively, our dust and gas mass tracers might behave differently here; for instance, environmental grain processing might be affecting the dust opacity (see Section~\ref{AppendixSubsection:Varying_Kappa}). 

Another noteworthy feature in the map for the LMC is the significantly depressed D/H to the southeast. This is coincident with the portion of the disc where low-metallicity material from the Magellanic Stream is being accreted onto the LMC \citep{Bekki2007L,Nidever2008B}. This accretion is leading to the enhanced star-formation of 30 Doradus and the surrounding area \citep{Fukui2017B,Tsuge2019A,Tsuge2020A}. Although the area of star-formation itself has fairly high D/H\,\textgreater\,$10^{-2.3}$, the southeastern fringes of the LMC have the lowest D/H {found in the} entire galaxy. Indeed, the southernmost tip of this region features D/H as low as $10^{-3.3}$, matching the typical D/H we find for the SMC.

The D/H map of the SMC shows it to have conspicuously lower D/H than any of {the} other galaxies in the sample, as we would expect from its 0.2\,${\rm Z_{\odot}}$ metallicity. Thanks to our new \hersc\ maps from \citetalias{CJRClark2021A}, and the sensitive $\Sigma_{H}$ data, we are able to trace D/H out to very large radii. Away from the main regions of star formation along the Bar and Wing, D/H falls steadily, to values of less than $10^{-4}$ in the most diffuse regions.

\needspace{2\baselineskip} \subsection{Evolution of D/H with $\Sigma_{H}$} \label{Subsection:DtH_versus_Sigma_H}

\begin{figure*}
\centering
\includegraphics[width=1.0\textwidth]{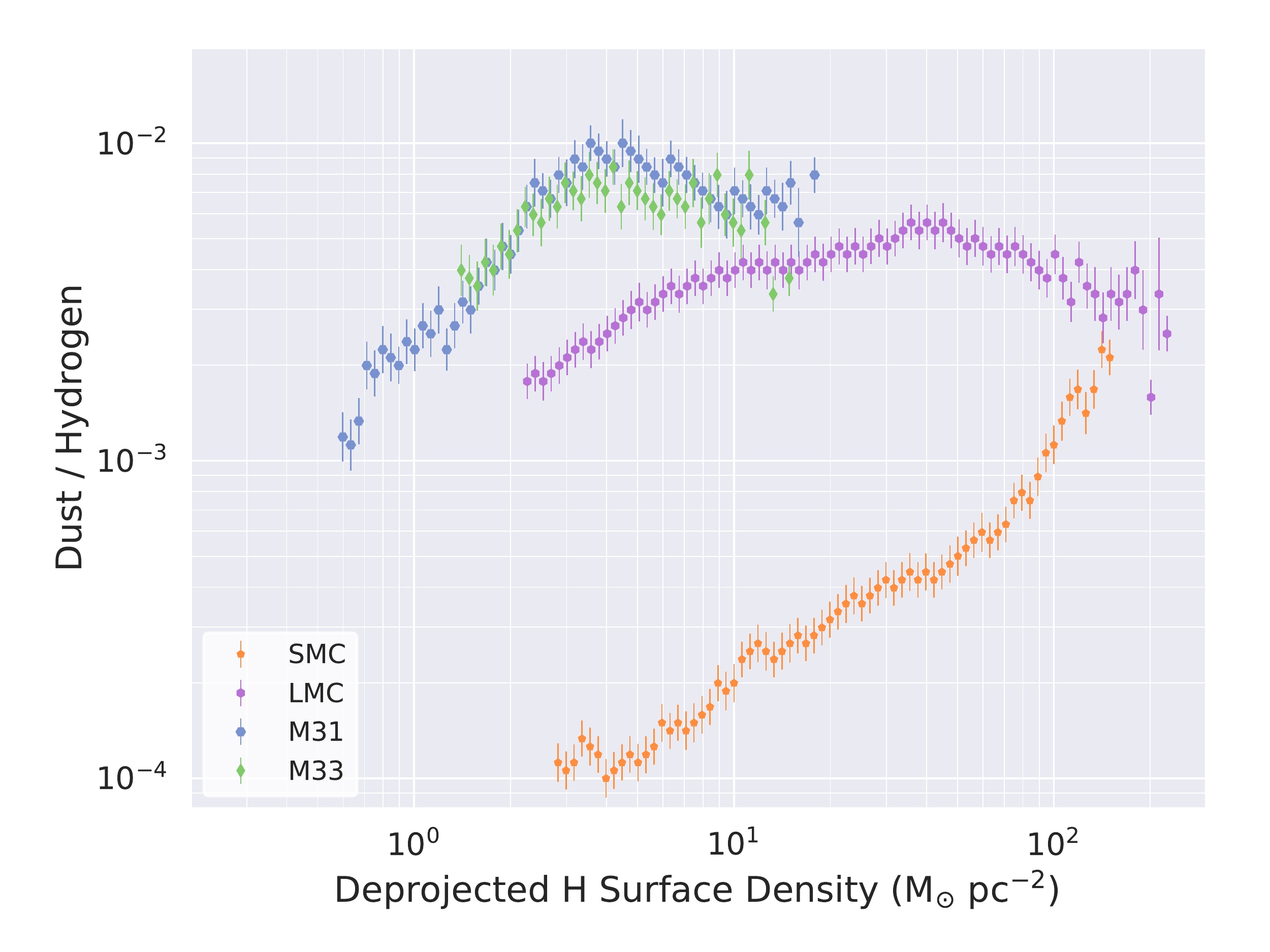}
\caption{Plot of dust-to-hydrogen ratio D/H against deprojected hydrogen surface density \SigmaDeproj, with values from individual pixels binned together into 0.025\,dex wide bins of $\Sigma_{H}$, for each our of sample galaxies. Points indicate bin medians, with error bars show uncertainty on those medians. Values of $\Sigma_{H}$ have had a deprojection correction applied to account for each galaxy's inclination.}
\label{Fig:DtH_vs_H}
\end{figure*}

In Figure~\ref{Fig:DtH_vs_H}, we present the central result of this paper -- the evolution of D/H with $\Sigma_{H}$ for our galaxies. As covered in Section~\ref{Section:Introduction}, we should expect D/H to increase with greater $\Sigma_{H}$, due to the increased efficiency of the accretion of gas-phase metals onto dust grains in higher-density ISM. To construct this plot, we binned the values from our D/H maps according to the $\Sigma_{H}$ of each pixel, into bins of width 0.025\,dex. The plotted points in Figure~\ref{Fig:DtH_vs_H} show the median value in each bin. The error bars incorporate the uncertainty on the median, calculated by performing 500 Monte-Carlo bootstrap resamples of all the D/H values in each bin, recomputing the median each time, then finding the standard deviation of those 500 bootstrapped medians; this was then added in quadrature to the 0.05\,dex uncertainty arising from the fact that our $\Sigma_{d}$ values were taken from a model grid with a 0.05\,dex step size (see Table~\ref{Table:DustBFF_Grid}), which limits the precision of any given D/H measurement. The scatter within each bin can be quite large; across all 4 galaxies, the average standard deviation of the D/H values contained within a bin is 0.8\,dex. Despite this, the {\it distribution} of values within each bin tends to be well-behaved and Gaussian; hence the average uncertainty on the bin medians, as measured via the bootstrap resampling, is only 0.02\,dex (ie, the uncertainty on the median for each bin is almost always dominated by the 0.05\,dex contribution of the uncertainty on the grid step size).

The x-axis surface density values in Figure~\ref{Fig:DtH_vs_H} have had a deprojection correction applied, to account {for the} effect of each galaxy's inclination\footnote{The same correction has also been applied to the x-axis $\Sigma_{H}$ values in Figures~\ref{Fig:DtH_vs_H_ChemEv}, \ref{Fig:LMC_DtH_vs_H_Offset}, \ref{Fig:SMC_DtH_vs_H_Offset}, \ref{AppendixFig:DtH_vs_H_alphaCOx}, \ref{AppendixFig:DtH_vs_H_1-Over-x}, \ref{AppendixFig:DtH_vs_H_Degraded}, \ref{AppendixFig:DtH_vs_H_Ionised}, and \ref{AppendixFig:DtH_vs_H_Kappa}.}. For an inclined disc galaxy like M\,31, the column of ISM sampled by a given pixel passes through a greater thickness of disc than it would for an equivalent face-on galaxy. This compromises our ability to treat observed surface density as a proxy for average volume density (as volume density is what will be dictating grain-growth, etc), and would prevent us from fairly comparing our galaxies' D/H at a given observed $\Sigma_{H}$. Therefore, for a given galaxy inclination $i$ (see Table~\ref{Table:Galaxy_Properties}), we apply a deprojection correction:

\begin{equation}
\Sigma_{H}^{\it(deproj)} = \cos{(i)}\ \Sigma_{H}
\label{Equation:Deproj}
\end{equation}

Hereafter, we refer to \SigmaDeproj\ in instances where comparison between galaxies makes the distinction between corrected and uncorrected $\Sigma_{H}$ values important. The $\cos{(i)}$ correction factor equates to 0.22, 0.56,  and 0.90 for M\,31, M\,33, and the LMC, respectively. No correction needs to be applied to the y-axis quantity of D/H, as this is a ratio of two quantities that experience the same projection, cancelling out its effect.

Because the SMC's highly disturbed morphology lacks a disc, inclination simply isn't an applicable concept, so we apply no correction. In addition, thanks to the SMC's extreme elongation along the line-of-sight \citep{Scowcroft2016B}, the link between observed surface density and actual volume density is further weakened in the case of the SMC; the reader should bear that the entire trend for the SMC in Figure~\ref{Fig:DtH_vs_H} could be shifted left or right by this poorly-constrained systematic. 

Figure~\ref{Fig:DtH_vs_H} shows clearly that there is very strong evolution in the dust-to-gas ratio with \SigmaDeproj\  for all of our sample galaxies. Indeed, M\,31 shows almost {\it an order of magnitude of evolution} in D/H between densities of 0.6--4\,${\rm M_{\odot}\,pc^{2}}$, while the SMC exhibits even more than that, with its D/H increasing by a factor of 20 between 3--150\,${\rm M_{\odot}\,pc^{2}}$. For all four galaxies, D/H increases steadily with density -- although only up until a point for M\,31, M\,33, and the LMC. All three of these all exhibit a turnover in D/H, after which it begins to {\it decrease} with increasing $\Sigma_{H}$ (this is explored in Section~\ref{Section:Causes_of_Turnover}). Regardless, the increase in D/H with \SigmaDeproj\ over most densities for all four galaxies {provides evidence for significant} density-dependent dust grain growth.

These results update our understanding of just how much D/H {\it can} vary within a galaxy, especially at fixed metallicity. Previous authors have found the dust-to-gas ratio varying by a factor of 3 in the LMC \citep{Roman-Duval2017B}, a factor of 5 in M\,31 \citep{Draine2014A}, a factor of 7 in the SMC \citep{Roman-Duval2014D,Roman-Duval2017B}, and a factor of 10 in M\,101 (\citealp{Chiang2018A,Chastenet2021A}; albeit over a factor of 5 variation in metallicity, whereas our variation is seen even with little-or-no metallicity variation, as per Section~\ref{Subsubsection:Effect_of_Metallicity}). We now find D/H spans a factor of 2.5 in M\,33, a factor of 3.5 in the LMC, a factor of 9.0 in M\,31, and a factor of 22.4 in the SMC.

\needspace{2\baselineskip} \subsubsection{Effect of Metallicity Gradients on Evolution in D/H} \label{Subsubsection:Effect_of_Metallicity}

If there were significant systematic variation in the ISM metallicity of our target galaxies, then an increase in D/H might not indicate grain growth -- instead, if regions of greater density also had higher metallicity, then D/H would be correspondingly greater even if the fraction of metals in dust grains was staying constant (in other words, even if there was no grain growth). However, this is not the case for our galaxies. 

Neither the ISM \citep{Pagel1978C,ToribioSanCipriano2017B} nor younger stellar clusters \citep{Grocholski2006B,Cioni2009B} in the LMC exhibit a significant radial metallicity gradient (\textless\,0.2\,dex, less than the intrinsic scatter). Although metallicity does appear depressed in its southeast periphery, where the LMC is interacting with low-metallicity gas of the Magellanic Stream \citep{Nidever2008B,Tsuge2019A,Tsuge2020A,Roman-Duval2021B}, this is true for both high- and low-density gas in this region, so should not systematically influence the D/H evolution profile of the LMC. For the SMC, there is no significant metallicity gradient observed in the ISM \citep{Pagel1978C,Cioni2009B,ToribioSanCipriano2017B}. 

There is a definite radial metallicity gradient in the ISM of M\,31, however it is only -0.56\,dex\,$R_{25}^{-1}$ \citep{Zurita2012A}\footnote{\label{Footnote:Z_gradient} The metallicity gradient in \citet{Zurita2012A} was reported in terms of dex\,kpc$^{-1}$; we have converted it to dex\,$R_{25}^{-1}$ to ease comparison between galaxies for the reader, as per the $R_{25}$ value in Table~\ref{Table:Galaxy_Properties}.}; so clearly this cannot account for the 0.9\,dex evolution in D/H in M\,31. Moreover, metallicity does not scale monotonically with $\Sigma_{H}$ in M\,31, thanks to the low ISM density of the inner $\sim$\,3\,kpc -- further reducing the potential for the radial metallicity gradient to give rise to increased D/H in regions of greater $\Sigma_{H}$. 

The ISM of M\,33 has an even smaller radial metallicity gradient, at -0.22\,dex\,$R_{25}^{-1}$ \citep{Magrini2016C}; similar to M\,31, this is unable to account for the 0.4\,dex evolution in D/H we see within this galaxy, especially because the gradient is comparable to the intrinsic scatter in metallicity at any given radius \citep{Magrini2016C}. 

We therefore are do not believe that metallicity effects are a primary driver of the evolution in D/H observed in Figure~\ref{Fig:DtH_vs_H}.

\needspace{2\baselineskip} \subsubsection{Effect of Dust Destruction at Lower Densities on Evolution in D/H} \label{Subsubsection:Effect_of_Dust_Destruction}

{Whilst there are compelling reasons to expect D/H to increase at higher densities due to more efficient interstellar grain growth, we should also expect dust grain destruction to become more efficient at lower densities. Therefore, a trend of D/H increasing with density could potentially be driven by either effect.

The primary forces of dust destruction in the ISM are expected to be sputtering due to \sne\ shocks, and photo-destruction by high-energy photons \citep{Jones2004B,Bocchio2014B,Slavin2015C}.

Generally, theoretical models expect the efficiency of grain destruction due to \sne\ shocks to only be relatively weakly dependent upon density. Typically, destruction efficiency is modeled to fall with increasing density, according to a power-law slope of around\,$-0.1$ \citep{Jones1994C,Dwek2007B,Temim2015B}. In contrast, all four of our galaxies show evolution in D/H with \SigmaDeproj\ with power law slopes ranging from 0.4 for the LMC, up to 1.0 for M\,31 (up until the D/H turnover; see Section~\ref{Section:Causes_of_Turnover}). This implies that either dust destruction due to \sne\ is not driving the vast majority of D/H evolution with density we find, or that the density-dependence of \sne\ dust-destruction efficiency has been under-estimated by theoretrical models.

The rate of interstellar dust destruction should also be expected to closely correlate with the density of young stars in a given environment. This is because young stars will produce high-energy photons capable of photo-destruction of grains, and because young stars indicate where dust-destroying core-collapse \sne\ should be occurring at the greatest rate.

UV emission traces young stars, that have formed within the past $\sim$100\,Myr \citep{Kennicutt1998H,Calzetti2005D,Buat2011C}. Therefore UV observations should allow us to trace the relative rate of dust destruction we should expect in different locations. The only high-quality high-resolution wide-area UV data available for all four of our sample galaxies comes from the UltraViolet/Optical Telescope (UVOT; \citealp{Roming2000D,Roming2004B}) on the {\it Neil Gehrels Swift Observatory} \citep{Gehrels2004F}. For the Magellanic Clouds, the {\it Swift}-UVOT data we use is the data presented in \citet{Hagen2017C}; for M\,31 and M\,33, we use reductions created according to the same method as in \citet{Hagen2017C}, with that data to be fully be presented in Decleir et al. ({\it in prep.}; for data access see Section~\ref{Section:Data_Products}). The data for M\,31 and M\,33 cover the full stellar discs of both galaxies; the SMC data covers the whole bar and part of the wing, whilst the LMC data covers the bar, 30 Doradus, and the surrounding regions.

To trace the young stars, we use the {\it Swift}-UVOT W2 band maps. Being the UVOT band with the shortest effective wavelength, at 192\,nm, W2 should best trace the higher-energy photons that will lead to dust destruction. We converted the {\it Swift}-UVOT W2 count-rate data to SI units, as per the AB magnitude zero points given in \citet{Breeveld2011G}, and thence to UV luminosity surface density, in L$_{\odot}$\,pc$^{-2}$. We then reprojected these maps to the same pixel grid as our D/H maps, and applied deprojection corrections.

\begin{figure}
\centering
\includegraphics[width=0.475\textwidth]{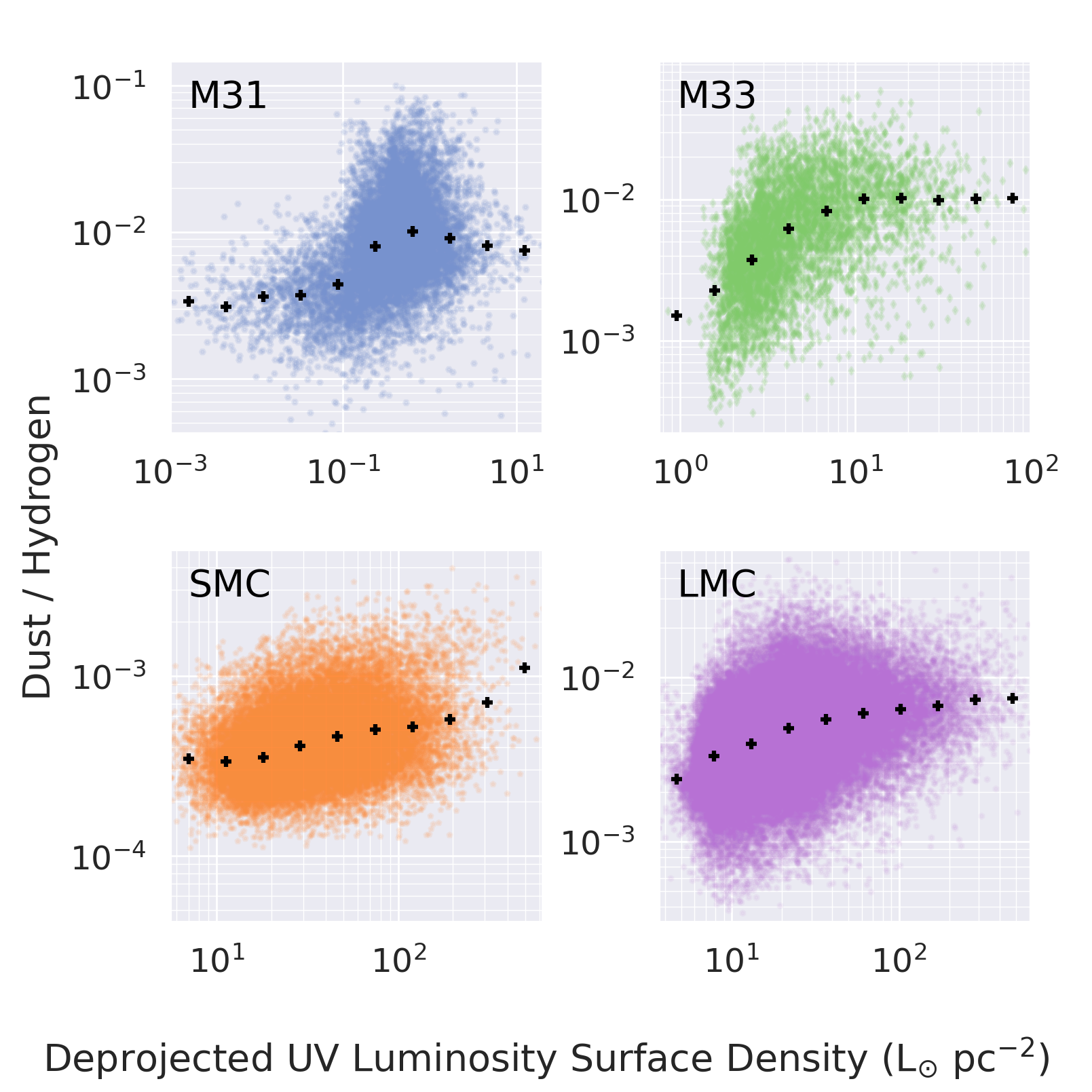}
\caption{Plots of D/H against deprojected {\it Swift}-UVOT W2 luminosity surface density, for each of our four sample galaxies. Binned median values are plotted with black crosses.}
\label{Fig:DtH_vs_UV}
\end{figure}

In Figure~\ref{Fig:DtH_vs_UV}, we plot D/H against UV luminosity density for each of our galaxies. To our surprise, we find that regions of great UV luminosity density tend to have {\it larger} D/H. This trend is significant for all four galaxies; in every case, Spearman rank correlation tests find $\mathcal{P}_{\it null} < 10^{-10}$, with correlation coefficients ranging from 0.37--0.50. On large scales, we might expect this sort of correlation, as the denser ISM where grain-growth can occur is also the material that should be fueling ongoing star formation. But in data with resolution as high as ours, as good as 14\,pc in the LMC, it is surprising to find that regions with recent star formation, and corresponding higher density of \sne\ shocks and hard radiation, still manage to have higher D/H. 

The origin of this surprising finding warrants further investigation in future work. One conceivable explanation is that we are tracing the creation of fresh dust by core-collapse \sne, which will be preferentially found in regions of recent star formation. If so, this has major implications for the role of \sne\ as net creators, versus destroyers, of dust \citep{Nozawa2006C,Preistley2021A}. Dust creation by asymptotic giant branch (ABG) stars, seems unlikely to be the cause. This is because D/H is visibly well correlated with the ISM structure in our galaxies; AGB stars, however, will have a distribution following that of each galaxy's evolved stellar population, which has a very different distribution in these galaxies.

We also note that Figure~\ref{AppendixFig:DtH_vs_H_Ionised} in Appendix~\ref{AppendixSubsection:Ionized_Gas} likewise suggests that greater ionized gas surface density, as traced by H$\alpha$ observations of the LMC and SMC, does not appear to be associated with reduced D/H, and any \SigmaDeproj. However, as this data is not available for M\,31 and M\,33, and with poorer resolution than provided by {\it Swift}-UVOT, we consider this a secondary line of evidence.

For all four galaxies, these results suggest that neither dust destruction by young stars (via hard radiation or shocks from core-collapse \sne), nor greater efficiency of dust destruction in lower-density ISM, nor the presence of ionized gas, are the primary driver of the evolution in D/H with \SigmaDeproj\ observed in Figure~\ref{Fig:DtH_vs_H}. This supports the explanation that this relationship is tracing increasing dust grain growth at higher densities.}

\needspace{2\baselineskip} \subsubsection{Variation in D/H Evolution Between Galaxies} \label{Subsubsection:DtH_Galaxy_Variation}

There are some striking similarities and differences between the relationships in how D/H evolves with \SigmaDeproj\ for each of the galaxies in Figure~\ref{Fig:DtH_vs_H}. The first thing of note is how extremely similar the profiles of M\,31 and M\,33 are to each other. Over the 0.6--30\,${\rm M_{\odot}\,pc^{-2}}$ range in \SigmaDeproj, they overlap near-perfectly\footnote{We are able to probe to lower densities for M\,31 than M\,33, primarily because of M\,31's greater inclination. Surface densities that would not be detectable in a less-inclined galaxy can be observed in M\,31, as the sampled ISM column is longer at higher inclination. Our deprojection corrections, discussed earlier in Section~\ref{Subsection:DtH_versus_Sigma_H}, allow us to compare galaxies directly despite this.}. They even experience their D/H turnovers at the same surface density, $\approx4\,{\rm M_{\odot}\,pc^{-2}}$. This would superficially suggest that the ISM in these galaxies has similar properties. This is, however, rather surprising, given that M\,31 has an average metallicity more than 2.5 times greater than M\,33, and a stellar mass more than 25 times greater. 

The representative dust SED parameters for M\,31 and M\,33 do indeed differ (see Table~\ref{Table:DustBFF_Params}), most notably in terms of $\beta$. M\,31 has higher $\beta$ values (ie, a steeper Rayleigh-Jeans slope), with minimal difference between $\beta_{1}$ and $\beta_{2}$. M\,33, on the other hand, as much lower values of $\beta$, more characteristic of lower-mass and dwarf galaxies \citep{Remy-Ruyer2013A}. Plus, M\,33 has a much more pronounced break to a shallower $\beta_{2}$ at longer wavelength; again more characteristic of dwarf galaxies \citep{Gordon2014B}. This is extremely interesting, as $\beta$ {can potentially trace} actual physical properties of the dust grains in question. In general, higher values of $\beta$ are expected from crystalline grains, silicate species, or coagulated grains; while lower values of $\beta$ are expected from metallic grains, amorphous grains, and carbonaceous species \citep{Tielens1987A,Kohler2015A,Jones2017A,Ysard2018A}. {The well-known temperature--$\beta$ degeneracy \citep{Shetty2009A,DJBSmith2013A} limits what we can infer from the parameters derived from any one pixel. However, because we sample very large numbers of pixels, and because the pixels are larger than the instrumental PSF (and hence statistically independent) the overall trends we find ought to be robust against the temperature--$\beta$ degeneracy \citep{MWLSmith2012B}. This is because the degeneracy does not impart a {\it bias} to the results of the fits; any trend observed over very large numbers of pixels cannot therefore be caused by the degeneracy}. Large-scale variation in $\beta$ therefore {hints at} different dust compositions (either because of their elemental compositions, or the environments in which the grains formed; see \citealp{Roman-Duval2022A}. But despite this, the global D/H ratios for M\,31 and M\,33 are effectively identical, being $10^{-2.11}$ versus $10^{-2.13}$ -- matching their similarity in D/H versus \SigmaDeproj\ profile.

In contrast, if any two galaxies in our sample might be {\it expected} to have ISM that behaves similarly, it would be M\,33 and the LMC. Their stellar masses, metallicities, and star-formation rates \citep{Harris2009A,Verley2009A} are very similar. Both galaxies are bound to massive spiral companions, with which they have undergone interactions (although M\,33 is more distant from its companion, and considerably less disturbed, with the last major encounter likely \textgreater\,1\,Gyr ago; \citealp{Bekki2008H,McConnachie2009D}). However, despite their similarity, the two galaxies show strikingly different profiles in their D/H evolution in Figure~\ref{Fig:DtH_vs_H}. The profile of the LMC would appear depressed compared to that of M\,33 by a factor of 1.5--3. And while both galaxies have a turnover in D/H, the location of that turnover is at a deprojected surface density of 40\,${\rm M_{\odot}\,pc^{-2}}$ for the LMC, compared to just 4\,${\rm M_{\odot}\,pc^{-2}}$ for M\,33 (and M\,31). Moreover, M\,33 (and M\,31) shows very steep increase in D/H with \SigmaDeproj\ before the turnover, with a power law index of 1.1 over the 0.6--4.0\,${\rm M_{\odot}\,pc^{-2}}$ range; whereas the D/H increase for the LMC happens much more gradually, with a power law index of only 0.39 over the 2.5--40\,${\rm M_{\odot}\,pc^{-2}}$ range.

The fact that two galaxies as superficially-similar as M\,33 and the LMC can have such starkly different D/H properties is significant, especially in the context of the common practice of inferring a galaxy's gas mass from observations of dust emission \citep{Eales2010C,Scoville2014B}. This is a standard method for estimating the ISM mass of galaxies for which direct measurements of CO and/or \HI\ are not available. It has been shown to be remarkably reliable for massive galaxies at high masses \citep{Scoville2014B,Scoville2016B}, but clearly extending the technique to more intermediate masses will have to be done with care, given that even fundamental galaxy properties like metallicity and stellar mass are not accurate predictors of a galaxy's D/H ratio. Moreover, while D/H for M\,33 and the LMC only differ by a factor of 2, it should be remembered that the method of estimating gas mass from dust emission tends not to use dust {\it mass}, but instead simply use dust luminosity in some longer-wavelength band such as 500 or 850\,\micron\ -- and the 500\,\micron\ luminosities of M\,33 and the LMC differ by a factor of \textgreater3. 

Lastly, the SMC clearly follows a very different evolutionary profile than the other galaxies. Not only does D/H continue to increase over the entire factor of 50 in surface density we sample, but moreover the gradient of the profile continues to get steeper, even up to the very highest density we are able to trace. Indeed, above 150\,${\rm M_{\odot}\,pc^{-2}}$, the highest \SigmaDeproj\ available for the SMC, the D/H profile for the SMC appears as though it may intersect that of the LMC. {This suggests that grain growth may have a particularly strong dependence with density in the SMC}. We can also be sure that the dust grains themselves have different properties in the SMC. Not only do most regions in the SMC appear to lack the 2175\,\textup{~\AA} extinction bump seen in higher-metallicity systems \citep{Gordon2003B,Murray2019C}, but our SED fitting finds the SMC to have a much lower $\beta$ than the other galaxies in our sample, along with the most significant flattening in $\beta$ at longer wavelengths (see Table~\ref{Table:DustBFF_Params} and Figure~\ref{Fig:SED_Params_Grid_2}), all indicating a different composition.

\needspace{2\baselineskip} \subsection{Modeling the Evolution in D/H} \label{Subsection:DtH_Modeling}

In order to aid our understanding of the D/H evolution profiles in Figure~\ref{Fig:DtH_vs_H}, we use the dust evolution model of \citet{Asano2013A} . This model traces how various galactic environmental parameters can affect accretion of metals onto dust grains -- such as the density of the ISM, the metallicity of the gas, the temperature of the existing dust grains, and the characteristic lifetime of the molecular clouds where grain growth occurs, and the average grain size. The model balances this with the rate at which supernov\ae\ can destroy dust grains, given the density and metallicity of the ISM. 

Using this model, the timescale for accretion of metals onto dust grains, $\tau_{\it acc}$, is given by:

\begin{multline}
\tau_{\it acc} = \\
2\times10^{7} \left(\frac{\overline{a}}{0.1\,{\rm \mu m}}\right) \left(\frac{n_{H}}{100\,{\rm cm^{-3}}}\right)^{-1} \left(\frac{T_{d}}{50\,{\rm K}}\right)^{-\frac{1}{2}} \left(\frac{Z}{0.02}\right)^{-1}
\label{Equation:Chemev_Tau_Accretion}
\end{multline}

\noindent where $\overline{a}$ is the average dust grain size in \micron, $n_{H}$ is the volumetric number density of hydrogen in ${\rm cm^{-3}}$, and $Z$ is the metallicity (in units of the metal mass fraction). With $\tau_{\it acc}$, one can then compute the fraction of the metal mass that is in dust grains (see \citealp{Zhukovska2008A}), using:

\begin{equation}
f_{d} = \frac{e^{t_{\it growth}/\tau_{\it acc}} f_{d_{0}}}{1 - f_{d_{0}} + (f_{d_{0}} e^{t_{\it growth}/\tau_{\it acc}})} 
\label{Equation:Chemev_Frac_Dust}
\end{equation}

\noindent where $t_{\it growth}$ is the average duration of episodes of grain-growth (if grain-growth happens predominantly in molecular clouds, then $t_{\it growth}$ corresponds to the average molecular cloud lifetime), and $f_{d_{0}}$ is the minimum fraction of metals that can be locked up in dust grains. Given this, the D/H for a given set of parameters is simply given by:

\begin{equation}
\frac{D}{H} = f_{d}\,Z
\label{Equation:Chemev_DtH}
\end{equation}

\noindent which naturally gives the result that D/H will peak at a value equal to the metallicity, when all metals are locked up in dust grains at $f_{d} = 1$. {This grain growth trend in the \citet{Asano2013A} model is qualitatively similar to that found in the \citet{Nanni2020C} model, also.}

Because we are observing surface densities, we are unable to directly measure $n_{H}$. We therefore need to incorporate a conversion factor $H_{\Sigma \Rightarrow n}$, in units of ${\rm cm^{-3} M_{\odot}^{-1} pc^{2}}$, such that: 

\begin{equation}
n_{H} = H_{\Sigma \Rightarrow n} \times \Sigma_{H}^{\it (deproj)} 
\label{Equation:Chemev_H_Conversion}
\end{equation}

Clearly, there are many potentially-tunable parameters involved. Fortunately, we are not necessarily interested in ascertaining the `true' values of these various parameters. Rather, we are concerned with identifying the general shape of the evolutionary trend we'd expect, and what relative differences in the parameters could potentially lead to the differences in D/H evolutionary profiles we find.

In order to find the best fit of this model framework to each of our galaxies, we use a model grid consisting of reasonable values of each of the parameters. The grid parameters are given in Table~\ref{AppendixTable:DtH_Evolution_Grid}; a full discussion of motivation of the parameter ranges is given in Appendix~\ref{AppendixSection:DtH_Params}. 

We assume an average grain radius of $\overline{a} = 0.1\,{\rm \mu m}$ \citep{Inoue2011C,Nozawa2011A,Asano2013A}. For each bin, we took the median of the median dust temperature from the SED-fitting of every pixel in the bin, and used this for that bin's $T_{d}$ value.

For each bin, we also assume a known metallicity. Given the lack of systematic variation in metallicity within the LMC and SMC, as discussed above, we use the fixed values of 0.5 and 0.2\,$\rm{Z_{\odot}}$, respectively, for all density bins in these galaxies. For M\,31 and M\,33, we also use fixed metallicities for all density bins, using the representative global metallicities, of 1.3 and 0.5\,$\rm{Z_{\odot}}$, respectively, even though these galaxies do have metallicity gradients. This is because each bin contains pixels representing a complex distribution of metallicities. For instance, in M\,31, there are low-density pixels with $\Sigma_{H} \approx 0.1\,{\rm M_{\odot}^{-1} pc^{2}}$ located in the galaxy's gas-poor center where $Z > 1.5\,{\rm Z_{\odot}}$, and in the galaxy's gas-poor outskirts where $Z < 0.75\,{\rm Z_{\odot}}$. The mixture of metallicities at each density varies considerably, and taking the average, for instance, of the metallicities within each bin leads to pathological model behavior. So instead, we use the fixed {metallicities}, which is in keeping with the understanding that this modeling is intended to be representative, as opposed to accurately capturing the specific physical conditions in these systems. 

\begin{figure}
\centering
\includegraphics[width=0.475\textwidth]{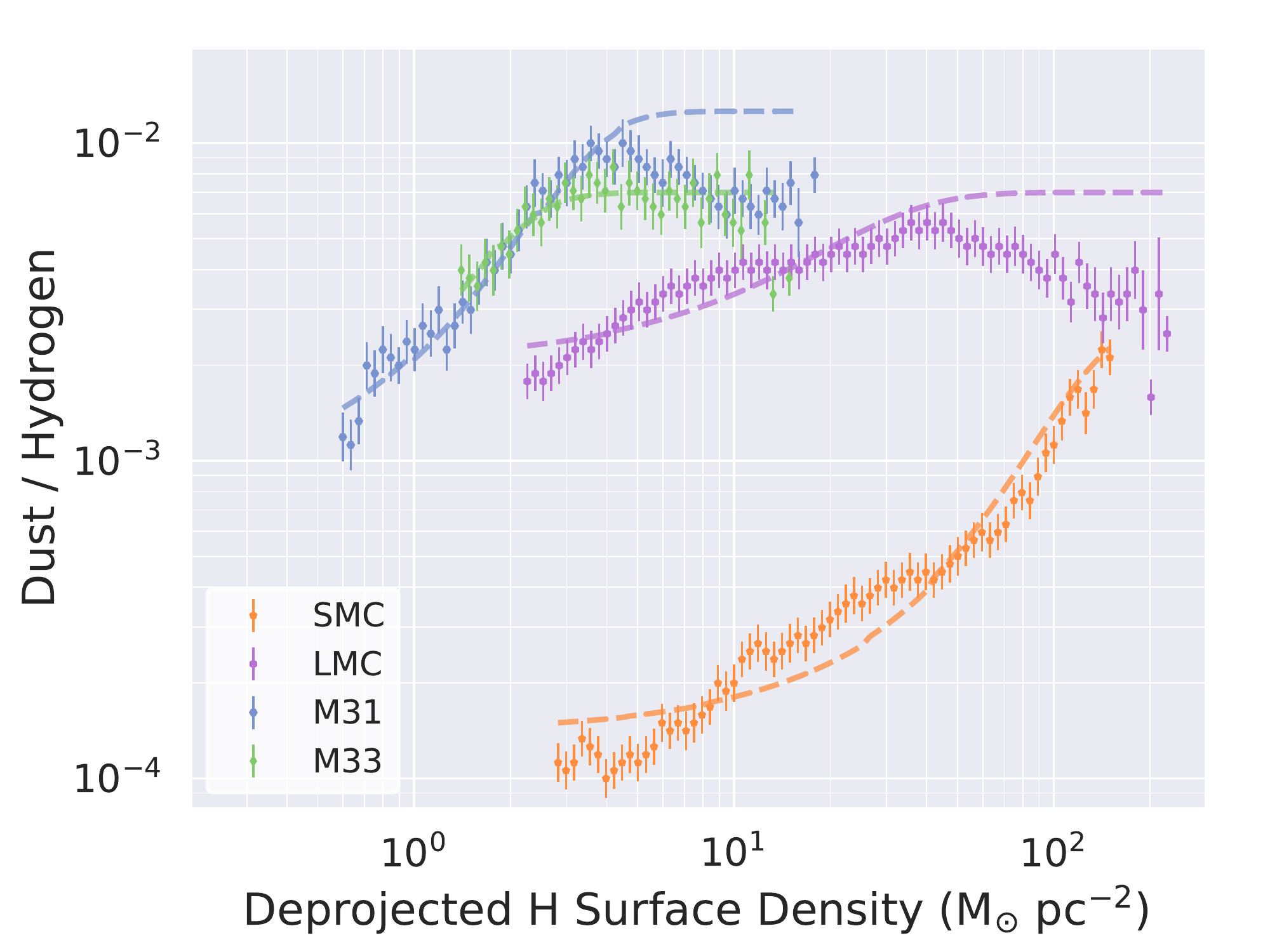}
\caption{Plot of D/H against \SigmaDeproj. Data points are the same as those shown in Figure~\ref{Fig:DtH_vs_H}, but now with the addition of the best-fit dust evolution model for each galaxy. The parameters that give the model for each galaxy are provided in Table~\ref{AppendixTable:DtH_Evolution_Best}.}
\label{Fig:DtH_vs_H_ChemEv}
\end{figure}

Using the given \SigmaDeproj, $Z$, $T_{d}$, and $\overline{a}$ values for each bin, we performed a $\chi^{2}$-minimizing grid search to find the best-fit parameters for each galaxy. Because the model will ultimately plateau at D/H = $Z$ above a certain density threshold, where all metals are found in dust grains, the model cannot incorporate the turnover in D/H observed for M\,31, M\,33, and the LMC; we therefore only fit the model to the points before the turnover in these cases (below 4\,${\rm M_{\odot}\,pc^{-2}}$ for M\,31 and M\,33, and below 40\,${\rm M_{\odot}\,pc^{-2}}$ for the LMC). A fuller discussion of this divergence will be conducted in Section~\ref{Section:Causes_of_Turnover}; however, we present the modeling here first, so that in the following sections we can then explore how to potentially reconcile the data with the models. The resulting best-fit parameter values are given in Table~\ref{AppendixTable:DtH_Evolution_Best}, with the models shown on a plot of \SigmaDeproj\ versus D/H in Figure~\ref{Fig:DtH_vs_H_ChemEv}.

Starting with M\,31 and M\,33, we can see in Figure~\ref{Fig:DtH_vs_H_ChemEv} that -- up until the turnover -- the \citet{Asano2013A} model almost perfectly traces the observed evolution in D/H with \SigmaDeproj. For M\,31, the model matches the observed profile across almost an order of magnitude in both D/H in \SigmaDeproj, from 0.6--4\,${\rm M_{\odot}\,pc^{-2}}$, and successfully replicates the increasing steepness of the D/H evolution, followed by a leveling-out. Our data do not probe down to surface densities quite as low as this for M\,33, but otherwise the agreement between model and data for $\Sigma_{H}^{\it(deproj)}\,<\,4\,{\rm M_{\odot}\,pc^{-2}}$ is similarly excellent.

For the LMC, the agreement between model and data in the 2.2--40\,${\rm M_{\odot}\,pc^{-2}}$ range, before the D/H turnover, is not quite as close a match, but nonetheless does broadly trace the increase in D/H, followed by a leveling-out.

For the SMC, the model does a similarly mixed job of fitting the data -- and the lack of a D/H turnover or plateau means that our model can be fit to the full range densities. The D/H of the SMC over the 8--50\,${\rm M_{\odot}\,pc^{-2}}$ range is somewhat elevated over what is expected from the best-fitting model. But otherwise, the model fits the rest of the profile for the SMC reasonably well, capturing the gentle slope at lower densities, and the steepening at higher densities.

So it would appear that the \citet{Asano2013A} model does a good job of capturing the broad strokes of evolution in the D/H.  Our data reflect the expectation in the \citet{Asano2013A} framework that grain-growth gets increasingly efficient as density increases; but only up until where all metals are locked into dust grains, where it plateaus.  For M\,31, M\,33, and the LMC in particular, we find it to be very reassuring that the levels of the model plateaus are all close to the levels of the peak D/H for each galaxy. Because the D/H plateau indicates when 100\% of ISM metals are locked up in dust grains, the D/H at which it occurs is not free to vary in the model, but rather happens at a fixed level of \mbox{D/H = $Z$}. For instance, in the case of the LMC, the plateau happens at D/H = $0.5\,{\rm Z_{\odot}} = 0.5 \times 0.014 = 7 \times 10^{-3}$, which is very close to the actual highest value of D/H we measure in the LMC, of $5.5 \times 10^{-3}$. The agreement is similarly reasonable for the other galaxies. Given that the values D/H we measure do not `know' the metallicity of their galaxy's ISM, and therefore the D/H at which the model plateau will occur, this agreement suggests that our measured D/H values are sensible.

Where the data and model diverge radically, however, is in the apparent turnover in D/H that three of our galaxies exhibit. That is the focus of the following section.

\needspace{2\baselineskip} \section{The Nature of the D/H Turnover} \label{Section:Causes_of_Turnover}

The most surprising feature in Figure~\ref{Fig:DtH_vs_H} is undoubtedly the fact that, for M\,31, M\,33, and the LMC, D/H does not plateau as expected from modeling, but rather turns over and starts to decrease above a certain value of \SigmaDeproj.

This is extremely surprising. As ISM density increases, the efficiency of dust grain growth should also increase. The greater the density of the ISM, the more frequently a dust grain will encounter gas phase metals, and therefore the more frequently that dust grain will increase its mass through accretion of those metals, driving up D/H. Moreover, higher-density ISM will provide superior shielding from the forces of grain destruction, such as shocks and high-energy radiation. Therefore, D/H should continue to increase with density, until grain-growth starts to saturate as fewer and fewer metals are left in the gas phase to accrete, ultimately reaching $f_{d} = 1$ -- hence the plateau predicted by the \citet{Asano2013A} model.

And there are further reasons to doubt whether this turnover could actually be present for our target galaxies. For instance, at \SigmaDeproj\,$>\,150\,{\rm M_{\odot}\,pc^{-2}}$, it appears that the D/H profile for the SMC will intersect the profile of the LMC. And it would be extremely surprising if the SMC were to have a D/H that exceeds that of the LMC at a given \SigmaDeproj, given the fact their metallicities differ by a factor of 2.5. Indeed, the peak D/H in the LMC is $\approx$2.5 times greater than that in the SMC.

In short, it is hard to conceive of any physical mechanism by which D/H could {\it actually} fall as density increases. So we examined the question of what could be causing this to {\it appear} to be the case in our data. We considered 6 possible reasons: Variations in $\alpha_{\it CO}$; noise-induced anticorrelation; physical resolution effects; dust destruction by supernovae and high-energy radiation in high-density environments due to star formation; the presence of dark gas, and varying dust mass opacity. Our full exploration of these possible explanations is presented in Appendix~\ref{AppendixSection:Turnover}. For readers not wishing to explore this investigation in full, the results are summarized as follows:

We are confident that we can rule out overestimation of $\alpha_{CO}$, or elevated dust destruction due to environmental effects, as causes of the turnover. It also appears that neither physical resolution limitations nor noise-induced anti-correlation could be causing the turnover, either. 

It seems likely that the presence of dark gas could be contributing to the appearance of the D/H bump and turnover for M\,31 and M\,33, and driving up D/H at the peak values in the LMC and SMC. However, at the same time, we are confident that dark gas could {\it not} be causing significant bias in the overall D/H evolution profile for the LMC, nor causing the turnover. This suggests that dark gas may not the only effect acting in M\,31 and M\,33, either.

The possibility that the dust mass absorption coefficient, $\kappa$, has a decreasing value at higher densities provides a viable solution to the apparent turnover, instead changing the high-density portion of the D/H evolution profile into a plateau, as we would expect based on dust evolution modeling. This agrees with the empirical result reported in \citet{CJRClark2019B}, but conflicts with predictions from physical dust grain models (in which $\kappa$ is expected to increase with rising density). 

Overall, we consider a combination of dark gas and varying $\kappa$ to be the most likely explanation, but with the expectation that the contributions of each, and of other possible factors, is likely varying between environments. 


\begin{figure*}
\centering
\includegraphics[width=0.9\textwidth]{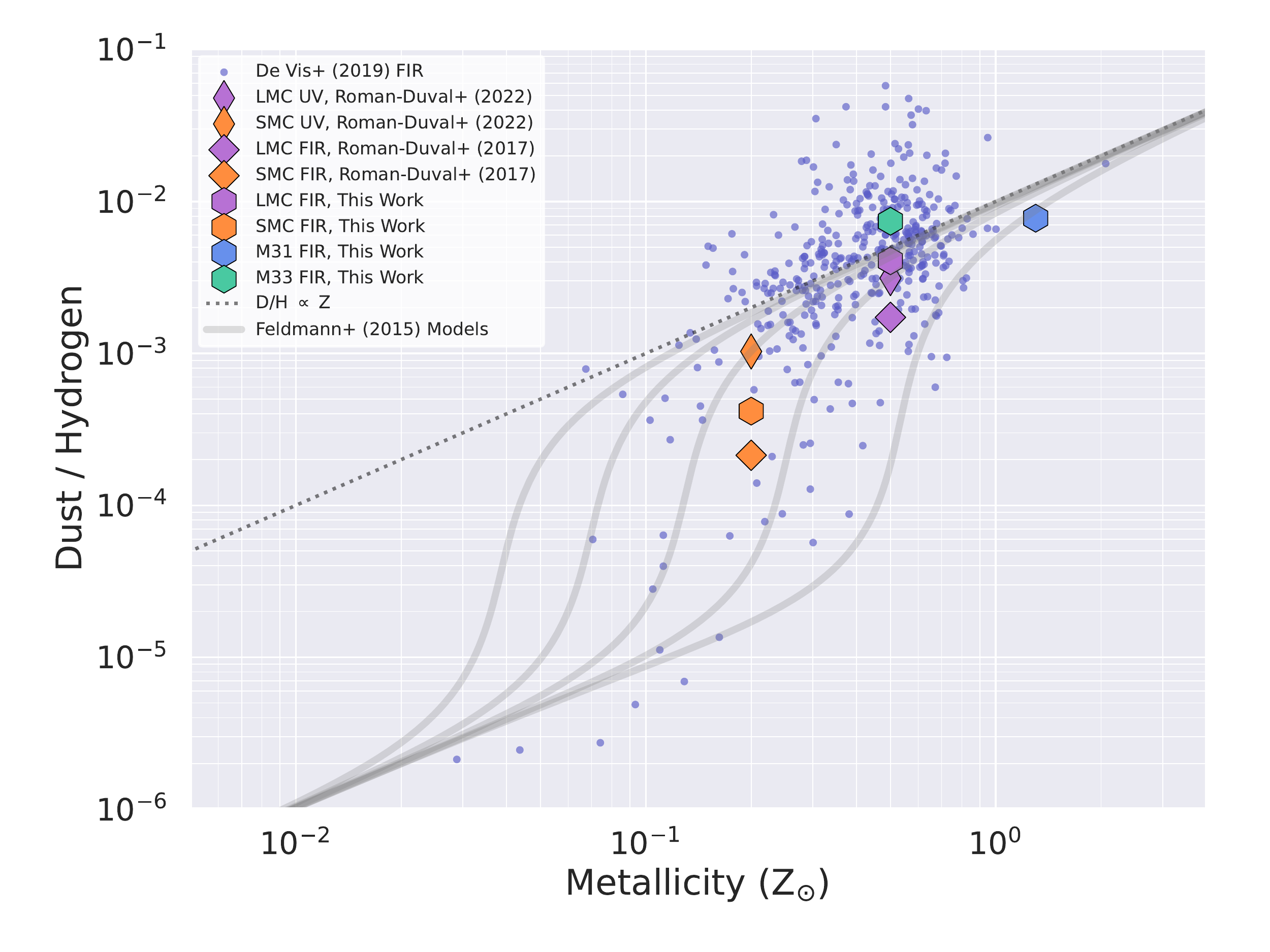}
\caption{The relationship between D/H and Z, showing observations and models from various sources. The large polygonal points indicate D/H values for the four galaxies in our sample, from this work (hexagons), from the previous low-resolution FIR measurements of \citeauthor{Roman-Duval2017B}\,(\citeyear{Roman-Duval2017B}; {wide diamonds}), and the UV absorption line measurements of elemental depletions from the {\it Hubble} METAL program \citeauthor{Roman-Duval2022A}\,(\citeyear{Roman-Duval2022A}; {narrow diamonds}). {The small circular points indicate values from the sample of \citeauthor{DeVis2019B} (\citeyear{DeVis2019B}; which incorporates and standardized measurements of galaxies from the samples of \citealp{Remy-Ruyer2014A,CJRClark2015A,DeVis2017A,Davies2017A,CJRClark2018A}), the largest sample to date exploring this parameter space}. The dotted line shows a trend of the D/H\,$\propto$\,Z (ie, a constant dust-to-metals ratio), passing through Z=Z${\rm _{\odot}}$, D/H=0.01. The thick gray lines show a selection of dust evolution models from \citet{Feldmann2015C}, {with a different location for each model of the `critical metallicity' at which D/H sharply increases with Z; specifically, we plot the \citet{Feldmann2015C} `equilibrium models' where the location of the critical metallicity is set by the ratio of molecular gas depletion timescale to the interstellar grain growth timescale, with lower ratios leading to higher critical metallicities (see their Figure~3).}}
\label{Fig:DtH_vs_Z}
\end{figure*}

\needspace{2\baselineskip} \section{The Discrepancy Between FIR to UV Dust-to-Gas Ratios} \label{Section:DtH_Reconcile} 

A major motivation for this work is the conspicuous disagreement of D/H measurements for the LMC and SMC derived from UV absorption line spectroscopy of elemental depletions, when compared to D/H measurements derived from FIR and radio observations. As mentioned in Section~\ref{Section:Introduction}, the FIR D/H estimates previously reported for the Magellanic Clouds are much smaller than those determined from depletions. This disagreement is clear in Figure~\ref{Fig:DtH_vs_Z}, with the previous FIR D/H measurements of \citet{Roman-Duval2017B} much lower than the UV D/H measurements found by the {\it Hubble} program Metal Evolution, Transport, and Abundance in the \textsc{Lmc} (METAL; \citealp{Roman-Duval2021B,Roman-Duval2022A})\footnote{Not plotted on Figure~\ref{Fig:DtH_vs_Z} are even earlier estimates of D/H from \citet{Roman-Duval2014D}. These used the older HERITAGE maps of the Magellanic Clouds, and employed the SED fitting results presented by \citet{Gordon2014B}. As discussed in Section~\ref{Subsubsection:Global_Values}, the \citet{Gordon2014B} dust masses are biased high, due to the large amount of missing flux in the HERITAGE \hersc-PACS 100 and 160\,\micron\ datam causing their SED fitting to output erroneously high dust temperatures.}. The \citet{Roman-Duval2017B} D/H measurements were performed using FIR data from \planck\ and IRAS. The disagreement between the D/H determined using the two methods is striking -- being a factor of $\sim$2 for the LMC, and a factor of $\sim$5 for the SMC.

The lower, FIR-derived value suggests the SMC has a significantly depressed D/H relative to its metallicity, and would indicate that the SMC is in the midst of the critical metallicity transition; the higher UV-derived value, however, suggests that the SMC has a D/H reasonably in line with the trend seen at higher metallicities, assuming a roughly constant dust-to-metals ratio. The situation is similar for the LMC; the lower FIR value would hint that the LMC is just starting to experience depressed D/H; the higher UV value would instead place the LMC on the constant dust-to-metals relation.

If the UV-derived D/H values are correct, that would mean that there are severe shortfalls in the dust masses determined for local galaxies, with the vast majority of the dust mass in the SMC being unaccounted for by previous studies such as \citet{Chastenet2017A} and \citet{Roman-Duval2017B}. On the other hand, it is possible that the UV-derived values are overestimates. 

While there should be less scope for such significant systematic error in the UV measurements, it is conceivable. For instance, \citet{Roman-Duval2022A} calculate the fractions of hard-to-observe carbon and oxygen (major dust constituents by mass) depleted into the dust phase in the Magellanic Clouds, by using the apparently-invariant relation between the depletion of these elements, and other more-easily-observed elements, with increasing H column \citep{Jenkins2009B}. It is possible that this relation becomes unreliable in the Magellanic Clouds, thereby affecting the fraction of metals in the dust phase at given densities.

We also note that directly comparing the $\Sigma_{H}$ measurements derived from UV, to those derived from the 21\,cm and CO radio data, can be troublesome. The target stars for the UV sightlines will each be located at a different depth in that galaxy's disc. On average, a given sightline will sample half of the thickness of the disc; we could therefore try applying a constant factor of 2 correction to the $\Sigma_{H}$ of each UV sightline, to attempt to correct for this. However, the true (unknown) correction will vary wildly between sightlines. This would place us in the undesirable situation of applying a `correction', where the uncertainty on that correction is much larger than the correction itself. Moreover, each UV sightline has the same radius as the target star, and so will therefore only be sampling a pencil beam of ISM tens of millions of kilometers across – in contrast to the tens of parsecs resolution of the 21\,cm and CO observations we use (the effects of this difference are analyzed in depth in Section~7 of \citealp{Roman-Duval2021B}). We therefore opt to only apply our standard deprojection correction to the surface densities derived from the UV data, and advise that the reader remains aware that the radio- and UV-derived $\Sigma_{H}$ values cannot be compared in a wholly `apples-to-apples' manner.

While the use of FIR all-sky survey data by \citet{Roman-Duval2017B} ensures that no diffuse emission was missed (in contrast to the old HERITAGE reductions of the Magellanic Cloud \hersc\ data), their use of \planck\ and IRAS data may have limited the accuracy of their results. For instance, they used IRAS data from the \textsc{Iras} Sky Survey Atlas (ISSA; \citealp{Wheelock1994A}), which suffers from a very non-linear detector response, that varies as a function of both the surface brightness and the angular scale of the emission observed (leading the ISSA explanatory supplement \citep{Wheelock1994A} to suggest 100\,\micron\ photometric uncertainty of up to 60\%), whereas our new feathered \hersc\ maps are pegged to the absolute calibration of the COsmic Background Explorer (COBE; \citealp{Boggess1992B}). Also, the peak of the dust temperature distribution lies in the \textgreater\,0.5\,dex gap in wavelength coverage between the 100\,\micron\ IRAS band and the 350\,\micron\ \planck\ band, likely limiting the accuracy of the \citet{Roman-Duval2017B} SED fitting, whereas our data samples this regime with the 160 and 250\,\micron\ bands. Plus, unlike us, \citet{Roman-Duval2017B} did not allow for a broken $\beta$ in their SED fitting, which may have compromised their results in areas with significant submm excess (see Figure~\ref{Fig:SED_Params_Grid_2}).

For both the LMC and SMC, the \citet{Roman-Duval2017B} relationships between D/H and \SigmaDeproj\ trace similar {\it slopes} to our own, over the range we have in common. However their trends are offset to much lower D/H (despite the fact they use the same $\kappa$ as we do). This is presumably due to the improvements in our FIR data and SED fitting. 

\begin{figure}
\centering
\includegraphics[width=0.475\textwidth]{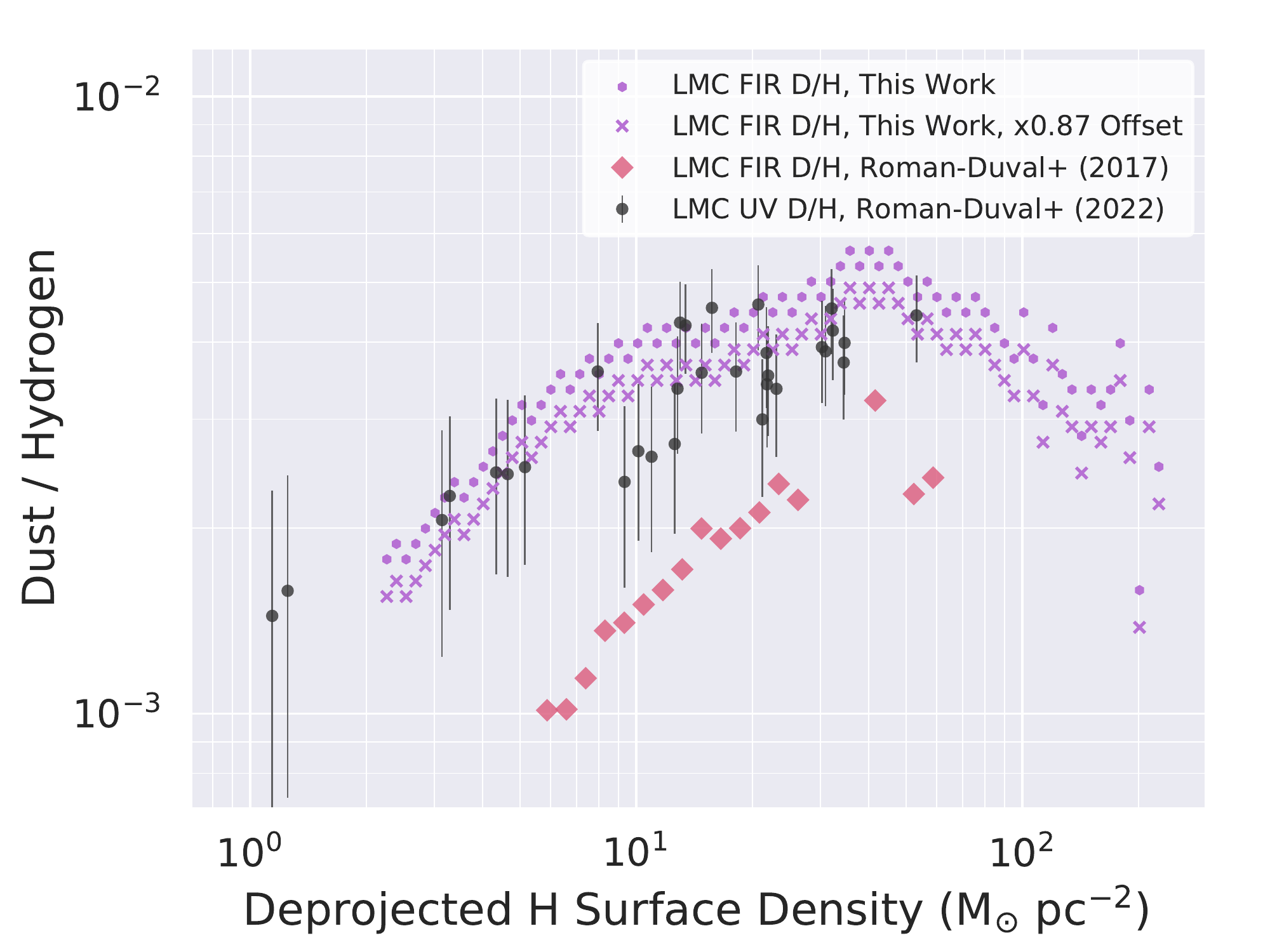}
\caption{Plots of D/H against $\Sigma_{H}$ for the LMC, comparing the measurements reported at various densities from several sources. Plotted in black circles are D/H values from UV absorption line measurements of elemental depletions by \citet{Roman-Duval2022A}. Plotted in pink diamonds are D/H values from IRAS--\planck\ FIR data by \citet{Roman-Duval2017B}. The values from this work are shown for each galaxy, both before and after having an offset in $\kappa$ applied, to make our FIR values match the UV values of \citet{Roman-Duval2022A}. Error bars on our points are omitted for clarity, but would be the same as in Figure~\ref{Fig:DtH_vs_H}.}
\label{Fig:LMC_DtH_vs_H_Offset}
\end{figure}

As a result of the consistently larger D/H values we find, there is now much less conflict between our FIR estimates of D/H, and the UV estimates of D/H from \citet{Roman-Duval2021B,Roman-Duval2022A}. For the LMC, the conflict has essentially been resolved entirely. Indeed, as can be seen in Figures~\ref{Fig:DtH_vs_Z} and \ref{Fig:LMC_DtH_vs_H_Offset}, our D/H are now slightly {\it greater} than those determined from the UV. To evaluate the scale of the remaining difference, we calculated what offset factor would have to be applied to our D/H profile to minimize the difference between it the UV measurements. For each UV measurement of D/H, we calculated the D/H implied by our profile at that same $\Sigma_{H}$ (by interpolating between the points in our profile that bracket the $\Sigma_{H}$ of the UV point), and found the difference between them. Having repeated this for every point, we then used a $\chi^{2}$-minimizing routine to find what offset factor, applied to our D/H profile, would minimize the overall difference between the two datasets.

For the LMC, we find that reducing our D/H values according to a factor of 0.87 would lead to the closest agreement between them and the UV measurements. Both the original and offset D/H profiles are shown in Figure~\ref{Fig:LMC_DtH_vs_H_Offset}. Moreover, not only does the absolute D/H level now match very well, but the {\it slope} of our D/H evolution profile is also an excellent match to that of the UV data.  Given the calibration uncertainties on both datasets, this 13\% disagreement between the UV and our FIR measurements of D/H is small enough that we feel confident in saying that the discrepancy can be deemed resolved in the case of the LMC.

Exceedingly frustratingly, however, the maximum surface density to which the {\it Hubble} data of \citet{Roman-Duval2021B,Roman-Duval2022A} could probe D/H corresponds exactly to the location of the D/H turnover. Their UV data does not probe above this density, and hence cannot help us ascertain the nature of the turnover. The {\it Hubble} data does not probe to higher densities than this because extinction increases with $\Sigma_{H}$, and above this density it becomes impossible to adequately detect the OB stars used as background sources for this kind of absorption line spectroscopy. Extending UV absorption line spectroscopic depletion measures to higher densities would require a large next-generation UV telescope, such as the Large UltraViolet Optical InfraRed concept (LUVIOR; \citealp{LUVOIR2019A}).

\begin{figure}
\centering
\includegraphics[width=0.475\textwidth]{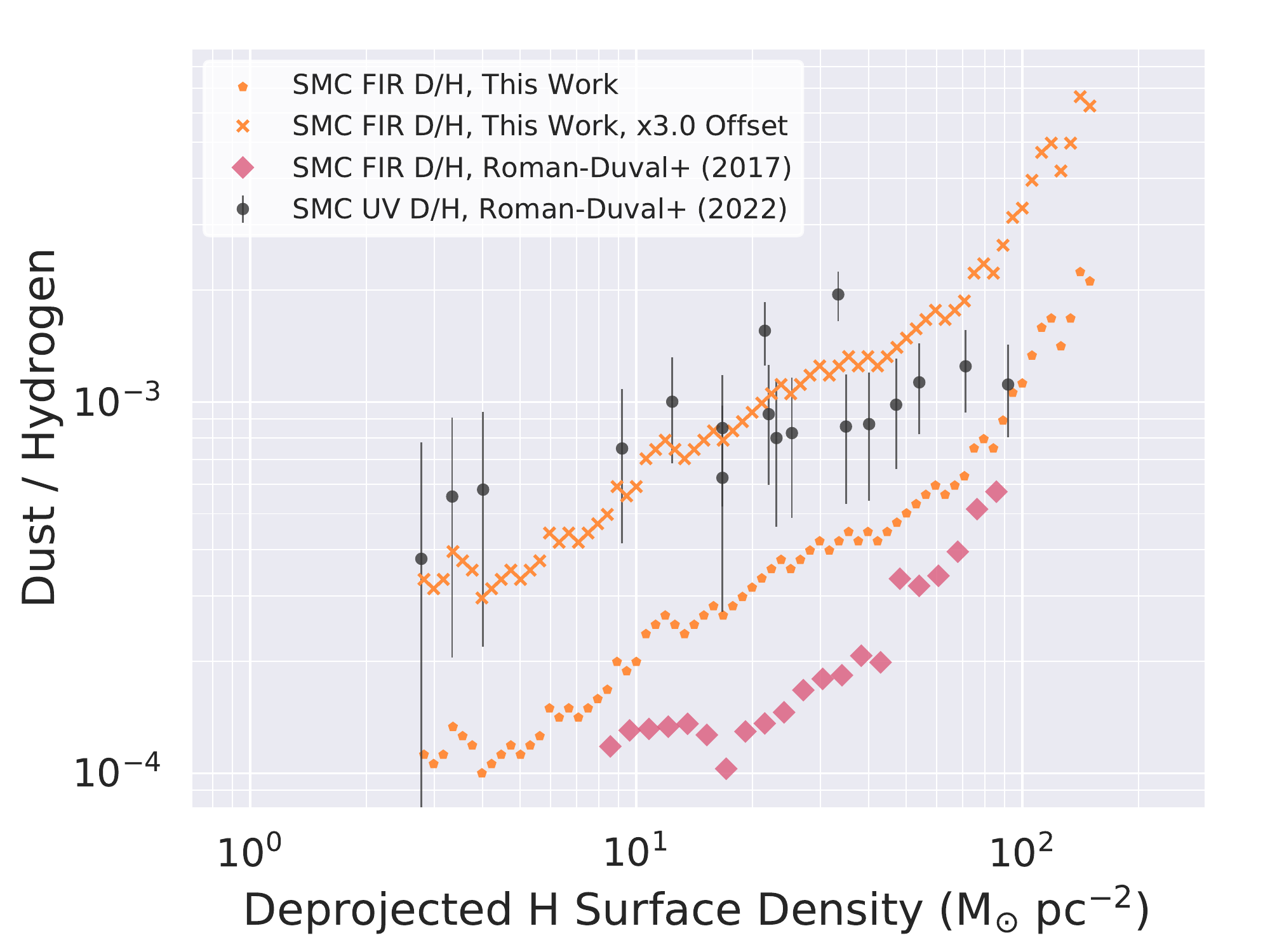}
\caption{Plots of D/H against $\Sigma_{H}$ for the SMC, comparing the measurements reported at various densities from several sources. As details as per Figure~\ref{Fig:LMC_DtH_vs_H_Offset}, except for SMC instead of LMC.}
\label{Fig:SMC_DtH_vs_H_Offset}
\end{figure}

The fact that the discrepancy in UV versus FIR measures of D/H has been resolved for the LMC makes it all the more interesting, however, that a significant discrepancy {\it persists} in the case of the SMC. As can be seen in Figure~\ref{Fig:SMC_DtH_vs_H_Offset}, our new data only partially closes the gap between FIR and UV estimates of D/H in the SMC. Previously, the difference was a factor of 5, as found by the \citet{Roman-Duval2017B} low-resolution FIR analysis; with our new data, this has been reduced to a factor of 3 (with this offset calculated in the same manner as for the LMC, above).

It is not immediately clear why our analysis would resolve the FIR `missing dust' problem for the LMC, but not the SMC. Both use the new \hersc\ data of \citetalias{CJRClark2021A}, reduced and feathered in exactly the same way. The processing and analysis also proceed identically throughout. The physical resolution for the SMC is 47\,pc, versus 14\,pc for the LMC; however, as shown in Section~\ref{AppendixSubsection:Physical_Resolution_Effects}, even a factor of 10 degradation in physical resolution causes no systematic shift in the D/H evolution profile; we would therefore not expect a much smaller change in resolution to lead to any significant bias -- let alone a bias as large as this.

As previously discussed, it is known that the dust properties in the SMC are fundamentally different from those in higher-metallicity galaxies in several ways. Most sightlines explored in the SMC lack the 2175\,\textup{~\AA} extinction bump seen in higher-metallicity systems \citep{Gordon2003B,Murray2019C}; the SMC has stronger submm excess emission \citep{Bot2010A,Planck2011XVII,Gordon2014B}; and the chemical composition of the dust in the SMC is different, too, with iron and carbon being more dominant than at higher metallicities \citep{Roman-Duval2022A}. 

Such large difference in physical dust properties should inevitably lead to some change in $\kappa$. A simple explanation for the persistent D/H offset for the SMC is that it has a dust mass absorption coefficient that is a factor of 2.5 smaller than that of the LMC -- and therefore also a factor of 2.5 smaller than that of the Milky Way, as the $\kappa_{160} = 1.24\,{\rm m^{2}\,kg^{-1}}$ value we use was itself calibrated using the emission and depletions of the Galactic cirrus \citet{Roman-Duval2017B}. Unlike the SMC, dust in both the LMC and Milky way exhibits the 2175\,\textup{~\AA} extinction bump \citep{Gordon2003B}, which meshes with the prospect of the SMC dust being different from Milky Way and LMC in other ways, too. However, dust in the SMC is more carbon-rich than the more silicate-rich dust of the Milky Way (and lesser extent of the LMC; \citealp{Roman-Duval2022A}), and carbon dust should be {\it more} emissive than silicate-dominated dust, not less \citep{Ysard2018A}. However, changes in grain morphology (such as porosity, size, shape, etc), for instance, could counteract the differences expected from composition alone. 

We do, however, wish to re-state that the link between actual volume density (the parameter that will drive grain growth, and hence D/H), and surface density $\Sigma_{H}$ (our observable proxy for volume density) potentially have a less direct relationship for the SMC than for the other galaxies of our sample, due to the complex elongation structure of the SMC along our line-of-sight \citep{Scowcroft2016B}. It is possible that this biases our measurements of D/H at different densities, but the specifics would depend entirely on the nature of three-dimensional structure of the SMC. 

However, even with the remaining discrepancy between our D/H for the SMC and the UV value, we can nonetheless rule out a steep drop in D/H at the metallicity of the SMC suggested by the older FIR value from \citet{Roman-Duval2017B}. This therefore indicates that the SMC is not in the critical metallicity regime. 

\begin{figure*}
\centering
\includegraphics[width=0.9\textwidth]{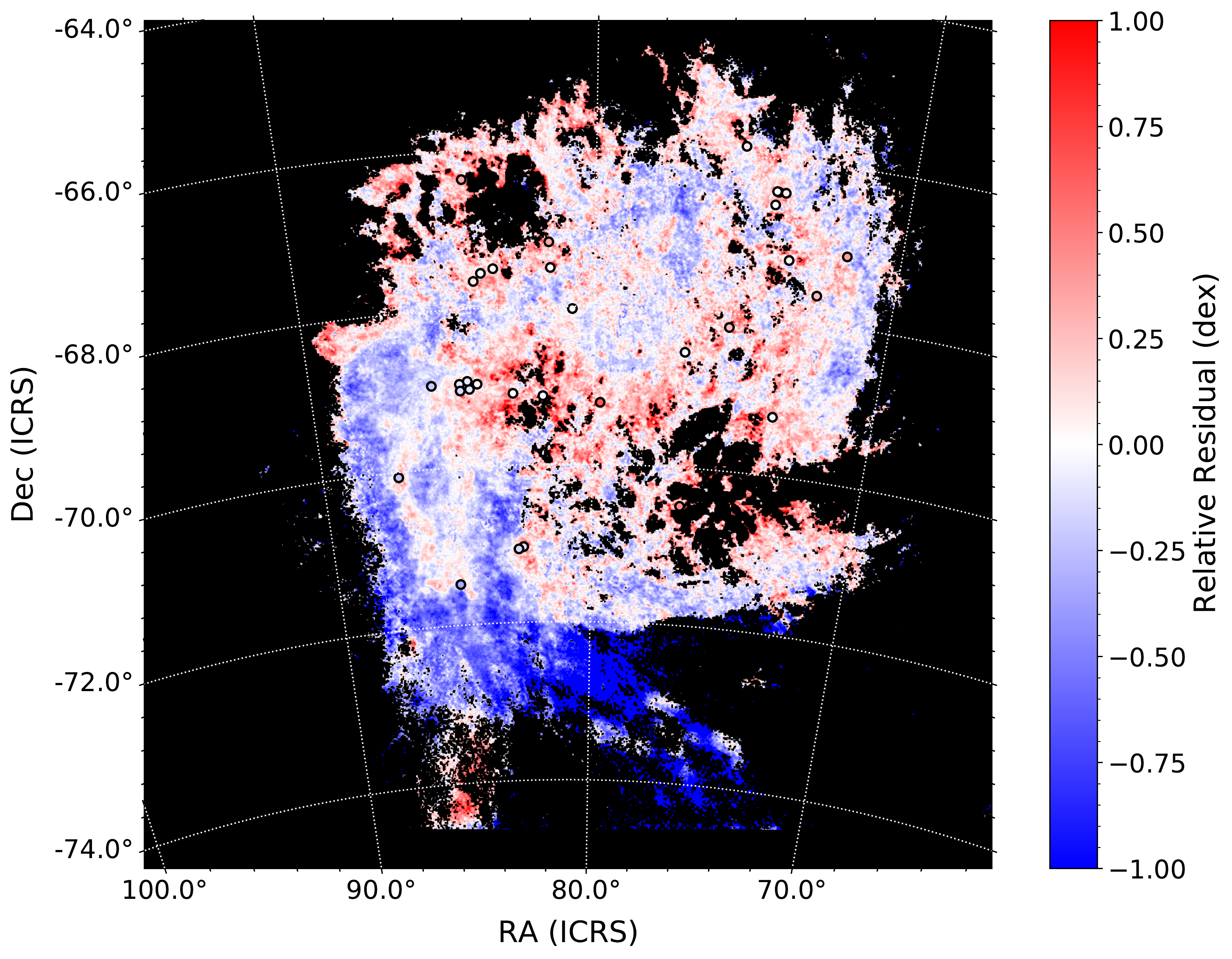}
\caption{A map of the residuals between the D/H measured for each pixel in the LMC, and the D/H we would {\it expect} for each pixel (as calculated using each pixel's $\Sigma_{H}$, and the relationship between $\Sigma_{H}$ and D/H in Figure~\ref{Fig:DtH_vs_H}). Red areas have a higher D/H than would be expected given their $\Sigma_{H}$; blue pixels have lower D/H than would be expected. The circles indicate the UV absorptions spectroscopy sightlines of \citet{Roman-Duval2021B}; the color with each circle indicates the residual between iron depletion (a proxy for D/H) calculated by \citet{Roman-Duval2021B} for that sightline based on its UV spectra, and the depletion they predicted based on the relationship they found between depletion and $\Sigma_{H}$.}
\label{Fig:LMC_Depletions_vs_Residuals_Map}
\end{figure*}

A final note regarding the D/H offsets: When comparing {\it global} D/H, as plotted in Figure~\ref{Fig:DtH_vs_Z}, the discrepancies between our values, and the UV values of \citet{Roman-Duval2021B,Roman-Duval2022A} are different than those in Figures~\ref{Fig:LMC_DtH_vs_H_Offset} and \ref{Fig:SMC_DtH_vs_H_Offset} (requiring an offset correction of 0.77 for the LMC, and 2.47 for the SMC). This change is due to the fact that \citet{Roman-Duval2022A} calculated their global D/H values by taking the relationship between $\Sigma_{H}$ and D/H over the range of densities sampled by their observations, and extrapolated it to higher and lower $\Sigma_{H}$. They could thereby assign an extrapolated D/H to the full range of $\Sigma_{H}$ in in LMC and SMC, and so estimate global D/H. We, on the hand, have measured D/H over the full range of $\Sigma_{H}$ in the data (and observe the D/H turnover at densities higher than those probed by the UV measurements, for example), hence the differences in our global D/H values. We consider the offset calculated from our binned D/H profile to be the best indicator of the difference between the two, as it is calculated over the range of densities for which both datasets have actual measurements.

\begin{figure}
\centering
\includegraphics[width=0.495\textwidth]{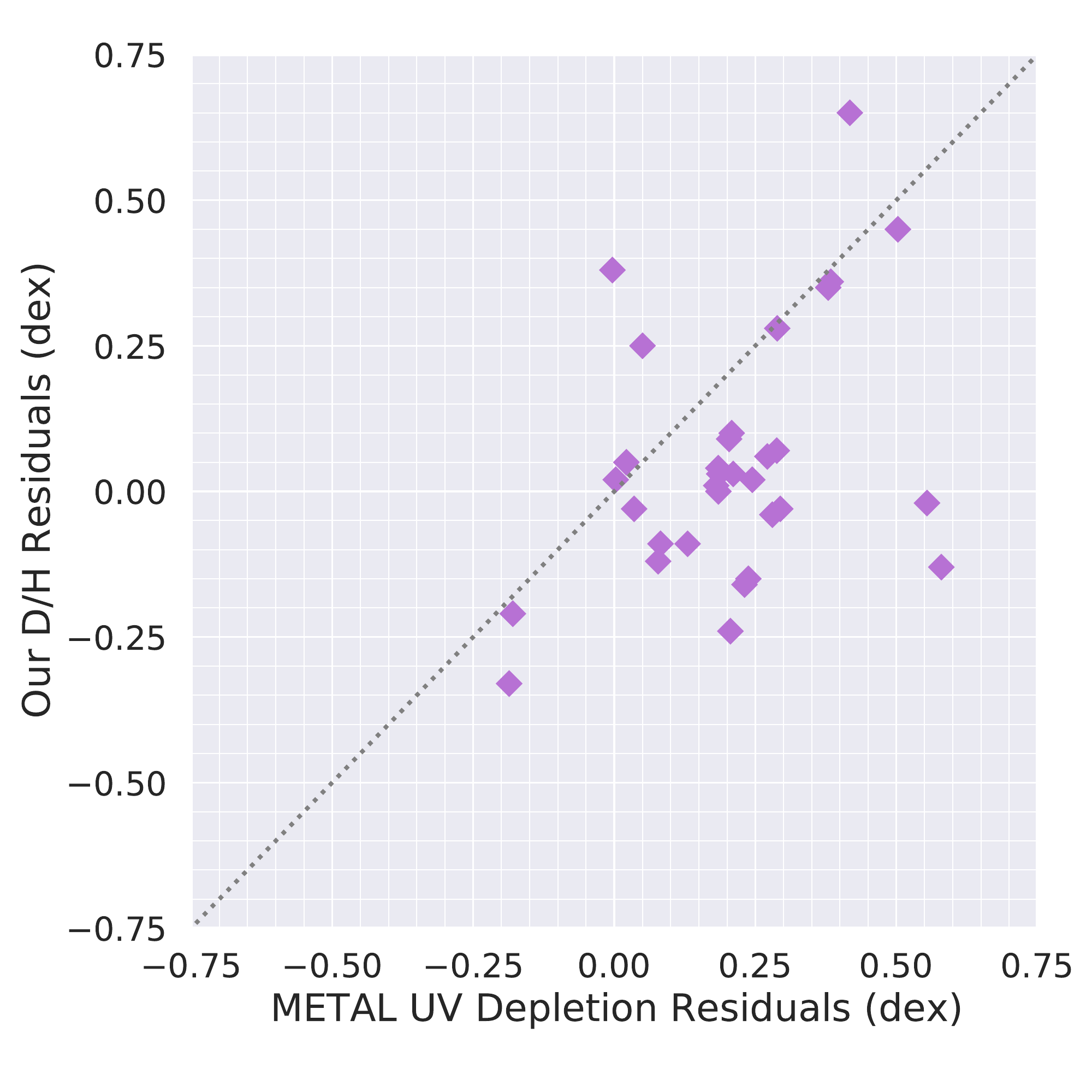}
\caption{Plot of the residuals found by \citet{Roman-Duval2021B}, between the iron depletion they measure for a given UV sightline (a proxy for D/H), and the depletion they would expect given that sightline's $\Sigma_{H}$ -- plotted against the residual between the D/H we find for that position of that sightline in our data, as compared to the D/H we would expect given that position's $\Sigma_{H}$ (given the relationship between $\Sigma_{H}$ and D/H we find in Figure~\ref{Fig:DtH_vs_H}). The dotted line shows the 1:1 relation.}
\label{Fig:LMC_Depletions_vs_Residuals_Plot}
\end{figure}

\needspace{2\baselineskip} \subsection{Comparing Local D/H Residuals Between FIR and UV Measurements} \label{Subsection:DtH_Residuals}

In \citet{Roman-Duval2021B}, the authors note an interesting trend in the residuals between the iron depletion (a proxy for the fraction of metals locked up in dust, which directly correlates with D/H at fixed metallicity) determined from each individual UV spectra, and the depletion that would be {\it expected} for that spectra, given that spectra's $\Sigma_{H}$. In general, \citet{Roman-Duval2021B} of course found that sightlines that sampled higher $\Sigma_{H}$ tended to have greater D/H (as inferred from the amount of iron depletion). However, they also found that in the southeastern portions of the LMC, sightlines would usually have lower D/H than would be expected for their $\Sigma_{H}$; conversely, in the northwestern portions of the SMC, sightlines often had higher D/H than would be expected from their $\Sigma_{H}$. \citet{Roman-Duval2021B} suggest that star-formation triggered by the gas accretion from the Magellanic Stream onto the southeast of the LMC could be causing an increased amount of grain processing and destruction through radiation, shocks, etc.

We were curious whether the same trend was visible in our data. Using our D/H versus $\Sigma_{H}$ relation for the LMC (as in Figure~\ref{Fig:DtH_vs_H}), we calculated the D/H we would {\it expect} for each pixel in the LMC, given its $\Sigma_{H}$. We then found the residual between the measured D/H in each pixel, and the predicted D/H. We plot a map of these residuals in Figure~\ref{Fig:LMC_Depletions_vs_Residuals_Map}.

Like \citet{Roman-Duval2021B}, we find that D/H is significantly depressed in the southeast of the LMC, relative to what we would expect based on $\Sigma_{H}$ alone. Figure~\ref{Fig:LMC_Depletions_vs_Residuals_Map} shows that this depression is most conspicuous along the edge of the \HI\ tail associated with the gas being accreted from the Magellanic Stream. Notably, the most depressed D/H is found furthest along this tail, to the southern edge of the LMC disc. This is further from the regions of enhanced star-formation being trigged by the infall, centered around 30 Doradus. This suggests that dust processing and destruction due to the effects of star formation is not the dominant cause of the lowered D/H in the southeast of the LMC; otherwise, we would expect D/H to be lower in areas closer to 30 Doradus, and the other areas of heightened star formation. However, grain processing due to gas collision and the resulting shocks, arising from the infall, could be the cause. The absence of a metallicity gradient in the young stars of the LMC argues against a metallicity effect in general \citep{Roman-Duval2021B}; however, metallicity will likely be depressed at larger radii, along the tail, where the infalling gas dominates \citep{Nidever2008B,Tsuge2019A,Tsuge2020A}.

The largest positive residuals in Figure~\ref{Fig:LMC_Depletions_vs_Residuals_Map} seem to be located along the edges of various large low-density features in the LMC (compare to Figure~\ref{Fig:LMC+SMC_Overview}), which correspond to known supershells \citep{Meaburn1980E}, carved by recent star formation. It is not immediately obvious why this might be the case. It is conceivable that the winds from young OB stars and recent supernovae are able to blow away the less-dense ISM more easily than the denser ISM where more grain-growth has occurred, preferentially leaving behind material with a greater dust content. Similarly, elevated D/H at the edge of star-forming regions, due to the evacuation of dust from the centers via wind-driven dust grain drift, is expected based on the work of \citet{Draine2011D}.

Figure~\ref{Fig:LMC_Depletions_vs_Residuals_Map} also shows the positions of the individual UV sightlines probed by \citet{Roman-Duval2021B}, and the residual they found for that sightline. We found the residual at each of those positions in our data by taking the mean of the values in our residual map in a 3$\times$3 pixel square aperture centered on the coordinates of each UV sightline\footnote{We used a 3$\times$3 pixel square aperture in order to sample multiple beams along each axis, and therefore reduce noise, while still sampling a small enough region to be comparable to the value at the specific location of the UV sightline}. The \citet{Roman-Duval2021B} D/H residuals correlate with those we find at the same positions. This is shown explicitly in Figure~\ref{Fig:LMC_Depletions_vs_Residuals_Plot}, which plots our D/H residuals against theirs. While the correlation is not strong, it would appear to be significant. According to a Kendall's Tau rank correlation test \citep{Kendall1990}, the probability of the null hypothesis, of no correlation, is $\mathcal{P}({\it null}) = 0.05$.

It is not surprising that there is not an especially tight correlation between the D/H residuals we find, and those of \citet{Roman-Duval2021B}. Their {\it Hubble} spectra sample a pencil beam only as wide as the star being measured, compared to the 15\,pc resolution (and therefore 45$\times$45\,pc aperture) of our data. Plus, each UV specta's pencil beam will only sample the ISM on the near side of the star being used -- and different stars in their sample will be located at different depths into the LMC disc, as viewed from Earth. Our data, on the other hand, sample the thickness of the entire LMC disc. The fact that our respective residuals do indeed seem to correlate -- and that there is most certainly large scale structure in these residuals, as clearly apparent in Figure~\ref{Fig:LMC_Depletions_vs_Residuals_Map} -- demonstrate that there are significant large-scale variations in the ISM properties of the LMC being driven by its ongoing interaction with the SMC through the Magellanic Stream.

\needspace{2\baselineskip} \section{Data Products} \label{Section:Data_Products}

Alongside this work, we are releasing the feathered \hersc\ maps presented in \citetalias{CJRClark2021A}. These are provided as Flexible Image Transport System (FITS; \citealp{Wells1981A, Hanisch2001F}) files, with one FITS file for each band, for each galaxy. These FITS files contain 4 extensions. Extension 1 (\texttt{IMAGE}) provides the standard feathered map. Extension 2 (\texttt{UNC}) provides the uncertainty map. Extension 3 (\texttt{MASK}) provides a binary mask map indicating the portion of the data where reliable, fully-feathered high-resolution coverage is available. Extension 4 provides the foreground-subtracted version of the feathered map (\texttt{FGND\_SUB}), the header of which also describes the uncertainty on that subtraction. All maps are in units of MJy\,sr$^{-1}$ (except for the binary masks). Full details of the creation, testing, and properties of these maps can be found in \citetalias{CJRClark2021A}.

We also provide data products from the analysis performed in this work. For the SED fitting, we provide maps of the median value of each parameter in each pixel (ie, the maps shown in Figures~\ref{Fig:SED_Params_Grid_1} and \ref{Fig:SED_Params_Grid_2}), and maps of the uncertainties on those medians (being the 68.3\% quantile around the median). This data is provided in FITS format; each of these FITS files contain 2 extensions. Extension 1 (\texttt{median}) provides the map of pixel parameter median values. Extension 2 (\texttt{uncert}) provides the map of uncertainties on those medians. Additionally, we provide the full posterior probability distribution for all SED parameters, consisting of 1000 posterior samples, for all pixels, in the form of a FITS file containing a 4-dimensional hypercube, with axes corresponding to right ascension, declination, parameters (in order: $\Sigma_{d}$, $T_{d}$, $\beta_{1}$, $\beta_{2}$, $\lambda_{\it break}$, and $e_{500}$), and samples. 

We also provide our maps of $\Sigma_{H}$, and of D/H. Note that none of the provided maps have had deprojection corrections applied

{Lastly, we provide the {\it Swift}-UVOT maps used in Section~\ref{Subsubsection:Effect_of_Dust_Destruction}. This data is provided for {\it Swift}-UVOT bands W1, W2, and M2. For each band, we provide a FITS file containing 3 extensions. Extension 1 (\texttt{SURF\_BRI}) provides the map of the surface brightness in MJy\,sr$^{-1}$ (converted using the {\it Swift}-UVOT zero points given in \citealp{Breeveld2011G}). Extension 2 (\texttt{RATE}) provides the map of the count rate (in photons\,sec$^{-1}$). Extension 3 (\texttt{EXP}) provides the map of the exposure time (in sec). The maps for the LMC and SMC are those presented in \citet{Hagen2017C}. The maps for M\,31 and M\,33 are were reduced following the same process as those in \citet{Hagen2017C}, and will be fully presented in Decleir et al. ({\it in prep.}), but are provided here for the purposes of reproducibility.}

This full dataset is available at: \url{https://doi.org/10.5281/zenodo.7392275} \citep{CJRClark2022B}. The feathered \hersc\ maps, as presented in \citetalias{CJRClark2021A}, can also be accessed at the NASA/IPAC Infrared Science Archive: \url{https://www.ipac.caltech.edu/doi/irsa/10.26131/IRSA545} \citep{CJRClark2021B}.

\needspace{2\baselineskip} \section{Conclusion} \label{Section:Conclusion}

In this paper, we have explored the relationship between dust and gas in the Local Group galaxies M\,31, M\,33, the LMC, and the SMC, a sample which spans a wide range in mass, metallicity, and other properties. Our investigation has taken advantage of new \hersc\ maps of these galaxies, presented in \citetalias{CJRClark2021A} of this series. Previous \hersc\ data for these galaxies suffered from severe filtering of extended emission (due to the \hersc\ data reduction process; \citealp{Meixner2013A,Roussel2013A,MWLSmith2017A,MWLSmith2021A}), systematically biasing that data's ability to detect diffuse dust, as well as compromising the scope for accurate foreground subtraction, along with other adverse effects. 

The new data from \citetalias{CJRClark2021A} combined the \hersc\ maps, in Fourier space, with data from \planck, IRAS, and COBE. Those other telescopes, while having resolution a factor of \textgreater10 worse than \hersc, did not filter out the diffuse emission. By merging the data, the \citetalias{CJRClark2021A} maps preserve the exquisite angular resolution of \hersc, while also recovering the previously-missed diffuse dust emission on large scales. By allowing us to probe the widest possible range of physical scales -- and hence {\it densities} -- in the ISM of our target galaxies, the \citetalias{CJRClark2021A} data has allowed us to investigate how dust evolves in the ISM. 

Previous work has found strong evidence that dust grains grow in denser regions of the ISM \citep{Fitzpatrick2005C,Jenkins2009B,Tchernyshyov2015B,Roman-Duval2014D,Roman-Duval2017B,Roman-Duval2021B}, but the specifics of how much grain growth happens in different environments, and especially at different metallicities, is very much an open question. In particular, galaxies with greater metallicities typically have higher D/H; however, below a certain metallicity, there is evidence that D/H drops sharply. This `critical metallicity' suggests that dust grain growth in the ISM only becomes efficient once metallicity reaches a certain level \citep{Asano2013,Feldmann2015C,Zhukovska2016A}. However, the location and nature of the critical metallicity transition remains unclear.

We used the new \hersc\ data from \citetalias{CJRClark2021A} to perform resolved fitting of the FIR dust SEDs of our galaxies. Comparing the resulting maps of dust mass surface density, to maps of the hydrogen surface density $\Sigma_{H}$ (derived from 21\,cm and CO observations), has allowed us to examine dust-to-gas ratio (which we quantified using the dust-to-hydrogen ratio, D/H) over an unparalleled 2.5 orders of magnitude in surface density for our sample of Local Group galaxies, and across the factor $\sim$6 variation in metallicity they represent. We have therefore been able to study the evolution of the dusty ISM of galaxies to a far greater level of detail than has been previously possible.

Our key findings are:

\begin{itemize}

\item The dust-to-gas ratio, D/H, shows very significant evolution with gas surface density, $\Sigma_{H}$, for all four of the galaxies in our sample. We find that D/H increases with density by a factor of 22.4 in the SMC, a factor of 9.0 in M\,31, a factor of 3.5 in the LMC, and a factor of 2.5 in M\,33. This is considerably more than found by any previous study of dust-to-gas variation within galaxies.

\item Because M\,31 and M\,33 have shallow metallicity gradients\footnote{The metallicity of M\,31 and M\,33 falls by only a factor of 3.6 per $R_{25}$ for M\,31, and a factor of 1.65 for M\,33 -- much less than the evolution in D/H.} and because the LMC and SMC have little-to-no metallicity gradient, we can be confident that this evolution in D/H is not simply due to a metallicity effect.

\item {We examine whether greater efficiency of dust destruction at lower densities could be driving the evolution in D/H. We consider: dust destruction by recently-formed stars (from hard radiation, and from the corresponding core-collapse \sne) as traced by UV emission; dust destruction due to ionized gas as traced by H$\alpha$ emission; and models of improved efficiency of dust destruction in lower density ISM. It does not appear that any of these can account for the observed evolution of D/H with ISM density. On the contrary, we are surprised that greater UV luminosity density, and ionized gas surface density, are correlated with {\it higher} D/H.}

\item {In light of the above, our favored explanation for the strong evolution in D/H with $\Sigma_{H}$ is that it is being driven by increasingly efficient dust grain growth at higher ISM densities.}

\item The D/H versus $\Sigma_{H}$ evolution profiles of M\,31 and M\,33 agree extremely well. The peak D/H for M\,31 of 0.01 is 20\% higher than the peak for M\,33, but otherwise they follow each other very closely, with integrated D/H differing by only 5\%. This is somewhat surprising, as M\,31 has 2.6 times higher metallicity, and 25 times more stellar mass, than M\,33.

\item Conversely, the large differences between the D/H evolution profiles of M\,33 and the LMC are very surprising, given these galaxies' close similarity in mass, metallicity, and star formation rate. Nonetheless, the peak D/H of M\,33 is 45\% greater than that of the LMC, and occurs at a surface density a factor of 10 smaller, with the integrated D/H of M\,33 being 82\% greater than the LMC's. This has implications for the common technique (often applied at high redshift) of estimating a galaxy's gas mass from its dust emission; while an observer would likely feel confident in applying the same conversion factor to two galaxies as apparently-similar as the LMC and M\,33, this would not in fact be reliable.

\item The D/H evolution profiles of M\,31, and M\,33, and the LMC share a confusing trait, whereby after steadily increasing with $\Sigma_{H}$, they then turn over and {\it decrease} at higher densities. There is no physical reason to expect this; dust evolution modeling predicts that D/H should plateau at higher densities, not turn over.

\item After extensive investigation, we rule out overestimation of $\alpha_{CO}$, elevated dust destruction due to star formation, physical resolution effects, and noise-induced anti-correlation, as the cause of the D/H turnover. We find that dark gas (atomic and/or molecular) could cause a turnover to appear for M\,31 and M\,33 (and artificially steepen the relationship between D/H and density in the SMC). However, while dark gas is probably elevating D/H in the LMC at intermediate densities, it could {\it not} be the driver of the turnover observed for the LMC. We find that a fall in the dust mass absorption coefficient ($\kappa$) with density to be a plausible explanation for the turnover in the LMC; if this is the case, we would expect this effect to also contribute to the turnover in M\,33 and M\,33.

\item Previous FIR-based estimates of D/H in the LMC and SMC disagreed with those derived from UV absorption line spectroscopy measurements of elemental depletions. This disagreement was a factor of 2 for the LMC, and a factor of 5 for the SMC, implying the existence of previously-missed dust in these galaxies. Our new D/H estimates resolve this tension for the LMC, but only reduce it to a factor of 2.5 disagreement for the SMC. Given the otherwise close agreement in the D/H evolution profiles between the FIR and UV results, we propose this suggests that the dust mass absorption coefficient, $\kappa$, is a factor of 3 lower in the SMC than for the sample's other galaxies, causing its D/H measurements (and dust mass) to be underestimated.

\end{itemize}

{\small

\acknowledgements

This research was improved by the insightful comments of the anonymous referee, whose input improved this work, and spurred additional lines of investigation that proved fruitful.

CJRC and JR-D acknowledge financial support from the National Aeronautics and Space Administration (NASA) Astrophysics Data Analysis Program (ADAP) grant 80NSSC18K0944.

This research benefitted throughout from the constructive, thoughtful, and friendly input (and reserach environment), provided by the ISM$*$@ST group\footnote{\url{https://www.ismstar.space/}}, whose help made this a better paper; with particular thanks to C. Murray, J. Wu, and C. Zucker.

This research made use of \texttt{Astropy}\footnote{\url{https://www.astropy.org/}}, a community-developed core \texttt{Python} package for Astronomy \citep{astropy2013,astropy2019}. This research made use of \texttt{reproject}\footnote{\url{https://reproject.readthedocs.io}}, an \texttt{Astropy}-affiliated \texttt{Python} package for image reprojection. This research made use of Photutils\footnote{\url{https://photutils.readthedocs.io}}, an \texttt{Astropy}-affiliated package for
detection and photometry of astronomical sources\citep{Bradley2020C}. This research made use of \texttt{NumPy}\footnote{\url{https://numpy.org/}} \citep{VanDerWalt2011B,Harris2020A}, \texttt{SciPy}\footnote{\url{https://scipy.org/}} \citep{SciPy2001,SciPy2020}, and \texttt{Matplotlib}\footnote{\url{https://matplotlib.org/}} \citep{Hunter2007A}. This research made use of the \texttt{pandas}\footnote{\url{https://pandas.pydata.org/}} data structures package for \texttt{Python} \citep{McKinney2010}. This research made use of \texttt{corner}\footnote{\url{https://corner.readthedocs.io}}, a python package for the display of multidimensional samples \citep{ForemanMackey2016D}. This research made use of \texttt{iPython}, an enhanced interactive \texttt{Python} \citep{Perez2007A}.

This research made use of sequential color-vision-deficiency-friendly colormaps from \texttt{cmocean}\footnote{\url{https://matplotlib.org/cmocean/}} \citep{Thyng2016A} and \texttt{CMasher}\footnote{\url{https://cmasher.readthedocs.io}} \citep{VanDerVelden2020A}

This research made use of \texttt{TOPCAT}\footnote{\url{http://www.star.bris.ac.uk/~mbt/topcat/}} \citep{Taylor2005A}, an interactive graphical viewer and editor for tabular data.

This research made use of the SIMBAD database\footnote{\url{https://simbad.u-strasbg.fr/simbad/}}; \citealp{Wenger2000D}) and the VizieR catalogue access tool\footnote{\url{https://vizier.u-strasbg.fr/viz-bin/VizieR}} \citep{Ochsenbein2000B}, both operated at CDS, Strasbourg, France. This research has made use of the {\sc Nasa/ipac} Extragalactic Database\footnote{\url{https://ned.ipac.caltech.edu/}} (NED), operated by the Jet Propulsion Laboratory, California Institute of Technology, under contract with NASA. This research has made use of the NASA/IPAC InfraRed Science Archive\footnote{\url{https://irsa.ipac.caltech.edu}} (IRSA), which is funded by NASA and operated by the California Institute of Technology. This research made use of the HyperLEDA database\footnote{\url{http://leda.univ-lyon1.fr/}} \citep{Makarov2014A}.
\newline
\newline

}

\facilities{Herschel, Planck, IRAS, COBE, Swift, Parkes, Arecibo, VLA, ATCA, Effelsberg}

\bibliography{ChrisBib}{}
\bibliographystyle{aasjournal}

\appendix
\restartappendixnumbering

\needspace{3\baselineskip} \section{D/H Evolution Model Parameters} \label{AppendixSection:DtH_Params}

\begin{table}
\centering
\caption{D/H evolution model grid parameter ranges and step sizes, for our implementation of the \citet{Asano2013A} model. The grid for $t_{\it growth}$ is spaced logarithmically, with step size therefore given in dex.}
\label{AppendixTable:DtH_Evolution_Grid}
\begin{tabular}{lrrr}
\toprule \toprule
\multicolumn{1}{c}{Parameter} &
\multicolumn{1}{c}{Minimum} &
\multicolumn{1}{c}{Maximum} &
\multicolumn{1}{c}{Step} \\
\cmidrule(lr){1-4}
$H_{\Sigma \Rightarrow n}$ (${\rm cm^{-3} M_{\odot}^{-1} pc^{2}}$) & 0.05 & 5.0 & 0.02 \\
$f_{d_{0}}$ & 0.0 & 0.30 & 0.002 \\
$t_{\it growth}$ (Myr) & 1 & 100 & 0.05\,dex \\
\bottomrule
\end{tabular}
\footnotesize
\justify
\end{table}

In Section~\ref{Subsection:DtH_Modeling}, we describe how we model the evolution of D/H with $\Sigma_{d}$, using the model framework of \citet{Asano2013A}. In Table~\ref{AppendixTable:DtH_Evolution_Grid}, we give the parameter grid we use to fit the model to our data. 

For the hydrogen surface-to-volume density conversion, $H_{\Sigma \Rightarrow n}$, we use a grid ranging from 0.05--5.0\,${\rm cm^{-3} M_{\odot}^{-1} pc^{2}}$. For context, assuming the disc of the LMC has a thickness of 100\,pc \citep{BGElmegreen2001A}, and taking from our $\Sigma_{H}$ maps a mean hydrogen surface density for the LMC disk of approximately $7\,{\rm M_{\odot}\,pc^{-2}}$, this implies a mean volume density of 2.8\,cm$^{-3}$, and hence $H_{\Sigma \Rightarrow n} = 0.4\,{\rm M_{\odot}^{-1} pc^{2}}$. Our 0.05--5.0\,${\rm cm^{-3} M_{\odot}^{-1} pc^{2}}$ grid, spanning over an order of magnitude, is therefore roughly centered on this value in log space. Our best fit parameters for $H_{\Sigma \Rightarrow n}$ tend to be a factor of {8--12} greater than this; we postulate that this may reflect the fact that dust grain-growth along a given sightline is driven not by the average volume density being sampled, but rather the areas of greater density in particular. {$H_{\Sigma \Rightarrow n}$ will be very dependent upon what fraction of the ISM along a given sightline is found in the denser molecular phase, versus the the more vacuous atomic phase. This will be especially true for a galaxy like the SMC, being highly elongated along our line-of-sight, such that a given sightline could have a very high observed surface density, but still have little-to-none of that material in environments of greater volume density. In practical terms, increasing $H_{\Sigma \Rightarrow n}$ has the effect of decreasing the density at which the D/H plateau is reached.}

For the minimum fraction of the metal mass locked up in dust grains, $f_{d_{0}}$, we use a grid ranging from 0.0--0.30. Given that the average value of $f_{d}$ is approximately 0.5 in high-metallicity spiral galaxies \citep{James2002,Jenkins2009B,Chiang2018A,Telford2019B}, and given that elemental depletions (and therefore $f_{d}$) vary with density by a orders of magnitude \citep{Jenkins2009B,Roman-Duval2021B}, it is unlikely that the very {\it lowest} value of $f_{d}$ in a galaxy would exceed 0.30. {In practical terms, $f_{d_{0}}$ sets the asymptote in D/H that is reached at low densities. This also has the effect of dictating the steepness of the relationship between $\Sigma_{H}$ and D/H. This is because the maximum possible D/H is set by metallicity (being when all metals are in dust), a lower $f_{d_{0}}$ requires a steeper evolution of D/H with $\Sigma_{H}$ (the range of densities from which D/H evolves from its minimum to maximum value is not adjustable by the free parameter. This appears to be the main reason why the LMC requires a larger best-fit value of $f_{d_{0}}$ than the other galaxies, as it has a shallower evolutionary trend between $\Sigma_{H}$ and D/H.} 

\begin{table}
\centering
\caption{Best-fit parameters for the D/H evolution model, for each of our galaxies. These are the parameters used for the models plotted in Figure~\ref{Fig:DtH_vs_H_ChemEv}.}
\label{AppendixTable:DtH_Evolution_Best}
\begin{tabular}{lrrrr}
\toprule \toprule
\multicolumn{1}{c}{Parameter} &
\multicolumn{1}{c}{M\,31} &
\multicolumn{1}{c}{M\,33} &
\multicolumn{1}{c}{LMC} &
\multicolumn{1}{c}{SMC} \\
\cmidrule(lr){1-5}
$H_{\Sigma \Rightarrow n}$ (${\rm cm^{-3} M_{\odot}^{-1} pc^{2}}$) & 4.71 & 4.09 & 3.37 & 4.71 \\
$f_{d_{0}}$ & 0.062 & 0.084 & 0.292 & 0.05 \\
$t_{\it growth}$ (Myr) & 45 & 18 & 2 & 6 \\
\bottomrule
\end{tabular}
\footnotesize
\justify
\end{table}

For the average duration of episodes of grain growth, $t_{\it growth}$, we use a grid ranging from 1--100\,Myr. If grain growth happens predominantly in molecular clouds, then this corresponds to the average molecular cloud lifespan. The average molecular cloud lifespan is thought to be around 10\,Myr, with estimates varying by a factor of a few \citep{Kruijssen2015A,Meidt2015C,Chevance2020A}. The fact our data show D/H increasing steadily over a wide range of \SigmaDeproj\ suggests that growth isn't just happening in the very densest regions (ie, not only in molecular clouds), and/or that dust destruction happens to different degrees over a wide range of densities. Observationally, an increased rate of dust destruction in less-dense environments would manifest in this model framework as a reduction in $t_{\it growth}$. So in practice, $t_{\it growth}$ encompasses the typical duration of episodes of grain growth, weighted by how efficiently destruction occurs when grains are not undergoing growth. Ultimately, our best-fit values of $t_{\it growth}$ are within a factor of a few of typical estimates of molecular cloud lifetimes. {In practical terms, $t_{\it growth}$ is degenerate with $H_{\Sigma \Rightarrow n}$, with greater values of $t_{\it growth}$ decreasing the density at which the D/H plateau is reached. Essentially, this is due to the fact that more grain growth can occur when episodes of grain growth last longer, and/or when the volume density of the ISM is greater.}

{Despite $t_{\it growth}$ being degenerate with $H_{\Sigma \Rightarrow n}$, we opt to keep both parameters in the model. The primary reason for this is that both could reasonably be expected to vary a great deal between between galaxies, and we don't want to force an entirely unphysical value of either paremeter to be adopted. This degeneracy also means that there are combinations of these parameters, other than those given in Table~\ref{AppendixTable:DtH_Evolution_Best}, which give fits that are effectively just as good,}

The model grid contains just over 1.2 million models. The best-fit parameters we found for each galaxy are provided in Table~\ref{AppendixTable:DtH_Evolution_Best}. We stress again that we do not suggest that these are the best estimates for the `true' values of these parameters. But rather, that they are intended to be reasonable values that yield models that fit the data as well as possible, and indicate the sort of D/H evolutionary profile we should expect.

\needspace{3\baselineskip} \section{Possible Causes of the D/H Turnover} \label{AppendixSection:Turnover}

\begin{figure*}
\centering
\includegraphics[width=1.0\textwidth]{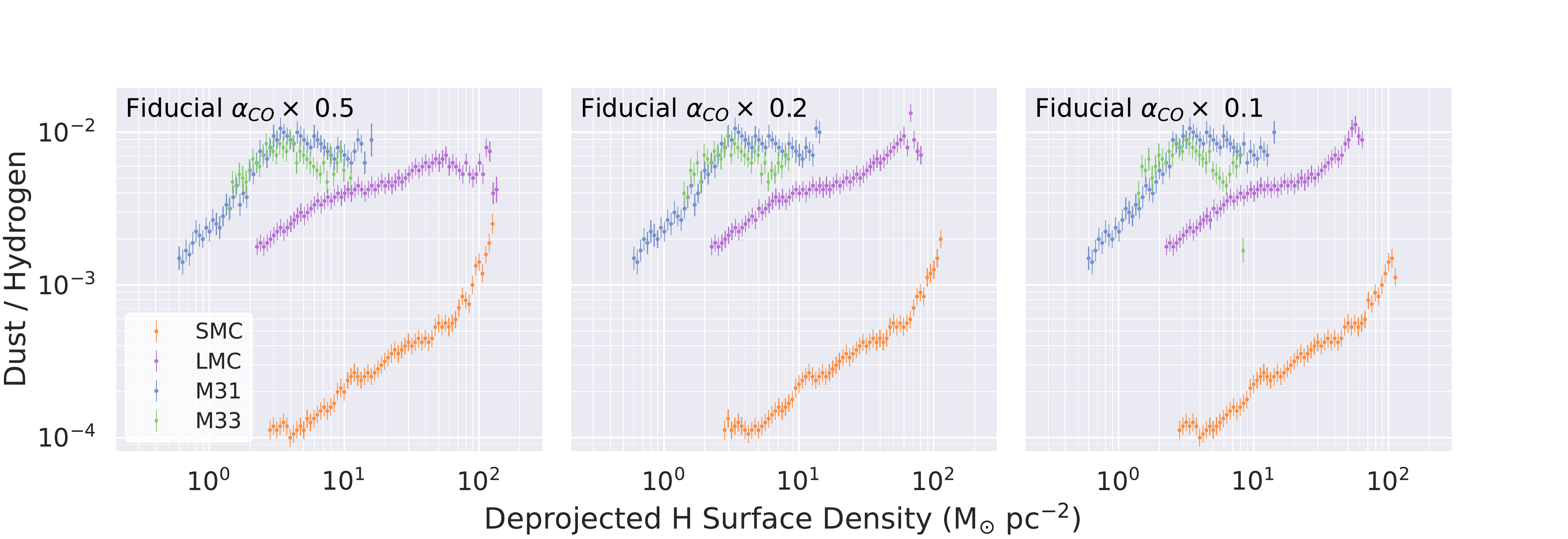}
\caption{Plots of D/H against $\Sigma_{H}$, for which molecular gas surface densities have been recomputed using values of $\alpha_{\it CO}$ that have been changed by factors of 0.5 ({\it left}), 0.2 ({\it center}), and 0.1 ({\it right}), as compared to each galaxy's fiducial value of $\alpha_{\it CO}$ used in Figure~\ref{Fig:DtH_vs_H}.}
\label{AppendixFig:DtH_vs_H_alphaCOx}
\end{figure*}

Here we present our in-depth investigation into the possible causes of the apparent turnover in the D/H at higher values of $\Sigma_{H}$ for the LMC, M\,31, and M\,33, as presented in Section~\ref{Section:Causes_of_Turnover}. Specifically, we explore 6 possible explanations for the apparent turnover: Differences in $\alpha_{\it CO}$; noise-induced anticorrelation; physical resolution effects; dust destruction by star formation; the presence of dark gas, and varying dust mass opacity.

\needspace{2\baselineskip} \subsection{Variations in $\alpha_{\it CO}$?} \label{AppendixSubsection:Differences_in_alpha-CO}

If we were to have overestimated $\alpha_{\it CO}$ for our galaxies, then we would be overestimating the amount of molecular gas present, which would mean that we would get erroneously low D/H values -- specifically at the higher densities where molecular gas represents a larger faction of the total dust budget.

Very suggestively, the surface densities at which molecular gas starts to dominate over atomic gas in our \SigmaDeproj\ data are 2.1, 3.6, and 35\,${\rm M_{\odot}\,pc^{-2}}$ for M\,31, M\,33, and the LMC respectively -- all very close to the densities at which their turnovers happen, being $\approx$\,4\,${\rm M_{\odot}\,pc^{-2}}$ for M\,31, M\,33, and $\approx$\,40\,${\rm M_{\odot}\,pc^{-2}}$ for the LMC. Meanwhile, for the SMC, which doesn't have a turnover, there is likewise no surface density at which the measured molecular gas surface density exceeds that of the atomic gas\footnote{While the SMC of course has regions that are molecular-gas dominated, there is no bin of \SigmaDeproj\ in which the mass contribution of molecular gas exceeds that of atomic gas. This is most likely due to a combination of our limiting resolution, and the extreme line-of-sight depth of the SMC. The contribution of a given molecular-gas-dominated region to the surface density of a given pixel can be exceeded by the contributions of very long columns of atomic gas in front of and behind of it, further diluted by the limits of our spatial resolution}.

We examined to what degree $\alpha_{\it CO}$ would need to be reduced, in order to get rid of the turnover. To do this, we repeated our D/H analysis, instead using values of $\alpha_{\it CO}$ that were modified by factors of 0.5, 0.2, and 0.1, relative to the fiducial values we use for each galaxy (namely 3.2\,${\rm K^{-1}\,km^{-1}\,s\,pc^{-2}}$ for M\,31 and M\,33, 6.4 \,${\rm K^{-1}\,km^{-1}\,s\,pc^{-2}}$ for the LMC, and 21\,${\rm K^{-1}\,km^{-1}\,s\,pc^{-2}}$ for the SMC; see Section 7.1 of \citetalias{CJRClark2021A} for specifics on our choice of $\alpha_{\it CO}$ values). The resulting D/H evolutionary profiles are shown in Figure~\ref{AppendixFig:DtH_vs_H_alphaCOx}.

The plots in Figure~\ref{AppendixFig:DtH_vs_H_alphaCOx} show that reducing $\alpha_{\it CO}$ by over a factor 2--5 is able to mostly remove the turnover in the case of the LMC. But even a factor of 10 reduction in $\alpha_{\it CO}$ cannot remove the turnover for M\,31 or M\,33; indeed in the case of M\,33, the turnover becomes {\it more} distinct. This may seem counter-intuitive, but the reasons for it are twofold.

Firstly, reducing $\alpha_{\it CO}$ will increase the D/H measured for a given pixel, while decreasing the \SigmaDeproj. This will have the effect of moving values upwards and leftwards on Figure~\ref{AppendixFig:DtH_vs_H_alphaCOx}, therefore tending to cause measurements to move {\it along} the post-turnover portion of a profile, as opposed to changing the {\it shape} of a profile.

Secondly, for the factor 0.2 and 0.1 changes to the fiducial $\alpha_{\it CO}$, we are effectively `removing' the vast majority of the molecular gas mass from these galaxies. This has the effect of making it very hard to accurately distinguish the relative gas surface densities of pixels. As an example, consider two pixels, for which the molecular gas content has been calculated using the fiducial $\alpha_{\it CO}$: an atomic-gas-dominated pixel with $\Sigma_{\it HI}=1.6$ and $\Sigma_{\it H_{2}}=0.4$; and a molecular-gas-dominated pixel with $\Sigma_{\it HI}=1.3$ and $\Sigma_{\it H_{2}}=3.0$. Here, the molecular-dominated pixel has a considerably larger total mass, while both pixels have similar atomic gas content, due to any higher-density gas tending to enter the molecular phase. Now consider the same pair of pixels when $\alpha_{\it CO}$ is multiplied by 0.1. Now, the originally atomic-dominated pixel will have $\Sigma_{\it HI}=1.6$ and $\Sigma_{\it H_{2}}=0.04$, while the formerly molecular-dominated pixel will have $\Sigma_{\it HI}=1.3$ and $\Sigma_{\it H_{2}}=0.3$, meaning that the two pixels would now be measured as having near-identical gas content. In short, because differences in atomic gas content between pixels become relatively smaller at higher densities, significantly reducing $\alpha_{\it CO}$ makes high-density pixels appear to have similar gas content to one another. As a result, whereas higher dust densities were previously very strongly associated with pixels that have higher gas densities, that relationship is now more mixed. The worse this mixing gets, the more the trend in D/H vs \SigmaDeproj\ will tend towards a gradient of $-1$, in a classic instance of noise-induced anti-correlation\footnote{For any plot of $\frac{a}{b}$ versus $b$, where there is no intrinsic correlation between $a$ and $b$, the gradient will tend towards $-1$ the as noise in $a$ and $b$ increases.}, which is indeed what we see happening for M\,33 in Figure~\ref{AppendixFig:DtH_vs_H_alphaCOx}.

Given that significantly reducing $\alpha_{\it CO}$ gives rise to this sort of spurious effect in our data, it would seem that reducing $\alpha_{\it CO}$ by such larger factors is likely unphysical. Which should not be a surprising; for instance, reducing the $\alpha_{\it CO}$ of the LMC by a factor of 5 would make it only 1.28 \,${\rm K^{-1}\,km^{-1}\,s\,pc^{-2}}$ -- less than half the standard Milky Way value of 3.2\,${\rm K^{-1}\,km^{-1}\,s\,pc^{-2}}$, and much less than would be expected for a galaxy with less than half Solar metallicity \citep{Bigiel2010D,Gratier2010C,Druard2014A}. And on top of this, reducing $\alpha_{\it CO}$ doesn't even manage to get rid of the D/H turnover; although it is possible that it may minimize the turnover partially in the case of the LMC, even for less aggressive reduced values of $\alpha_{\it CO}$.

As such we are confident that overestimation of $\alpha_{\it CO}$ is not the cause of the D/H turnover observed for our galaxies.

\needspace{2\baselineskip} \subsection{Noise-Induced Anti-Correlation?} \label{AppendixSubsection:Noise_Anticorrelation}

As seen in Section~\ref{AppendixSubsection:Differences_in_alpha-CO}, noise-induced anti-correlation drove negative gradients in D/H evolution when using very low values of $\alpha_{\it CO}$. So it is clearly worth checking whether noise-induced anti-correlation could also be giving rise to the turnover in the first place. To test this, we constructed simulated versions of our data, to explore how noise with different behaviors can affect the trends we observe. 

\begin{figure}
\centering
\includegraphics[width=0.475\textwidth]{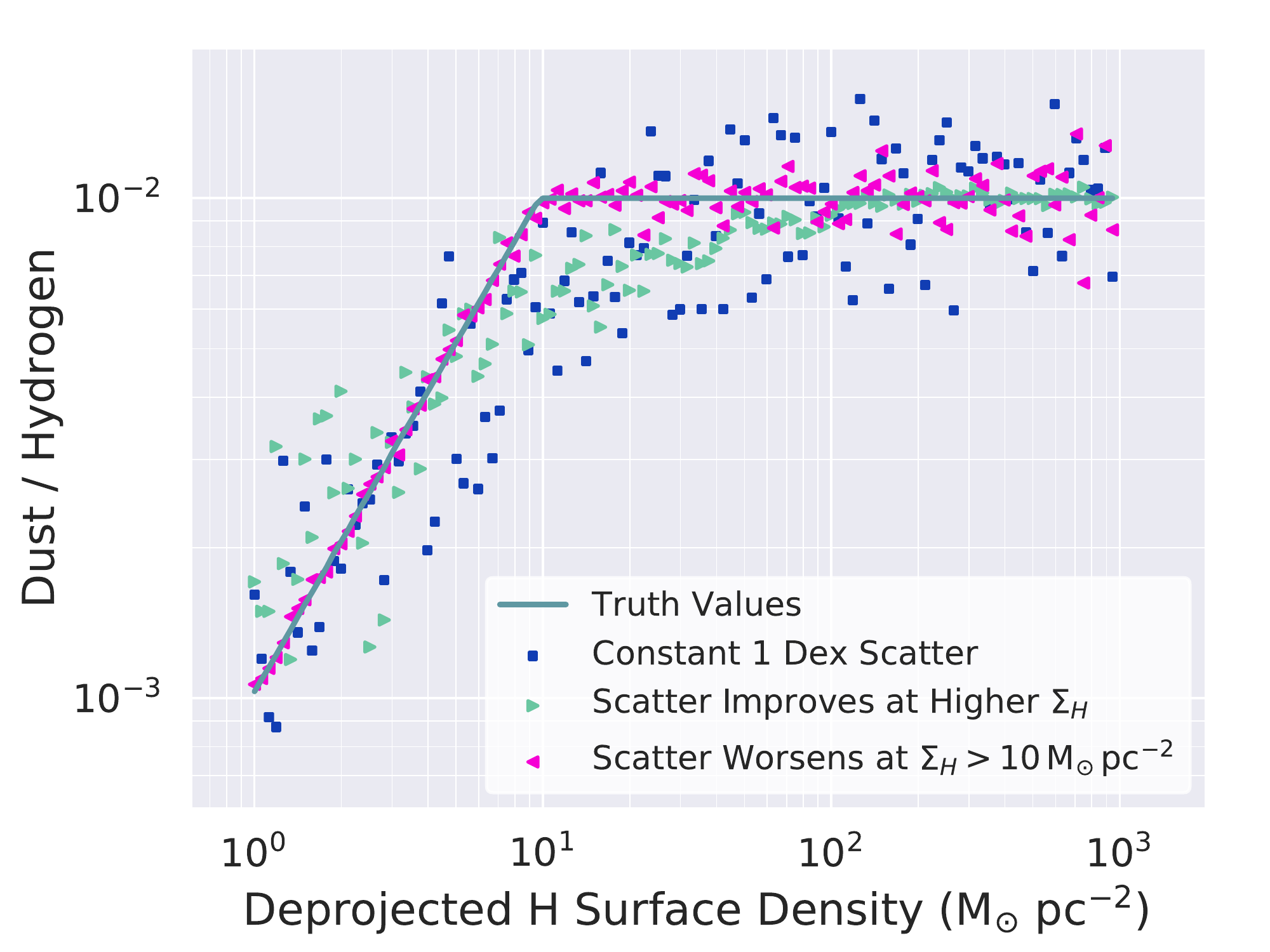}
\caption{Plot of simulated trends in D/H against $\Sigma_{H}$, to examine whether noise can introduce artificial anti-correlation at high surface densities. The gray line shows the underlying truth value for our toy model. The blue squares show the medians of each 0.025\,dex bin, when using artificial observed data, where each measurement has 1\,dex of Gaussian scatter in both \SigmaDeproj\ and $\Sigma_{d}^{\it (deproj)}$. The green right-pointing triangles show bin medians where the scatter gets smaller as \SigmaDeproj\ increases. The pink left-pointing triangles show bin medians where the scatter is constant up until $\Sigma_{H}^{\it(deproj)} = 10\,{\rm M_{\odot}\,pc^{-2}}$, then steadily increases at higher densities.}
\label{AppendixFig:DtH_vs_H_1-Over-x}
\end{figure}

To start with, we created a noiseless toy model in which D/H evolves from 0.001 at $\Sigma_{H}^{\it (deproj)} = 1\,{\rm M_{\odot}\,pc^{-2}}$, to 0.01 at $\Sigma_{H}^{\it (deproj)} = 10\,{\rm M_{\odot}\,pc^{-2}}$, at which D/H then plateaus. This toy model is plotted in Figure~\ref{AppendixFig:DtH_vs_H_1-Over-x}, and is intended to be the simplest possible version of the increase-then-plateau trend in dust evolution we might expect \citep{Asano2013A}.

First, we assess the impact that generally large scatter could have when observing this trend. We created simulated data for which $10^{6}$ data points, evenly distributed in ${\rm log_{10}}(\Sigma_{H}^{\it (deproj)})$, were drawn from the toy model over a $10^{-2} < \Sigma_{H}^{\it (deproj)} < 10^{5}$ interval\footnote{This interval is larger than the $10^{0} < \Sigma_{H}^{\it (deproj)} < 10^{3}$ range which we plot and discuss. The reason being that if we only drew values from within the range of interest, then some of the values would be scattered out of interval when we apply noise, biasing the binned averages at either end.}. We then applied 1\,dex of Gaussian noise to the \SigmaDeproj and $\Sigma_{d}$ of each randomly-drawn value, recomputed D/H for these, then placed them in 0.025\,dex bins, as per the real data in Figure~\ref{Fig:DtH_vs_H}, etc. A noise level of 1\,dex should do a good job of probing the effects of significant scatter on the observed trend\footnote{Amongst the bins for our actual data for the sample galaxies, 70\% of bins' values have standard deviations of \textless\,1\,dex, and 92\% have standard deviations of \textless\,1.5\,dex (all bins for the LMC and SMC have standard deviations of \textless\,1.41\,dex).}. The points generated by introducing this scatter are shown as blue squares in Figure~\ref{AppendixFig:DtH_vs_H_1-Over-x}, and as can be seen, they still trace the underlying trend. The knee at $\Sigma_{H}^{\it (deproj)} = 10\,{\rm M_{\odot}\,pc^{-2}}$ is no longer as sharp, having been smoothed out by the scatter, but otherwise there are no adverse effects present, and certainly no sign of a spurious D/H turnover.

Next, we perform a test designed to more closely simulate the scatter present in our actual data, which tends to steadily decrease for bins at higher densities\footnote{Roughly speaking, typical scatter in bins falls from $\approx$\,1.5\,dex at $\Sigma_{H}^{\it (deproj)} = 1\,{\rm M_{\odot}\,pc^{-2}}$, down to $\approx$\,0.2\,dex at $\Sigma_{H}^{\it (deproj)} = 50\,{\rm M_{\odot}\,pc^{-2}}$.}. We proceed as above, but now, instead of being a constant 1\,dex, the scale of the Gaussian scatter on both \SigmaDeproj\ and $\Sigma_{d}$ is a function of \SigmaDeproj, falling as density increases. Scatter in dex, $\sigma$, is given by $\sigma = -0.5{\rm log_{10}}(\Sigma_{H}^{\it (deproj)}) + 1.0$, with a floor value of $\sigma = 0.1$\,dex imposed at the highest densities where $\sigma$ would otherwise fall lower than this. The result of this simulation is shown with the green triangles in Figure~\ref{AppendixFig:DtH_vs_H_1-Over-x}. Introducing this form of scatter behavior does not appear to introduce any pathologies into the D/H evolution profile (other than the knee once again being smoothed out). This is reassuring, given that this should roughly emulate the way the scatter evolves with density in the real data.

Lastly, we test a `plausible worst-case scenario' for the evolution of scatter with density. For this, we hold the scale of the Gaussian noise at a constant low level of 0.05\,dex in \SigmaDeproj\ and 0.2\,dex in $\Sigma_{d}$ for values up to $\Sigma_{H}^{\it (deproj)} = 10\,{\rm M_{\odot}\,pc^{-2}}$; above this, the noise grows sharply with $\Sigma_{H}$ according to a power law, increasing to 0.15\,dex in \SigmaDeproj\ and 1.2\,dex in $\Sigma_{d}$ at $\Sigma_{H}^{\it (deproj)} = 1000\,{\rm M_{\odot}\,pc^{-2}}$. By keeping noise low before the knee, where D/H is increasing, then making noise increase considerably after the knee (ie, in the density regime where we observe the turnover in the real data), we should be maximizing the potential for noise-induced anti-correlation, or some other artifact, to create a spurious trend in D/H at higher densities. The binned medians for this test are shown with pink triangles in Figure~\ref{AppendixFig:DtH_vs_H_1-Over-x}. Even with this test, we find no suspicious behavior in the resulting D/H evolution profile. 

Following these tests, we are aware of no way in which noise-induced anti-correlation could give rise to an artificial D/H turnover in our data.

\needspace{2\baselineskip} \subsection{Physical Resolution Effects?} \label{AppendixSubsection:Physical_Resolution_Effects}

The physical resolution our data is able to achieve for M\,31 and M\,33 (137 and 147\,pc, respectively) is up to 10 times worse that what it can achieve for the LMC and SMC (15 and 47\,pc, respectively). It is therefore noteworthy that the two galaxies with the much poorer physical resolution also exhibit the sharper turnovers in D/H. So it is conceivable that limited physical resolution could give rise to spurious trends in our results. {In particular, temperature mixing could cause the dust mass determined from SED fitting to be biased low. The impact of temperature mixing will be expected to increase as physical resolution worsens and as density increases -- because a single pixel may then contain not only the cold, dense dust in giant molecular clouds, but also warm dust being heated by recent-formed stars, all blended together. This might cause dust mass to be underestimated progressively more as density increases, driving down D/H in a manner resembling our observed turnover.}

To test this possibility, we reprocessed our \hersc\ and \SigmaDeproj\ data for the LMC and SMC, degrading the observations so that the Magellanic Clouds appeared as they would were they at the same distance as M\,31 and M\,33. Specifically, we degraded the data for the LMC and SMC to produce physical resolution of 142\,pc -- the equivalent of 36\arcsec\ angular resolution at a distance of 815\,kpc. As 815\,kpc is the mid-point between the distances to M\,31 and M\,33, this maximizes our ability to compare fairly to both\footnote{We felt no need to make any adjustments to the data for M\,31 and M\,33, given how similar their distances are already. Indeed, the nearest parts of the disc of M\,33 are closer to us than the furthest parts of the disk of M\,31, so degrading the LMC and SMC data to match the midpoint between the two should allow for an entirely fair comparison.}. If limited physical resolution is causing spurious trends in our results, then degrading the physical resolution of the LMC and SMC by a factor of $\sim$10 should lead to significant changes in their D/H evolution profiles.

\begin{figure}
\centering
\includegraphics[width=0.475\textwidth]{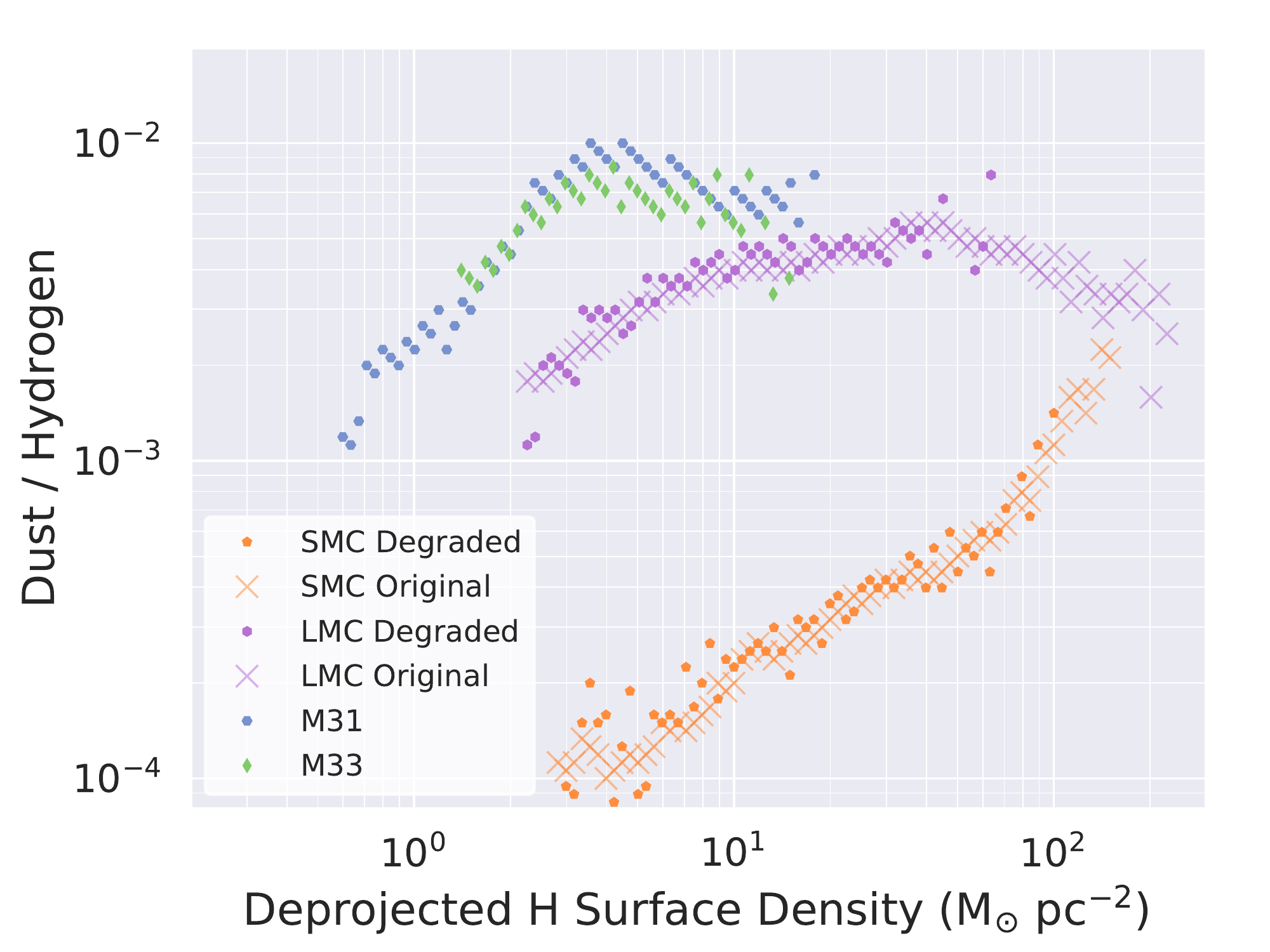}
\caption{Plot of D/H against \SigmaDeproj\, in which values for the LMC and SMC have been produced using maps degraded to recreate how the Magellanic Clouds would appear in our analysis were they at the same distance as M\,31 and M\,33, thereby giving all 4 galaxies the same physical resolution. The original data for the LMC and SMC (as plotted in Figure~\ref{Fig:DtH_vs_H}) is indicated by the crosses, for comparison. Error bars have been omitted to assist clarity, but are not significantly different for the degraded data than for the original, as plotted in Figure~\ref{Fig:DtH_vs_H}.}
\label{AppendixFig:DtH_vs_H_Degraded}
\end{figure}

However, as can be seen in Figure~\ref{AppendixFig:DtH_vs_H_Degraded} the D/H versus \SigmaDeproj\, the profiles followed by the LMC and SMC are almost completely unchanged by being degraded to the same physical resolution as M\,31 and M\,33. The only significant difference is that, as would be expected, the degraded data cannot trace to surface densities as high as the full-resolution data\footnote{The reason why the LMC and SMC D/H profiles still probe to greater \SigmaDeproj\ than for M\,31 and M\,33, despite their data being degraded to the same effective distance, is because of the larger deprojection corrections applied to M\,31 and M\,33, due to their greater inclination. For instance, were the factor 0.22 correction not applied to M\,31, it would appear to probe up to densities of $80\,{\rm M_{\odot}\,pc^{-2}}$, closely matching the highest densities probed by the LMC and SMC in their degraded data.}. Otherwise, the degraded data simply traces the same D/H evolution profile as the original data, albeit with somewhat more scatter. 

We are in fact pleasantly surprised by how well the D/H evolution profiles for the LMC and SMC have been preserved. In particular, reducing physical resolution from 14\,pc to 142\,pc, in the case of the LMC, will lead to vastly more temperature mixing, and other merging of dust populations within each pixel \citep{Galliano2011B}. The fact that dust properties can still be recovered well enough to leave the D/H evolution profile effectively unchanged is a better outcome than we might have expected. {This gives us further confidence that temperature mixing is not significantly biasing the results of our SED fits, in addition to the lack of residuals described in Section~\ref{Subsection:SED_Fitting_Results}.}

The D/H evolution profiles for the LMC and SMC are not at all biased by degrading their data to the effective distance of M\,31 and M\,33. This suggests that the observed profiles for M\,31 and M\,33 are similarly not significantly biased versus how they would appear were they observed at a closer distance to us. We therefore conclude that the effects of physical resolution limits are probably not causing the appearance of the D/H turnover {in the case of these two galaxies.} 

{For the LMC, the degraded data is not able to probe to the highest densities, and only traces up to $\approx\,60\,{\rm M_{\odot}\,pc^{-2}}$. We therefore cannot make a strong statement either way about the possibility of a spurious turnover being caused by temperature mixing at densities greater than this. However, we note that of all four galaxies we consider, the SMC should be the one vulnerable to the greatest impact from temperature-mixing. The reason for this is that its extreme elongation along the line-of-sight significantly increases the likelihood of different dust populations, heated to different temperatures by different environmental conditions, being present along a shared column, in a single pixel. However, instead of displaying a turnover, the SMC in fact shows the D/H relation getting {\it steeper} at the highest densities. This remains true even for the degraded data in Figure~\ref{AppendixFig:DtH_vs_H_Degraded}, where we would expect the impact of any temperature mixing to be exaggerated, thanks to each pixel sampling 9$\times$ more area. This gives us some additional reason to think that temperature mixing is not seriously biasing our D/H values at the highest densities.}

\needspace{2\baselineskip} \subsection{Dust Destruction by Environmental Effects at Higher Densities?} \label{AppendixSubsection:Ionized_Gas}

\begin{figure}
\centering
\includegraphics[width=0.475\textwidth]{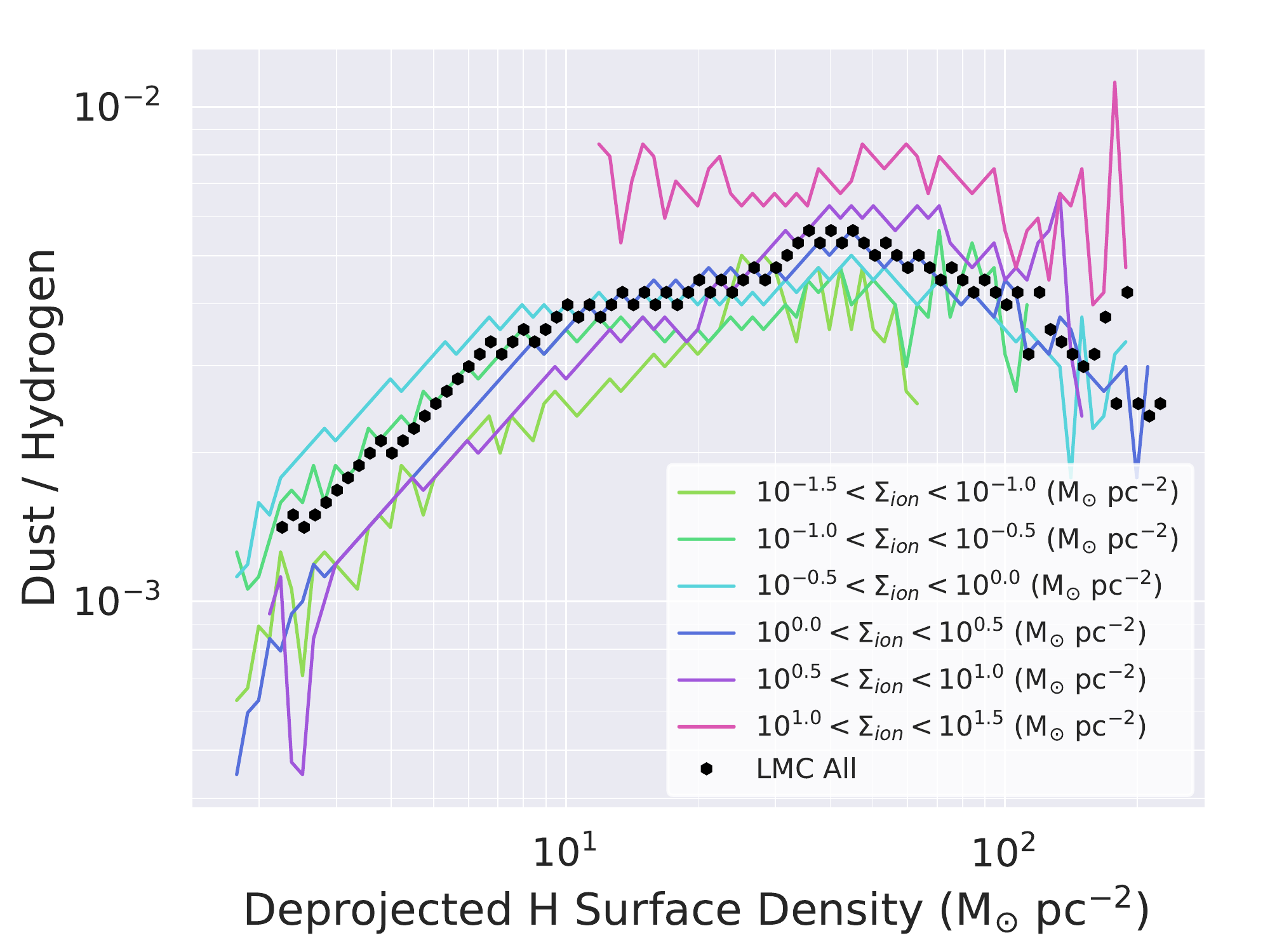}
\includegraphics[width=0.475\textwidth]{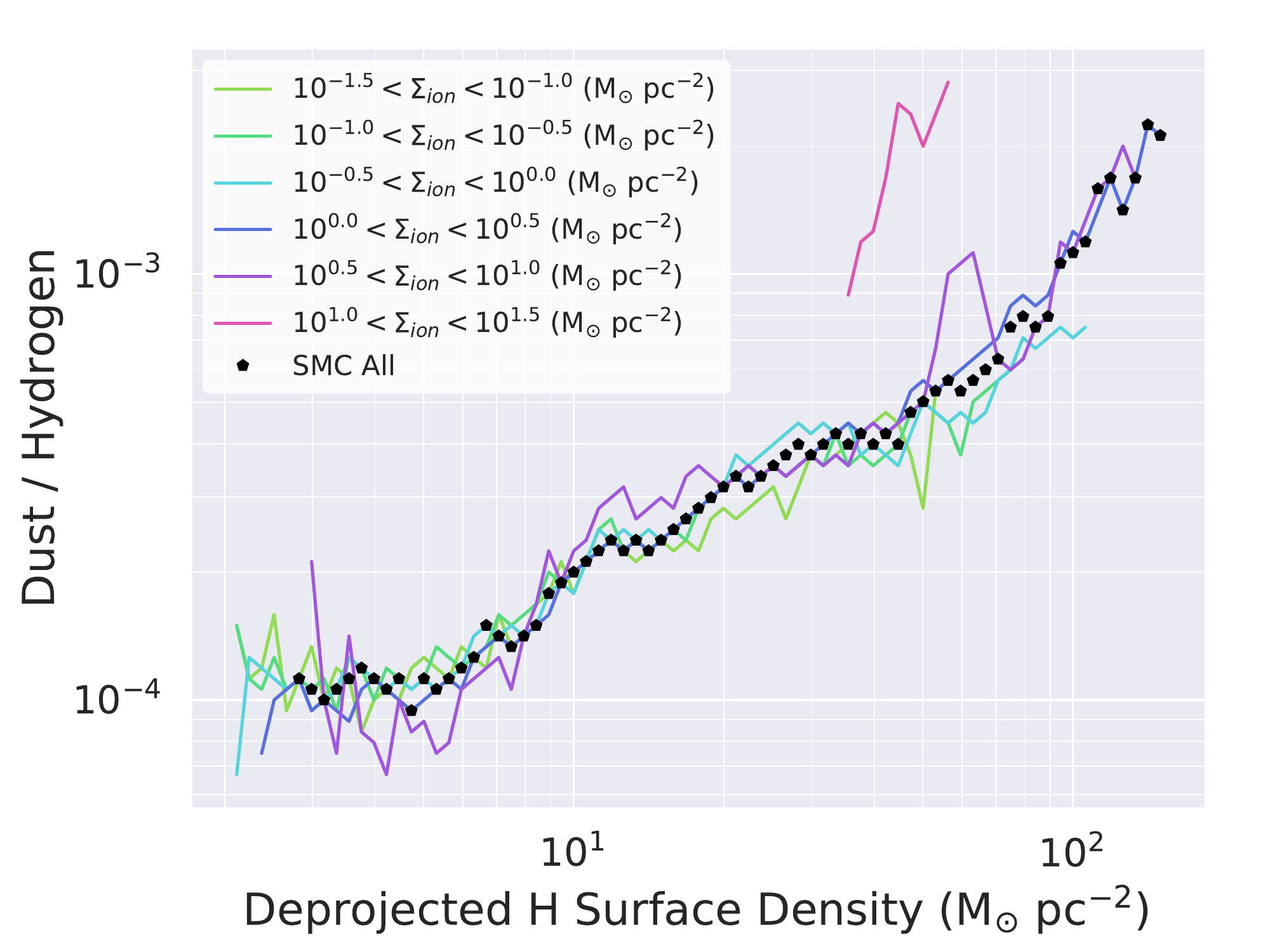}
\caption{Plot of D/H against \SigmaDeproj\ for the LMC ({\it upper}) and SMC ({\it lower}), with each line showing the relationship for pixels that lie within a certain range of ionized gas surface densities, $\Sigma_{\it H+}^{\it (deproj)}$. Also plotted, in black points, is the relationship when all pixels are counted. Note that for these plots, \SigmaDeproj\ (and therefore D/H) incorporates the contribution of $\Sigma_{\it H+}^{\it (deproj)}$ (in contrast to Figure~\ref{Fig:DtH_vs_H}, etc).}
\label{AppendixFig:DtH_vs_H_Ionised}
\end{figure}

Various processes can destroy dust grains in the ISM. Many of these processes arise due to star formation. For instance, by the supernova shocks following the deaths of recently-formed massive stars; or by direct photodestruction of grains by UV and X-ray photons; or by thermal sputtering of grains in the hot ionized gas produced by young massive stars \citep{Bocchio2014B,Slavin2015A,Jones2017A,Galliano2018C}. 

It is therefore conceivable that, in areas of higher-density ISM where more star formation occurs \citep{Kennicutt1998H}, the dust destructive processes associated with recent star formation lead to D/H being depressed, manifesting as the turnover we observe. If this is the case, then for a given \SigmaDeproj\ we would expect to find lower D/H in areas with more star formation, than in areas with less star formation at the same \SigmaDeproj.

To test this, we use the H$\alpha$ maps of \citet{Gaustad2011B}, which cover both the LMC and SMC at a resolution of 48\arcsec; well matched to the resolution of our dust and gas data. We use these H$\alpha$ maps to create maps of the surface density of ionized gas, $\Sigma_{\it H+}^{\it (deproj)}$, following the prescription of \citet{Paradis2011B}. These maps of ionized gas  are an ideal way of tracing where dust destruction due to star formation is likely to be happening. Not only is ionized gas a proxy for star formation, but the ionized gas itself is the environment in which the resultant dust destruction will occur.

We repeated our analysis of D/H versus \SigmaDeproj\ for the LMC and SMC, splitting the data into 6 bins of ionized gas surface density, from $\Sigma_{\it H+} ^{\it (deproj)} = 10^{-1.5}\,{\rm M_{\odot}\,pc^{-2}}$ to $\Sigma_{\it H+}^{\it (deproj)} = 10^{1.5}\,{\rm M_{\odot}\,pc^{-2}}$, with each bin having a width of 0.5\,dex. In Figure~\ref{AppendixFig:DtH_vs_H_Ionised}, the D/H evolution profile within each of these $\Sigma_{\it H+}^{\it (deproj)}$ bins is plotted, for the LMC and SMC. Note that for these plots, \SigmaDeproj, including the denominator of D/H, has been revised to also include the contribution of $\Sigma_{\it H+}^{\it (deproj)}$, to ensure internal consistency\footnote{We do note include the contribution of $\Sigma_{\it H+}^{\it (deproj)}$ when considering $\Sigma_{H}$ elsewhere in this paper, because in general we do not expect the ionized gas to be co-local with the cold dense ISM where grain growth should occur. Therefore including $\Sigma_{\it H+}^{\it (deproj)}$ would be expected to weaken our ability to trace evolution in D/H. Moreover, the D/H evolution profile does not change significantly when $\Sigma_{\it H+}^{\it (deproj)}$ is included -- compare the black points in Figure~\ref{AppendixFig:DtH_vs_H_Ionised} to the corresponding points in Figure~\ref{Fig:DtH_vs_H}}. We also plot the global D/H evolution profile for both galaxies in Figure~\ref{AppendixFig:DtH_vs_H_Ionised}, incorporating the ionized gas into \SigmaDeproj.

Figure~\ref{AppendixFig:DtH_vs_H_Ionised} shows that the D/H evolution profile behaves very similarly for regions of high $\Sigma_{\it H+}^{\it (deproj)}$ as it does in regions of low $\Sigma_{\it H+}^{\it (deproj)}$. The only exception to this is for the regions of the very highest ionized gas surface density, $10^{1.0} < \Sigma_{\it H+} < 10^{1.5} {\rm M_{\odot}\,pc^{-2}}$, where in both the LMC and SMC the D/H is conspicuously elevated above the general trend. This argues strongly against the hypothesis that the effects of nearby star formation could be depressing D/H. And most importantly of all, the D/H turnover in the LMC {\it is still present} for all the bins of $\Sigma_{\it H+}^{\it (deproj)}$ that trace that range of \SigmaDeproj\ (although the profile for the highest $\Sigma_{\it H+}^{\it (deproj)}$ bin does get noisy at $\Sigma_{H} > 100\,{\rm M_{\odot}\,pc^{-2}}$).

{Moreover, Figure~\ref{Fig:DtH_vs_UV} and the corresponding discussion in Section~\ref{Subsubsection:Effect_of_Dust_Destruction} also show that more intense UV radiation fields are associated with higher D/H, not lower, in all four of our sample galaxies.}

These observations strongly suggest that dust destruction due to elevated star formation in regions of \SigmaDeproj\ is not the cause of the D/H turnover.

\needspace{2\baselineskip} \subsection{Dark Gas?} \label{Subsection:Dark_Gas}

To measure the gas content of our galaxies, we use 21\,cm and CO observations as tracers of the atomic and molecular gas components. However, if these tracers miss some fraction of the gas in certain environments, we would find  incorrect D/H values, thereby introducing errors into the D/H evolution profiles.

In particular, the presence of optically-thick \HI\ \citep{Fukui2015A,Murray2018B}, or CO-dark ${\rm H}_{2}$ \citep{Reach1994C,Grenier2005D,Wolfire2010A}, could lead us to under-estimate the amount of gas present in given environment, and therefore artificially inflate D/H. If this were to preferentially happen over a specific $\Sigma_{H}$ regime, then what ought to be a plateau in D/H could instead incorrectly manifest as a bump. If such a bump happened at the same surface density regime where D/H evolution transitioned from growth to plateau (see Section~\ref{Subsection:DtH_Modeling}), then we would observe a spurious turnover in D/H.

\needspace{2\baselineskip} \subsubsection{Dark Gas in the LMC?} \label{AppendixSubsubsection:Dark_Gas_LMC}

First, we consider this possibility for the case of the LMC. The D/H for the LMC has a peak value of 0.0056, at $\Sigma_{H}^{\it (deproj)} = 40\,{\rm M_{\odot}\,pc^{-2}}$. Then, as density increases, D/H falls to an average of 0.0024 for $\Sigma_{H}^{\it (deproj)} \geq 10\,{\rm M_{\odot}\,pc^{-2}}$. If the true value of the D/H plateau is 0.0024, and the peak to D/H\,=\,0.0056 is being caused artificially by dark gas, this would require 57\% of the gas content at densities around $40\,{\rm M_{\odot}\,pc^{-2}}$ to be dark, not traced by 21\,cm or CO emission. 

In general, CO-dark molecular gas would be expected to be more prevalent in regions with the lowest molecular gas densities \citep{Wolfire2010A,Glover2010B}, while optically-thick \HI\ is most likely to occur in regions with higher atomic gas density \citep{Lee2015B}. Therefore the contribution of dark gas is likely to be greatest at intermediate densities, where these two regimes overlap -- which should broadly correspond to the density where \SigmaDeproj\ transitions from being atomic-gas dominated to molecular-gas dominated. 

As already discussed in Section~\ref{AppendixSubsection:Differences_in_alpha-CO}, this atomic--molecular transition region, at $\Sigma_{H}^{\it (deproj)} = 35\,{\rm M_{\odot}\,pc^{-2}}$ (see also \citealp{Roman-Duval2014D}), well matches the surface density regime where the D/H turnover starts for the LMC. At first glance, this is highly suggestive that dark gas could be contributing to the apparent turnover in LMC. 

However, dark gas cannot {\it depress} D/H measurements, only raise them. So, given that D/H falls to \textless\,$3 \times 10^{-3}$ at the highest measured LMC densities of $\Sigma_{H}^{\it (deproj)} = 200\,{\rm M_{\odot}\,pc^{-2}}$, this means that for dark gas to causing the turnover, all D/H measurements above $3 \times 10^{-3}$ would have to be artificially inflated by dark gas, and in fact should be no more than $3 \times 10^{-3}$. However, the bins with D/H greater than $3 \times 10^{-3}$ span a very wide range of density, 5--200$\,{\rm M_{\odot}\,pc^{-2}}$. If we have to assume D/H is in reality no more than $3 \times 10^{-3}$ over this entire range, then this would require D/H to be essentially flat over 1.6\,dex in density --  tantamount to saying that there is essentially no D/H evolution with density in the LMC. We rule out this possibility, especially because D/H estimates for the LMC determined via UV absorption line depletion measurements from \citet{Roman-Duval2021B} agree excellently with our own for densities of $2 < \Sigma_{H}^{\it (deproj)} < 40 {\rm M_{\odot}\,pc^{-2}}$; they find a factor $\approx$3 increase in D/H over this range, with a peak D/H of 0.0046, within 18\% our own. This lets us be confident that D/H is not flat over this range, and therefore the D/H turnover is not predominantly a dark gas artifact. We perform a detailed comparison with the \citet{Roman-Duval2021B} UV measurements of D/H in Section~\ref{Section:DtH_Reconcile}.

Even if the D/H turnover for the LMC cannot be explained by dark gas, it is nonetheless worth considering what other effect dark gas is having upon the LMC D/H evolution profile. It remains highly suggestive that D/H peaks at nearly the same \SigmaDeproj\ where atomic-to-molecular transition occurs in the LMC. Moreover, D/H shows a conspicuous `bump' at densities between 30--60\,${\rm M_{\odot}\,pc^{-2}}$. This bump occurs at a very similar range of surface densities as those where the simulations of \citet{Glover2011D} find CO-dark gas to be significant. At average visual extinction of $A_{V} < 1$\,mag, \citet{Glover2011D} find that H$_{2}$ makes up a negligible fraction of the total H mass; while at $A_{V} > 3$\,mag, reduced photodissociation means that effectively all gas-phase carbon is found in CO, causing the CO-to-H$_{2}$ conversion stabilize. Given the \citet{Glover2011D} conversion between column density and $A_{V}$, being $5.348 \times 10^{-22}\,\frac{Z}{\rm Z_{\odot}}\,{\rm mag\,cm^{2}}$, and given that $1\,{\rm M_{\odot}\,pc^{-2}} = 1.25 \times 10^{20}\,{\rm cm^{-2}}$, the $1 < A_{V} < 3$\,mag regime where \citet{Glover2011D} find CO-dark H$_{2}$ to be significant corresponds to a surface density range of 30--90\,${\rm M_{\odot}\,pc^{-2}}$ -- a close match to the D/H bump in the LMC. If this is indeed the origin of this bump feature, it suggests that approximately 30\% of the gas in the  30--60\,${\rm M_{\odot}\,pc^{-2}}$ density range is CO-dark.

So far we have not considered dark, optically-thick \HI\ here. However, the contribution of dark \HI\ would be expected to occur {\it below} the densities at which H$_{2}$ starts to dominate (and be traced by CO). Therefore dark \HI\ is not a likely explanation for the D/H turnover in the LMC, and it occurs at densities above those at which molecular gas begins to dominate. 

So, while dark gas cannot be the driver of the high-density D/H turnover in the LMC, it can make it appear even more conspicuous, by elevating D/H at these intermediate densities.

\needspace{2\baselineskip} \subsubsection{Dark Gas in M\,31 and M\,33?} \label{AppendixSubsubsection:Dark_Gas_M31_M33}

Next we consider whether dark gas could be causing a spurious D/H turnover in the case of M\,31 and M\,33. Because the striking similarity in how D/H evolves with \SigmaDeproj\ for both galaixes, including the turnover, we focus this analysis on M\,31, as it has more data, over a wider range of densities -- under the expectation that explanations for our observations of M\,31 will also be valid for M\,33.

The peak D/H for M\,31 is 0.01, at $\Sigma_{H}^{\it (deproj)} = 4\,{\rm M_{\odot}\,pc^{-2}}$. Then D/H falls to an average of 0.0067 for $\Sigma_{H}^{\it (deproj)} > 10\,{\rm M_{\odot}\,pc^{-2}}$. If the true value of the D/H plateau is 0.0067, and the peak to D/H\,=\,0.01 is  being caused artificially by dark gas, this would require 33\% of the gas content at densities around $4\,{\rm M_{\odot}\,pc^{-2}}$ to be dark, not traced by 21\,cm or CO emission. This is fairly plausible.  A dark gas fraction of 33\% falls well within the range of estimates proposed by various authors \citep{Grenier2005D,Braun2009A,Abdo2010D,Planck2011XIX,Paradis2012B}. Additionally, the turnover for M\,31 happens at the density where the gas is transitioning from atomic- to molecular-gas dominated, the regime where dark gas is most likely to have an effect\footnote{Unfortunately, our poorer spatial resolution for M\,31 (and M\,33) means we cannot distinguish the highest density regimes at $\Sigma_{H}^{\it (deproj)} > 15\,{\rm M_{\odot}\,pc^{-2}}$, preventing us from performing the detailed \SigmaDeproj-to-$A_{V}$ comparison to the \citet{Glover2011D} CO-dark H$_{2}$ models that we could test for the LMC.}. 

On the other hand, studies focused on M\,31 have suggested that it hosts a very minimal CO-dark molecular gas component \citep{MWLSmith2012B,Athikkat-Eknath2021A}. Additionally, because dark gas is expected to be more common in lower-metallicity galaxies \citep{Genzel2012A,Madden2020A}, any dark gas artefact should be {\it more} prominent for the LMC than for M\,31 --  whereas in Section~\ref{AppendixSubsubsection:Dark_Gas_LMC} above, we have just established that the influence of dark gas in the LMC appears most likely limited to a range of intermediate densities, and not driving the D/H turnover at higher densities.

As such, we are unable to say with confidence whether dark gas is a major driver of the turnover in M\,31 and M\,33, but it seems a plausible explanation.

\needspace{2\baselineskip} \subsubsection{Dark Gas in the SMC?} \label{AppendixSubsubsection:Dark_Gas_SMC}

Although D/H does not turn over for the SMC, we nonetheless briefly consider what effect dark gas could be having on our measurements for this system, too.

In particular, we note that the relationship between D/H and density in the SMC gets abruptly steeper at $\Sigma_{H}^{\it (deproj)} \approx 50\,{\rm M_{\odot}\,pc^{-2}}$. Once again considering the \citet{Glover2011D} simulations of CO-dark gas in different environments, we note that $\Sigma_{H}^{\it (deproj)} = 50\,{\rm M_{\odot}\,pc^{-2}}$ corresponds to $A_{V} = 0.85$\,mag, once again very close to point at which they predict CO-dark H$_{2}$ to become significant. This also matches closely with the density at which \citet{Roman-Duval2014D} find the atomic-to-molecular transition begins, at $80\,{\rm M_{\odot}\,pc^{-2}}$. 

If the steepening in D/H at $> 50\,{\rm M_{\odot}\,pc^{-2}}$ in the SMC is indeed an artifact of CO-dark gas, then it is conceivable that the true evolution of D/H at these higher densities in fact follows the shallower trend found at $< 50\,{\rm M_{\odot}\,pc^{-2}}$. If this is the case, then that suggests that \textgreater\,60\% of the gas at $\Sigma_{H}^{\it (deproj)} > 100\,{\rm M_{\odot}\,pc^{-2}}$ is dark. If the \citet{Glover2011D} simulations are correct, then we would expect the $A_{V} > 3$\,mag regime (where CO reliably traces H$_{2}$) to start at 150--200\,${\rm M_{\odot}\,pc^{-2}}$, above the densities we are able to probe with our spatial resolution. Nonetheless, the effect of dark gas could even be masking the beginning of a D/H plateau at the highest densities we can sample.

\needspace{2\baselineskip} \subsection{Varying Dust Mass Opacity?} \label{AppendixSubsection:Varying_Kappa}

\begin{figure*}
\centering
\includegraphics[width=0.32\textwidth]{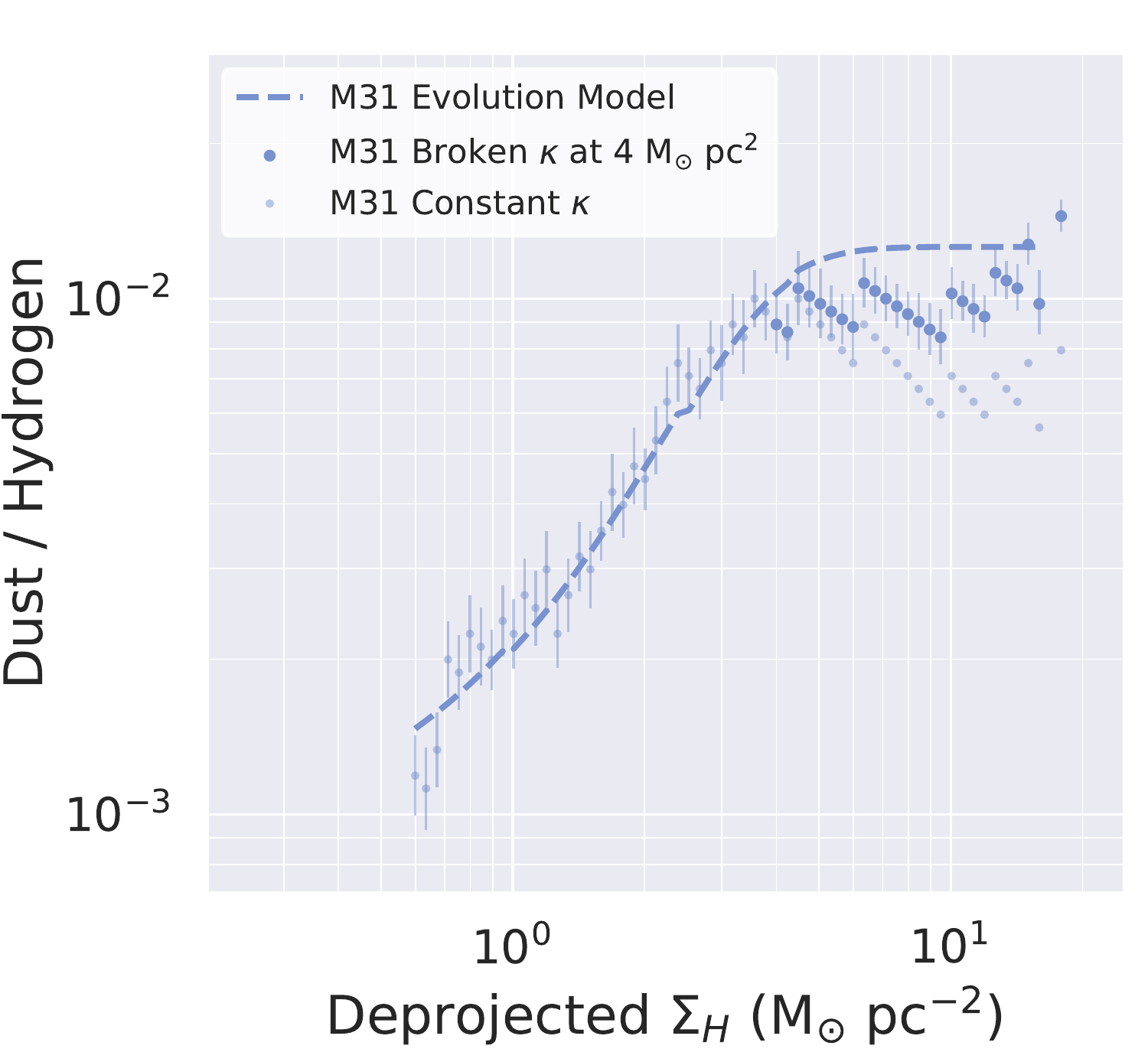}
\includegraphics[width=0.32\textwidth]{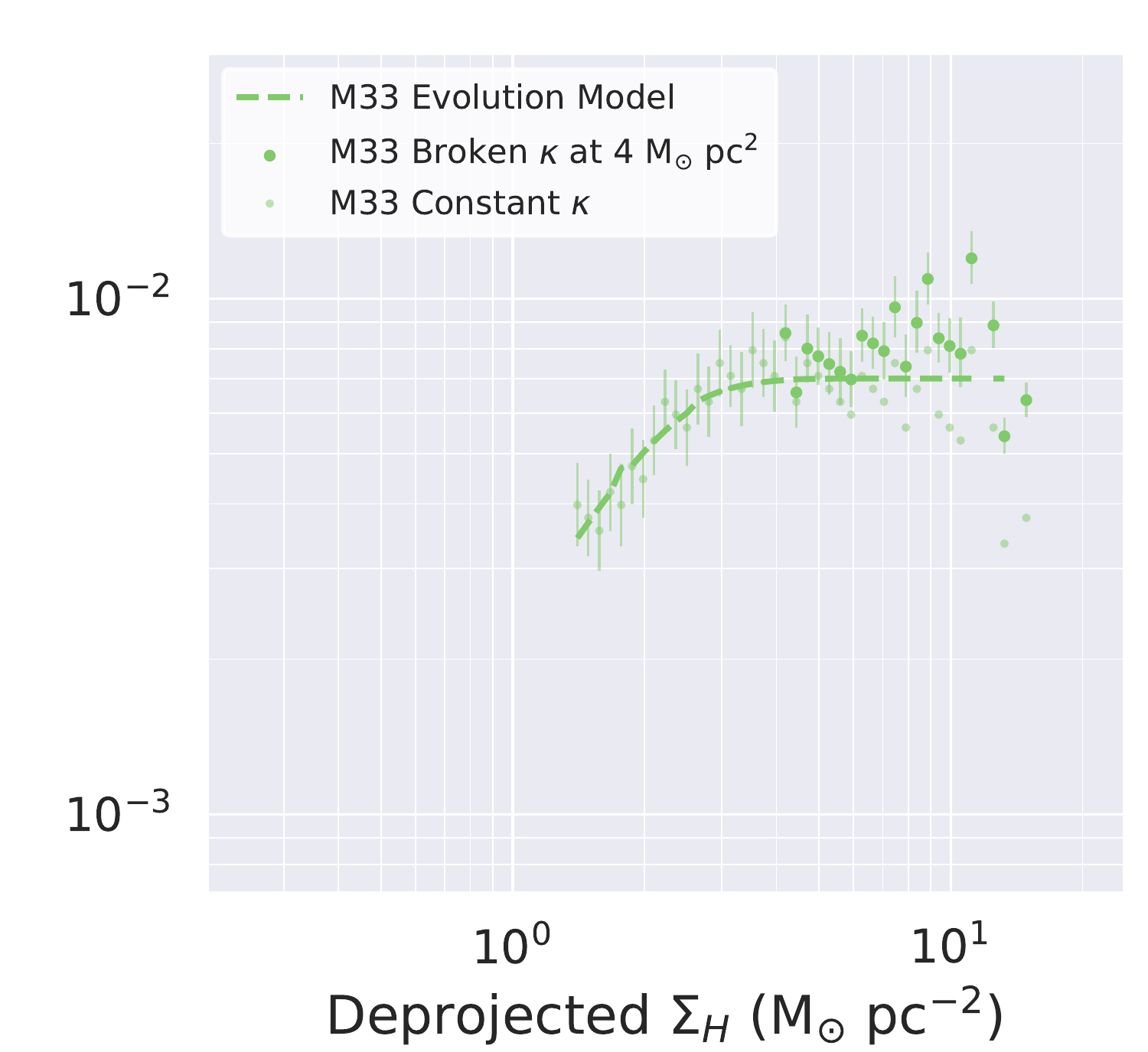}
\includegraphics[width=0.32\textwidth]{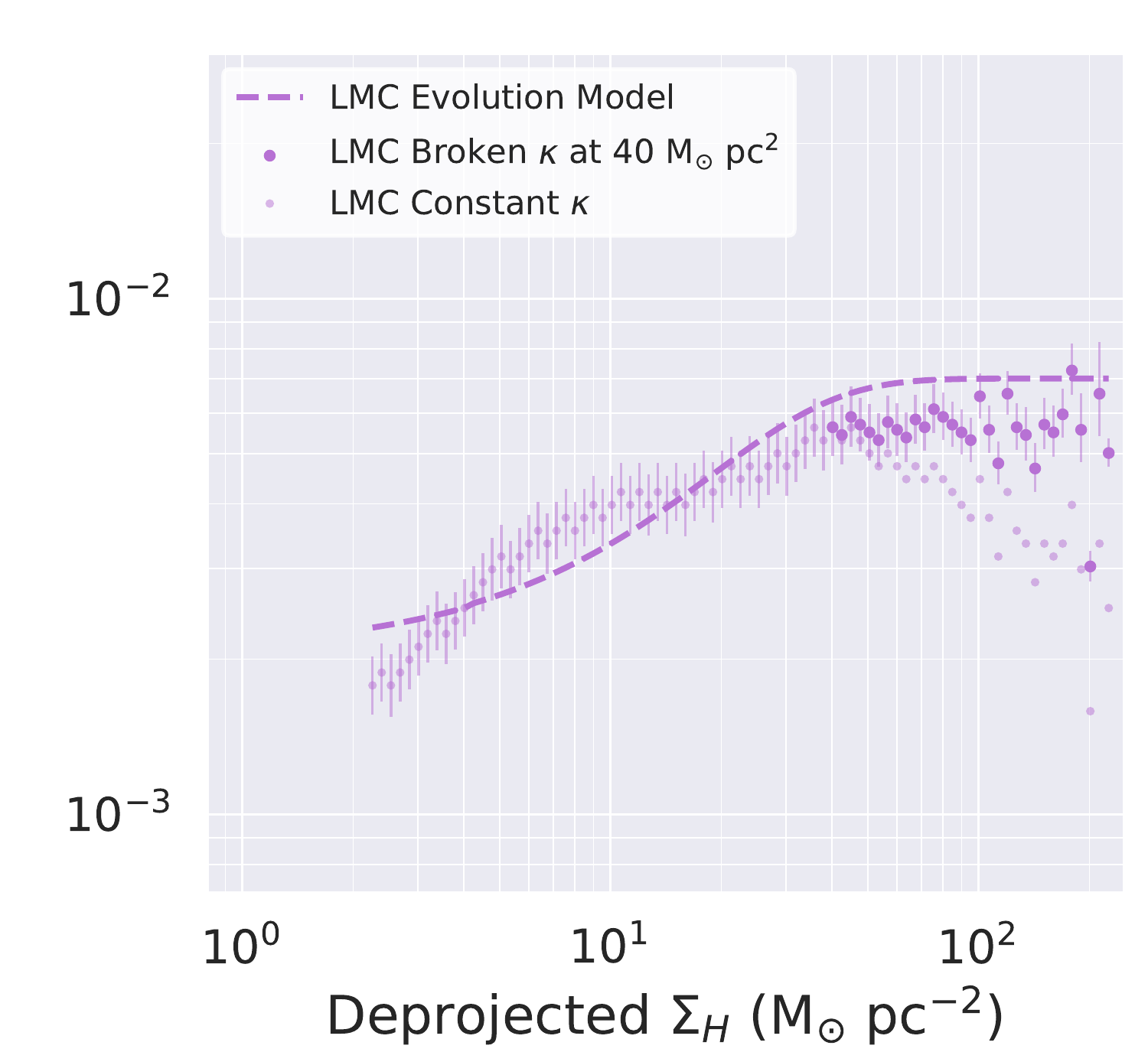}
\caption{Plots of D/H against $\Sigma_{H}$, for M\,31 ({\it left}), M\,33 ({\it center}), and the LMC ({\it right}), where we examine the effect of decreasing the dust mass absorption coefficient, $\kappa$, above a certain break density, according to a power law slope of $-0.4$, as per \citet{CJRClark2019B}.}
\label{AppendixFig:DtH_vs_H_Kappa}
\end{figure*}

The apparent turnover in D/H can be explained if we are systematically underestimating the dust mass at higher densities. This could occur if the intrinsic luminosity of the dust changes at different densities -- ie, if the mass mass absorption coefficient, $\kappa$, varies as a function of \SigmaDeproj. 

It is essentially guaranteed that $\kappa$ varies with density. We know that the composition of dust varies considerably with density, with different elements depleting from the gas phase onto dust grains at different rates in different environments \citep{Jenkins2009B,Jenkins2017A,Roman-Duval2021B,Roman-Duval2022A}. Similarly, the dust grain-size distribution is believed to evolve with density \citep{Hirashita2013F,Aoyama2020A}, as is the structure of the grains, as they accrete layers and coagulate in different phases of the ISM \citep{Cuppen2007A,Jones2016A,Jones2017A,Jones2018A}. In light of all of this, it is near-inconceivable that $\kappa$ wouldn't evolve with density to some degree. 

For our SED fitting, we assumed a constant value of $\kappa$. If, however, $\kappa$ were to decrease at higher \SigmaDeproj, then this would lead to us underestimate D/H at those higher densities, and could cause a plateau in D/H to instead manifest as a turnover. 

Is it plausible that $\kappa$ decreases in higher-density ISM? The exact nature of how $\kappa$ evolves with density is mostly unconstrained at present. However, models consistently predict that $\kappa$ should {\it increase} in higher-density ISM, due to grains developing a `fluffy' structure as they coagulate \citep{Ossenkopf1994B,Li2003C,Jones2018A}. On the other hand, \citet{CJRClark2019B} used an empirical method to construct resolved maps of $\kappa$ within nearby spiral galaxies M\,74 and M\,83, and found evidence that $\kappa$ {\it decreases} with increasing $\Sigma_{H}$. Relatedly, \citet{Bianchi2019A} and Bianchi et al. ({\it subm.}) find that higher ISM surface densities and higher molecular-to-atomic gas ratios are associated with greater dust surface brightness per unit gas surface density (resolved within 9 nearby spiral galaxies), and with greater dust luminosity per gas mass (for integrated measurements of 204 late-type galaxies), also suggesting dust becomes less emissive at higher ISM densities.

Specifically, \citet{CJRClark2019B} found that $\kappa$ falls with $\Sigma_{H}$ according to a power law index of $-0.4$. Because M\,73 and M\,83 were relatively poorly resolved in that study (590 and 330\,pc\ resolution respectively, compared to 14--147\,pc in this work), and because those galaxies are highly star-forming (with correspondingly high gas surface densities) the lowest $\Sigma_{H}$ to which \citet{CJRClark2019B} could probe in M\,74 and M\,83 was $3\,{\rm M_{\odot}\,pc^{-2}}$ (with the highest well-sampled $\Sigma_{H}$ being $303\,{\rm M_{\odot}\,pc^{-2}}$). The much greater distances to M\,74 and M\,83 (as compared to the Local Group) also make it likely that temperature mixing will be influencing the trends observed in those galaxies \citep{Priestley2020B}. Nonetheless, the $\Sigma_{H}$ range for which \citet{CJRClark2019B} report a fall in $\kappa$ with $\Sigma_{H}$ corresponds to the surface densities above which we start to see the D/H turnovers in our target galaxies.

We examined whether a $\kappa$ that falls with increasing density, as suggested by \citet{CJRClark2019B}, could explain our apparent D/H turnover. Specifically, we tried a broken $\kappa$ model, where $\kappa$ remains constant at $\kappa_{160} = 1.24\,{\rm m^{2}\,kg^{-1}}$ (see Section~\ref{Section:SED_Fitting}) up until a break density, after which $\kappa$ falls with $\Sigma_{H}$ according to a $-0.4$ power law, as per \citet{CJRClark2019B}. For the location of the $\kappa$ break in each galaxy, we used the surface density at which D/H peaks: $\Sigma_{H} = 4\,{\rm M_{\odot}\,pc^{-2}}$ for M\,31 and M\,33, and $\Sigma_{H} = 40\,{\rm M_{\odot}\,pc^{-2}}$ for the LMC. The results of this are plotted in Figure~\ref{AppendixFig:DtH_vs_H_Kappa}.

In Figure~\ref{AppendixFig:DtH_vs_H_Kappa}, we see that introducing this broken $\kappa$ model does an excellent job of resolving the D/H turnover for M\,31, M\,33, and the LMC. In each case, the D/H evolution profile at higher densities now conforms well to the plateau we would expect, based on dust evolution modeling (see Section~\ref{Subsection:DtH_Modeling}). We therefore think it is certainly plausible that decreasing in $\kappa$ at higher $\Sigma_{H}$ could be the cause of the D/H turnover. It is particularly interesting that the $-0.4$ power law rate of decrease of $\kappa$ versus $\Sigma_{H}$, as found in \citet{CJRClark2019B}, is also the rate of decrease required to resolve the turnover in our D/H evolution profiles. That said, the fact that physical dust grain models consistently predict the opposite behavior -- that $\kappa$ is expected to increase with density -- should count against this hypothesis. Nonetheless the fact that this explanation works so well makes us consider it very plausible\footnote{Given that the lead author of this work was also lead author of the \citet{CJRClark2019B} study, they feel it worth mentioning that they did not enter into this current study with any intention to search for anti-correlation between $\kappa$ and $\Sigma_{H}$, nor did they have any expectation that they would find evidence for it. Rather, the fact that the \citet{CJRClark2019B} result resolves the D/H turnover found here came as a surprise.}.

Why the $\Sigma_{H}$ at which the D/H turnover happens -- and hence the $\Sigma_{H}$ at which $\kappa$ breaks -- would be an order of magnitude greater in the LMC than in M\,31 and M\,33 is unclear, as is the lack of break in the SMC. We tested changing the break density for the LMC to $\Sigma_{H} = 4\,{\rm M_{\odot}\,pc^{-2}}$, to match M\,31 and M\,33, but this pushes the D/H plateau for the LMC to a level of D/H\,\textgreater\,0.015, even higher than the peak D/H in M\,31, in excess of the metals available in the LMC for forming dust, so this is an unlikely scenario. As discussed in Section~\ref{AppendixSubsection:Differences_in_alpha-CO}, the different turnover surface densities do correspond to the point at which the ISM transitions from atomic- to molecular-gas dominated in these galaxies, which happens in LMC at a $\Sigma_{H}$ that is an order of magnitude higher than in M\,31 and M\,33. So there are processes in the ISM of these galaxies happening at the different densities at which their turnovers happen.

Separately, in Section~\ref{Section:DtH_Reconcile}, we find evidence for a {\it constant offset} in $\kappa$ in the SMC, relative to the LMC. While separate from any variation in $\kappa$ with density, this nonetheless further highlights the fact that we really should not expect $\kappa$ to remain constant throughout.



\end{document}